\documentclass[journal]{IEEEtran}
\IEEEoverridecommandlockouts
\usepackage{lineno}
\usepackage{hyperref}
\usepackage{cite}
\usepackage{amsmath,amssymb,amsfonts}
\usepackage{amsmath}
\usepackage{amsthm}

\usepackage{graphicx}
\usepackage{textcomp}
\usepackage{xcolor}
\usepackage{graphicx}
\usepackage{float}
\usepackage{subfigure}
\usepackage{amsmath}
\usepackage{amsfonts,amssymb}
\usepackage{mathrsfs}
\usepackage{mathtools}
\usepackage{algorithm}
\usepackage{algorithmicx}
\usepackage{algpseudocode}
\usepackage{bm}
\usepackage{multirow}
\usepackage{array}
\usepackage{amssymb}
\usepackage{amsmath}
\usepackage{cite}
\usepackage{url}
\usepackage{xcolor}
\usepackage{cite,graphicx,amsmath,amssymb}
\usepackage{subfigure}
\usepackage{fancyhdr}
\usepackage{mdwmath}
\usepackage{mdwtab}
\usepackage{caption}
\usepackage{amsthm}
\usepackage{setspace}
\usepackage{bm}
\usepackage{algorithm}
\usepackage{algpseudocode}
\usepackage{mathtools}
\usepackage{dsfont}
\usepackage{bbm}
\newtheorem{remark}{Remark}
\newtheorem{theorem}{Theorem}

\newtheorem{lemma}{Lemma}

\newtheorem{corollary}{Corollary}

\makeatletter
\newcommand{\biggg}{\bBigg@{3}}
\newcommand{\Biggg}{\bBigg@{3.5}}
\makeatother
\def\BibTeX{{\rm B\kern-.05em{\sc i\kern-.025em b}\kern-.08em
    T\kern-.1667em\lower.7ex\hbox{E}\kern-.125emX}}
\expandafter\def\expandafter\normalsize\expandafter{%
    \normalsize%
    \setlength\abovedisplayskip{4pt}%
    \setlength\belowdisplayskip{4pt}%
    \setlength\abovedisplayshortskip{2pt}%
    \setlength\belowdisplayshortskip{2pt}%
}
\begin{document}
\title{Diversity and Multiplexing for Continuous-Aperture Array (CAPA)-Based Communications}
\author{Chongjun Ouyang, Zhaolin Wang, Xingqi Zhang, and Yuanwei Liu\vspace{-10pt}
\thanks{C. Ouyang and Z. Wang are with the School of Electronic Engineering and Computer Science, Queen Mary University of London, London, E1 4NS, U.K. (email: \{c.ouyang, zhaolin.wang\}@qmul.ac.uk).}
\thanks{X. Zhang is with Department of Electrical and Computer Engineering, University of Alberta, Edmonton AB, T6G 2R3, Canada (email: xingqi.zhang@ualberta.ca).}
\thanks{Y. Liu is with the Department of Electrical and Electronic Engineering, The University of Hong Kong, Hong Kong (email: yuanwei@hku.hk).}}
\maketitle
\begin{abstract}
A general fading model for multipath channels between two non-parallel continuous-aperture arrays (CAPAs) is proposed. Building on this model, the performance of diversity and multiplexing achieved by CAPAs over fading channels is analyzed. \romannumeral1) For multiple-input single-output (MISO) and single-input multiple-output (SIMO) channels, Landau's eigenvalue theorem is applied to analyze the autocorrelation of the spatial response. Closed-form expressions are derived for the outage probability (OP) and ergodic channel capacity (ECC). Asymptotic analyses in the high signal-to-noise ratio (SNR) regime are conducted to reveal the maximal achievable diversity and multiplexing gains. The diversity-multiplexing trade-off (DMT) is characterized, along with the array gain within the DMT framework. \romannumeral2) For multiple-input multiple-output (MIMO) channels, a wavenumber-domain-based transmission framework is proposed to leverage the spatial degrees of freedom offered by CAPAs. Asymptotic approximations for the OP and ECC are derived, and the DMT is explored. The performance of CAPAs is further compared with that of conventional spatially-discrete arrays (SPDAs). Analytical and numerical results demonstrate that: \romannumeral1) CAPAs achieve a lower OP and higher ECC than SPDAs; \romannumeral2) CAPAs achieve the same DMT as SPDAs with antenna spacing no larger than half a wavelength while attaining a higher array gain; and \romannumeral3) CAPAs outperform SPDAs with antenna spacing greater than half a wavelength in terms of DMT.
\end{abstract} 
\section{Introduction}
Multiple-antenna technology is a cornerstone in the evolution of modern cellular networks. At its core lies the principle of utilizing an increased number of antenna elements to enhance spatial degrees of freedom (DoFs) and boost channel capacity. The number of spatial DoFs in a multiple-antenna system is inherently limited by the number of antennas it incorporates. To expand the available DoFs, integrating more antennas into a confined space has emerged as an effective approach. This trend toward larger aperture sizes and denser antenna deployments is reflected in cutting-edge array architectures, such as reconfigurable intelligent surfaces \cite{liu2021reconfigurable}, holographic multiple-input multiple-output (MIMO) \cite{pizzo2020spatially}, dynamic metasurface antennas \cite{shlezinger2021dynamic}, among others.

The ultimate evolution of existing multiple-antenna systems is expected to manifest as a spatially-continuous electromagnetic (EM) aperture. This concept, referred to as a \emph{continuous-aperture array (CAPA)}, features an uncountable infinity of antennas separated by infinitesimal distances. CAPAs signify a paradigm shift from conventional spatially-discrete arrays (SPDAs) and represent a significant leap forward in maximizing spatial DoFs. This advancement unlocks substantial gains in system performance. The transition from discrete to continuous-aperture arrays enables CAPAs to enhance signal resolution and provide more flexible control over EM wave propagation \cite{ouyang2024primer}. This innovative approach paves the way for next-generation antenna technologies, promising to revolutionize wireless communication infrastructures with their advanced spatial processing capabilities \cite{liu2024capa}.
\subsection{Prior Works}
Recent years have seen growing research interest in the design and analysis of CAPA-based wireless communications. For example, the authors in \cite{sanguinetti2022wavenumber} proposed a wavenumber-division multiplexing framework to enable multi-stream data transmission between two linear CAPAs. This framework was later extended to downlink and uplink CAPA-based multiuser channels \cite{zhang2023pattern,qian2024spectral}, where corresponding beamforming policies were developed. A calculus of variations-based approach was introduced in \cite{wang2024beamforming,wang2024optimal} to optimize downlink CAPA beamforming with reduced complexity compared to \cite{zhang2023pattern,qian2024spectral}. In addition to these initial efforts, researchers have explored the integration of CAPAs with emerging technologies, including wireless energy transfer \cite{huang2024holographic}, deep learning \cite{guo2024multi}, and integrated sensing and communications \cite{zhao2025downlink}, where beamforming is customized for CAPAs in each scenario.

Beyond beamforming design, researchers have also analyzed the fundamental performance limits of CAPA-based wireless systems. For instance, the authors in \cite{ouyang2024impact} analyzed the array gain achieved by a CAPA. In \cite{gruber2008new}, the Shannon information capacity of space-time wireless channels formed by a pair of CAPAs was studied. This work was extended to account for non-white EM interference in \cite{wan2023mutual}. Further extensions to the capacity region and channel capacity of CAPA-based multiuser uplink and downlink channels are presented in \cite{zhao2024continuous,ouyang2024performance}. In addition to information-theoretic capacity limits, the analysis of the number of spatial DoFs in CAPA-based MIMO channels has attracted significant research attention \cite{poon2005degrees,pizzo2022landau,pizzo2022spatial,pizzo2022nyquist}. While the aforementioned studies primarily focus on communications, the authors in \cite{chen2024near} analyzed the Ziv-Zakai bound for CAPA-based positioning. For more recent advances in the field of CAPAs, readers are encouraged to refer to the comprehensive surveys \cite{liu2025near,gong2024holographic} and tutorials \cite{liu2023near-field,ouyang2024primer}.
\subsection{Motivations and Contributions}
Despite significant progress in CAPA-related research, most existing studies have focused on line-of-sight (LoS) channels. While LoS propagation simplifies theoretical investigations into fundamental performance limits, practical wireless communication systems must account for multipath scattering and small-scale fading. In this context, it is essential to analyze the performance of CAPA-based communications in multipath fading environments. For example, the works in \cite{pizzo2022landau} analyzed the spatial DoFs of CAPA-based multipath channels. Beyond spatial DoFs, the works in \cite{smith2024continuous,smith2024performance,zhang2023fundamental} derived approximations for the ergodic channel capacity (ECC) and outage probability (OP) achieved by continuous or semi-continuous arrays to measure the spectral efficiency (SE). However, these studies did not explore system design insights in depth. Specifically, the trade-off between diversity gain, multiplexing gain, and array gain---defined based on ECC and OP in the high signal-to-noise ratio (SNR) regime---remains underexplored. To date, a comprehensive investigation into these insights for CAPA-based fading channels is still lacking.

To provide deeper insight into CAPA’s performance in spatial multiplexing and diversity, this paper analyzes the ECC and OP of CAPA-based \emph{isotropic fading} channels. Our main contributions are summarized as follows.
\begin{itemize}
  \item We extend existing fading models for CAPA-based MIMO channels from a parallel setup to an arbitrarily non-parallel configuration. Using this model, we apply \emph{Landau's eigenvalue theorem} to prove that the eigenvalues of the autocorrelation function for CAPA-based multiple-input single-output (MISO) and single-input multiple-output (SIMO) channels exhibit a step-like behavior, with the number of significant eigenvalues determined by the aperture size and the wavelength. We then closed-form expressions for ECC and OP in MISO/SIMO channels. To provide further insights, we perform asymptotic analyses in the high-SNR regime to characterize the maximal multiplexing gain, maximal diversity gain, and the diversity-multiplexing trade-off (DMT) achievable by CAPAs.      
  \item We then analyze the eigenvalues of the autocorrelation function of CAPA-based MIMO fading channels using \emph{Landau's eigenvalue theorem}. We prove that the eigenvalues exhibit a step-like behavior, where the number of significant eigenvalues depends on the product of the transmit and receive aperture sizes. To fully exploit the spatial DoFs provided by these significant eigenvalues, we use the linear combination of two Fourier bases \cite{sanguinetti2022wavenumber,pizzo2022fourier} to approximate the spatial channel response and propose a wavenumber-domain transmission framework to modulate the data information into source currents. Using this framework, we derive high-SNR asymptotic expressions for the ECC and OP in CAPA-based MIMO channels and characterize the DMT and array gain.
  \item We further compare the performance of conventional SPDAs and CAPAs in terms of array gain and DMT. We prove that CAPAs achieve a larger array gain than SPDAs. In terms of DMT, we show that while an SPDA with half-wavelength antenna spacing achieves the same DMT as a CAPA, an SPDA with larger antenna spacing results in a worse DMT. This discrepancy arises because SPDAs with antenna spacing greater than half-wavelength cannot fully capture wavenumber-domain information.
  \item We present computer simulation results to verify the accuracy of our derived results and to evaluate the performance of the wavenumber-domain transmission framework for CAPA-based MIMO fading channels. The numerical results demonstrate that: 1) CAPAs achieve a lower OP and higher ECC than SPDAs across all SNR ranges; 2) CAPAs attain the same DMT as SPDAs with half-wavelength and sub-half-wavelength antenna spacing while offering a larger array gain; and 3) CAPAs outperform SPDAs with antenna spacing larger than half a wavelength in terms of DMT. These findings highlight the superiority of CAPAs over conventional SPDAs in SE performance.
\end{itemize}
\subsection{Organization and Notations}
The remainder of this paper is organized as follows. Section \ref{Section: System Model} introduces the system model for CAPAs. Section \ref{Section: Channel Modeling} describes the fading channel model between two non-parallel CAPAs. In Sections \ref{Section: MISO/SIMO Channels} and \ref{Section: MIMO Channels}, we analyze the ECC and OP of CAPA-based MISO/SIMO and MIMO channels to investigate the diversity gain, multiplexing gain, and array gain. Section \ref{Section_SPDA} compares the performance of SPDAs and CAPAs. Section \ref{Section: Numerical Results} presents numerical results to validate the accuracy of the derived findings. Finally, Section \ref{Section: Conclusion} concludes the paper.
\subsubsection*{Notations}
Throughout this paper, scalars, vectors, and matrices are denoted by non-bold, bold lower-case, and bold upper-case letters, respectively. For a matrix $\mathbf{A}$, $[\mathbf{A}]_{i,j}$, ${\mathbf{A}}^{\mathsf{T}}$, ${\mathbf{A}}^{*}$, and ${\mathbf{A}}^{\mathsf{H}}$ denote the $(i,j)$th entry, transpose, conjugate, and conjugate transpose of $\mathbf{A}$, respectively. For a square matrix $\mathbf{B}$, ${\mathbf{B}}^{-1}$, ${\mathbf{B}}^{\frac{1}{2}}$, $\mathsf{tr}({\mathbf{B}})$, and $\det(\mathbf{B})$ denote the inverse, principal square root, trace, and determinant of $\mathbf{B}$, respectively. The notations $\lvert a\rvert$ and $\lVert \mathbf{a} \rVert$ represent the magnitude of scalar $a$ and the norm of vector $\mathbf{a}$, respectively. The identity matrix of size $N\times N$ is denoted by $\mathbf{I}_N$, and the zero matrix is represented by $\mathbf{0}$. The sets $\mathbbmss{C}$, $\mathbbmss{R}$, and $\mathbbmss{Z}$ denote the complex, real, and integer spaces, respectively. For a set $\mathcal{X}$, $\mu(\mathcal{X})$ denotes the Lebesgue measure\footnote{For subsets within Euclidean $n$-spaces of lower dimensions---specifically for $n = 1$, $2$, or $3$---the Lebesgue measure coincides with the standard measure of length, area, or volume, respectively.}, and $\lvert\mathcal{X}\rvert$ gives its cardinality. The flooring operator is denoted by $\lfloor\cdot\rfloor$, $\overset{d}{=}$ denotes equivalence in distribution, and $\mathbbmss{E}\{\cdot\}$ represents the mathematical expectation. The Dirac delta function and the Kronecker delta are denoted by $\delta(\cdot)$ and $\delta_{i,j}$, respectively. Finally, ${\rm{Exp}}(\lambda)$ denotes the exponential distribution with rate parameter $\lambda$, and ${\mathcal{CN}}({\bm\mu},\mathbf{X})$ denotes the circularly symmetric complex Gaussian distribution with mean $\bm\mu$ and covariance matrix $\mathbf{X}$.

\begin{figure}[!t]
 \centering
\setlength{\abovecaptionskip}{0pt}
\includegraphics[height=0.22\textwidth]{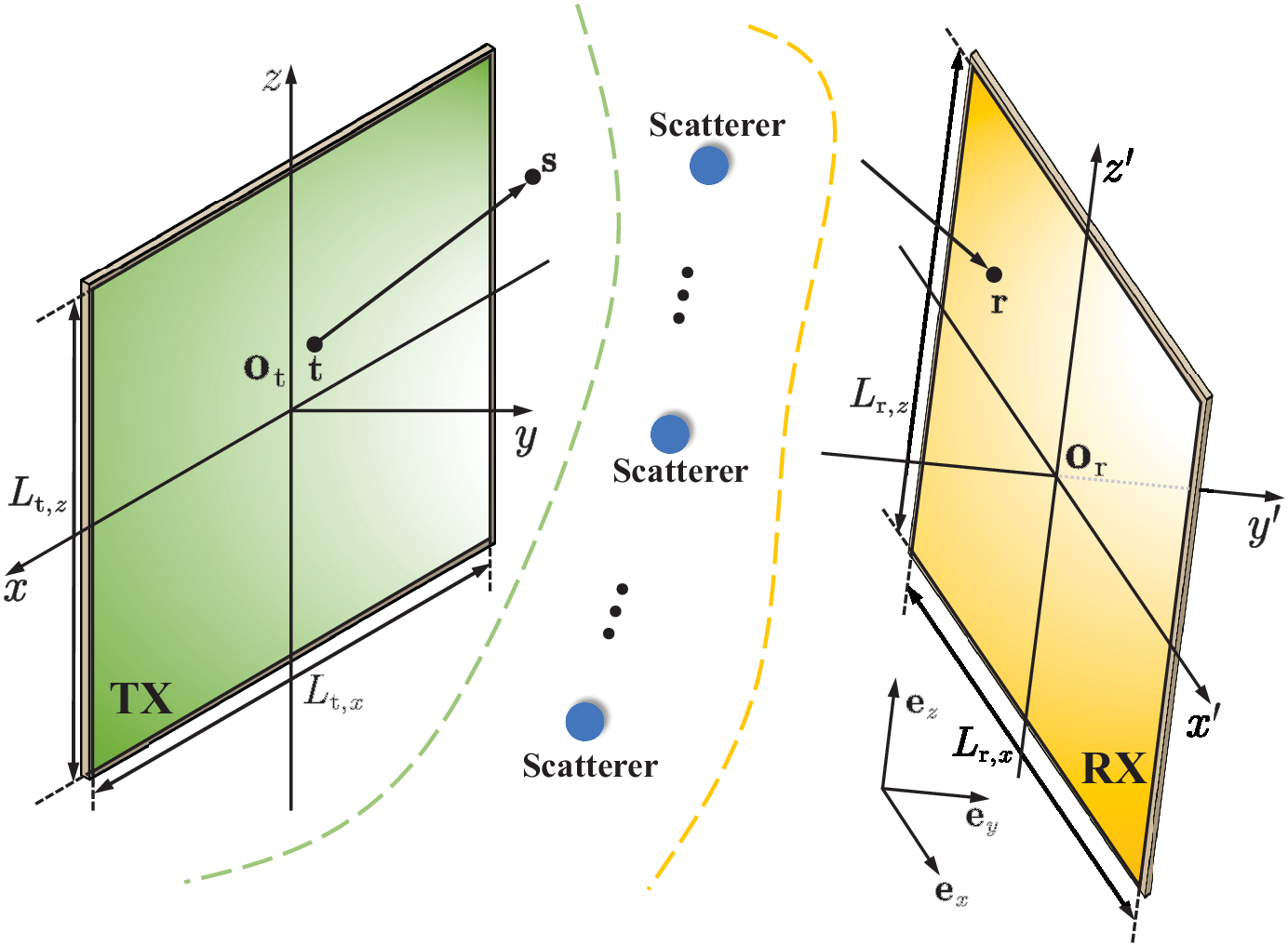}
\caption{Illustration of a CAPA-based channel.}
\label{Figure: System_Model}
\vspace{-15pt}
\end{figure}

\section{System Model}\label{Section: System Model}
We consider a point-to-point wireless communication system where both the transmitter (TX) and the receiver (RX) are equipped with a uni-polarized planar CAPA, as illustrated in {\figurename} {\ref{Figure: System_Model}}. The TX array is positioned on the $x$-$z$ plane and centered at the origin ${\mathbf{o}}_{\mathsf{t}}=[0,0,0]^{\mathsf{T}}$, with physical dimensions $L_{{\mathsf{t}},x}$ and $L_{{\mathsf{t}},z}$ along the $x$- and $z$-axes, respectively. Unlike existing studies \cite{pizzo2022fourier}, we assume a scenario where the RX array is not necessarily parallel to the TX array. 

The RX CAPA is centered at ${\mathbf{o}}_{\mathsf{r}}=[u_{x},u_{y},u_{z}]^{\mathsf{T}}\in{\mathbbmss{R}}^{3\times1}$ and has a physical aperture of $L_{{\mathsf{r}},x}\times L_{{\mathsf{r}},z}$. As shown in {\figurename} {\ref{Figure: System_Model}}, the principal axes of the RX CAPA are denoted by $\mathbf{e}_{x}\in{\mathbbmss{R}}^{3\times1}$ and $\mathbf{e}_{z}\in{\mathbbmss{R}}^{3\times1}$, and its orientation is aligned along $\mathbf{e}_{y}\in{\mathbbmss{R}}^{3\times1}$. The edges of the RX CAPA are parallel to $\mathbf{e}_{x}$ and $\mathbf{e}_{z}$, while the array itself is perpendicular to $\mathbf{e}_{y}$. Notably, $\{\mathbf{e}_{x},\mathbf{e}_{y},\mathbf{e}_{z}\}$ forms an orthonormal basis in ${\mathbbmss{R}}^{3\times1}$, satisfying ${\mathbf{E}}^{\mathsf{T}}{\mathbf{E}}={\mathbf{E}}{\mathbf{E}}^{\mathsf{T}}={\mathbf{I}}_3$, where ${\mathbf{E}}\triangleq [\mathbf{e}_{x},\mathbf{e}_{y},\mathbf{e}_{z}]\in{\mathbbmss{R}}^{3\times3}$. Using this basis, a three-dimensional (3D) $x'y'z'$ Cartesian coordinate system can be constructed, with its origin at ${\mathbf{o}}_{\mathsf{r}}$ and its axes $x'$-, $y'$-, and $z'$ aligned with $\mathbf{e}_{x}$, $\mathbf{e}_{y}$, and $\mathbf{e}_{z}$, respectively, as shown in {\figurename} {\ref{Figure: System_Model}}. The matrix ${\mathbf{E}}$ serves as a rotation matrix that transforms a point’s coordinates from the $x'y'z'$ system to the $xyz$ system. Specifically, let ${\mathbf{r}}'=[r_x',r_y',r_z']^{\mathsf{T}}\in{\mathbbmss{R}}^{3\times1}$ represent a point's coordinates in the $x'y'z'$ system. Then, its coordinates in the $xyz$ system are given by
\begin{align}\label{Transformation_Coordinate}
{\mathbf{r}}=\left[\begin{smallmatrix}r_x\\r_y\\r_z\end{smallmatrix}\right]={\mathbf{o}}_{\mathsf{r}}+{\mathbf{E}}{\mathbf{r}}'
=\left[\begin{smallmatrix}u_x\\u_y\\u_z\end{smallmatrix}\right]+[\mathbf{e}_{x},\mathbf{e}_{y},\mathbf{e}_{z}]
\left[\begin{smallmatrix}r_x'\\r_y'\\r_z'\end{smallmatrix}\right].
\end{align}  

For clarity, we denote the apertures of the TX and RX arrays as ${\mathcal{A}}_{\mathsf{t}}=\{[x,0,z]^{\mathsf{T}}|x\in[-\frac{L_{{\mathsf{t}},x}}{2},\frac{L_{{\mathsf{t}},x}}{2}],z\in[-\frac{L_{{\mathsf{t}},z}}{2},\frac{L_{{\mathsf{t}},z}}{2}]\}$ and ${\mathcal{A}}_{\mathsf{r}}=\{[x,0,z]^{\mathsf{T}}|x\in[-\frac{L_{{\mathsf{r}},x}}{2},\frac{L_{{\mathsf{r}},x}}{2}],z\in[-\frac{L_{{\mathsf{r}},z}}{2},\frac{L_{{\mathsf{r}},z}}{2}]\}$, respectively. Without loss of generality, we assume that the physical dimensions of each CAPA are integer multiples of the wavelength $\lambda$, unless specified otherwise. That is, $L_{{\mathsf{t}},x}=N_{{\mathsf{t}},x}\lambda$, $L_{{\mathsf{t}},z}=N_{{\mathsf{t}},z}\lambda$, $L_{{\mathsf{t}},x}=N_{{\mathsf{t}},x}\lambda$, and $L_{{\mathsf{t}},z}=N_{{\mathsf{t}},z}\lambda$, where $N_{{\mathsf{t}},x},N_{{\mathsf{t}},z},N_{{\mathsf{r}},x},N_{{\mathsf{r}},z}\in{\mathbbmss{Z}}$.

We now turn to the signal model for the considered CAPA system. Specifically, in a frequency-nonselective fading channel, the transmit and receive signals at a particular time are related by \cite{poon2005degrees,pizzo2022spatial}
\begin{align}\label{CAPA_Basic_Signal_Model_No_Trans}
y(\mathbf{r})=\int_{{\mathcal{A}}_{\mathsf{t}}}h({\mathbf{r}},{\mathbf{t}})x({\mathbf{t}}){\rm{d}}{\mathbf{t}}+z(\mathbf{r}).
\end{align}
The transmit signal $x(\cdot)$ delivered by the TX is a scalar field on ${\mathbbmss{R}}^{3\times1}$, which assigns each point ${\mathbf{t}}=[t_x,t_y,t_z]^{\mathsf{T}}\in{\mathbbmss{R}}^{3\times1}$ of the TX CAPA to $x({\mathbf{t}})$. Similarly, $y(\cdot)$ is the receive scalar field. The \emph{spatial channel response} $h(\cdot,\cdot)$ is a complex integral kernel whose domain is the set of transmit scalar fields and whose range is the set of receive scalar fields. The spatial response $h({\mathbf{r}},{\mathbf{t}})$ gives the channel gain between the transmit position ${\mathbf{t}}\in{\mathcal{A}}_{\mathsf{t}}$ and the receive position ${\mathbf{r}}={\mathbf{o}}_{\mathsf{r}}+{\mathbf{E}}{\mathbf{r}}'$ with ${\mathbf{r}}'\in{\mathcal{A}}_{\mathsf{r}}$. Additionally, $z(\mathbf{r})$ accounts for thermal noise. The noise field $z(\mathbf{r})$ is modeled as a Gaussian random process with ${\mathbbmss{E}}\{z(\mathbf{r})z^{*}(\mathbf{r}')\}=\sigma^2\delta({\mathbf{r}}-{\mathbf{r}'})$ and $z(\mathbf{r})\sim{\mathcal{CN}}(0,\sigma^2)$, where $\sigma^2$ represents the noise strength. Inserting \eqref{Transformation_Coordinate} into \eqref{CAPA_Basic_Signal_Model_No_Trans} gives
\begin{align}\label{CAPA_Basic_Signal_Model}
{\mathsf{y}}(\mathbf{r}')=\int_{{\mathcal{A}}_{\mathsf{t}}}{\mathsf{h}}({\mathbf{r}}',{\mathbf{t}})x({\mathbf{t}}){\rm{d}}{\mathbf{t}}+{\mathsf{z}}(\mathbf{r}'),
\end{align}
where ${\mathsf{y}}(\mathbf{r}')\triangleq y({\mathbf{o}}_{\mathsf{r}}+{\mathbf{E}}{\mathbf{r}}')$, ${\mathsf{h}}({\mathbf{r}}',{\mathbf{t}})\triangleq h({\mathbf{o}}_{\mathsf{r}}+{\mathbf{E}}{\mathbf{r}}',{\mathbf{t}})$, and ${\mathsf{z}}(\mathbf{r}')\triangleq z({\mathbf{o}}_{\mathsf{r}}+{\mathbf{E}}{\mathbf{r}}')$.

\section{Channel Modeling}\label{Section: Channel Modeling}
We assume that no direct path exists between the TX and RX arrays due to the presence of scatterers with arbitrary shapes and sizes. In the subsequent analysis, we adopt the methodology outlined in \cite{pizzo2022spatial} to characterize this multipath scattering environment. This study focuses on the effects of multipath fading on CAPAs while excluding EM mutual coupling (MC). Investigating MC remains an important direction for future research, with preliminary results available in \cite{pizzo2025mutual}.
\subsection{Transmitted Field}
\begin{figure}[!t]
\centering
    \subfigbottomskip=0pt
	\subfigcapskip=-5pt
\setlength{\abovecaptionskip}{0pt}
    \subfigure[$xyz$.]
    {
        \includegraphics[height=0.15\textwidth]{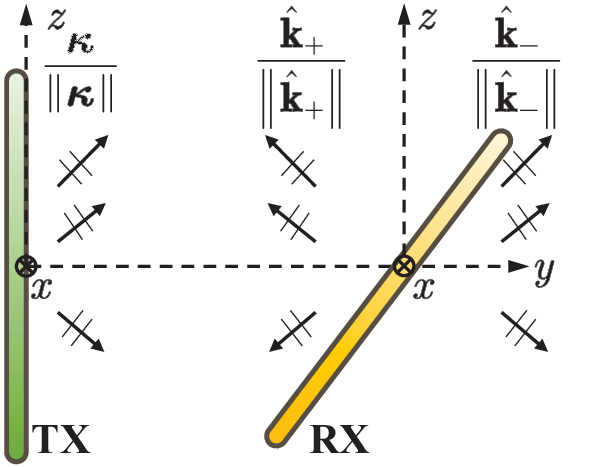}
	   \label{fig_direction1}	
    }
    \subfigure[$xyz$-$x'y'z'$.]
    {
        \includegraphics[height=0.15\textwidth]{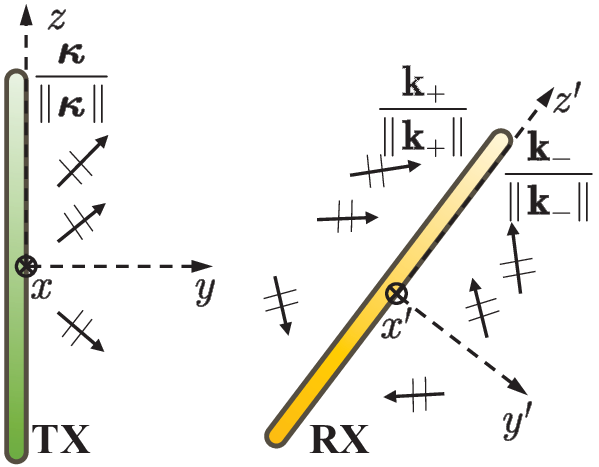}
	   \label{fig_direction2}	
    }
\caption{A side view of the CAPA-based channel.}
\label{Figure_direction}
\vspace{-15pt}
\end{figure}

First, we evaluate the \emph{transmitted field} excited by the source at an intermediate point ${\mathbf{s}}=[s_x,s_y,s_z]^{\mathsf{T}}\in{\mathbbmss{R}}^{3\times1}$, placed before any interaction with scatterers occurs ($s_y>t_y$), as shown in {\figurename} {\ref{Figure: System_Model}}. This transmitted field is expressed as follows:
\begin{align}\label{Transmit_Field_Basic}
e_{\mathsf{t}}({\mathbf{s}})=\int_{{\mathcal{A}}_{\mathsf{t}}}h_{\mathsf{LoS}}({\mathbf{s}},{\mathbf{t}})x({\mathbf{t}}){\rm{d}}{\mathbf{t}},
\end{align}  
where $h_{\mathsf{LoS}}({\mathbf{s}},{\mathbf{t}})$ represents the free-space EM propagation from $\mathbf{t}$ to $\mathbf{s}$. Specifically, $h_{\mathsf{LoS}}({\mathbf{s}},{\mathbf{t}})$ is given by \cite{ouyang2024primer,ouyang2024impact}
\begin{align}\label{Green_Function}
h_{\mathsf{LoS}}({\mathbf{s}},{\mathbf{t}})=\frac{{\rm{j}}k_0\eta{{\rm{e}}^{-{\rm{j}}k_0\lVert{\mathbf{s}}-{\mathbf{t}}\rVert}}}{4\pi\lVert{\mathbf{s}}-{\mathbf{t}}\rVert},
\end{align}
and equivalently, using Weyl's identity \cite{weyl1919ausbreitung}:
\begin{align}\label{Green_Function_Sub2}
h_{\mathsf{LoS}}({\mathbf{s}},{\mathbf{t}})=\frac{k_0\eta}{8\pi^2}\int_{-\infty}^{+\infty}\int_{-\infty}^{+\infty}\frac{{\rm{e}}^{-{\rm{j}}{\hat{\bm{\kappa}}}^{\mathsf{T}}({\mathbf{s}}-{\mathbf{t}})}}{\hat{\gamma}(\kappa_x,\kappa_z)}{\rm{d}}\kappa_x{\rm{d}}\kappa_z,
\end{align}
where $k_0=\frac{2\pi}{\lambda}$ is the wavenumber, $\lambda$ denotes the wavelength, $\eta=120\pi$ (in ohms, [$\Omega$]) is the impedance of free space, and ${\hat{\bm{\kappa}}}=[\kappa_x,{\hat{\gamma}}(\kappa_x,\kappa_z),\kappa_z]^{\mathsf{T}}\in{\mathbbmss{C}}^{3\times1}$. The function ${\hat{\gamma}}(\kappa_x,\kappa_z)$ is defined as follows:
\begin{align}
{\hat{\gamma}}(\kappa_x,\kappa_z)=\left\{\begin{matrix}
\sqrt{k_0^2-\kappa_x^2-\kappa_z^2}&\kappa_x^2+\kappa_z^2\leq k_0^2\\
-{\rm{j}}\sqrt{\kappa_x^2+\kappa_z^2-k_0^2}&\kappa_x^2+\kappa_z^2> k_0^2
\end{matrix}\right..
\end{align}
Equation \eqref{Green_Function_Sub2} can be interpreted as an integral summation of plane waves propagating in all directions $\frac{\hat{\bm\kappa}}{k_0}$, including evanescent waves. For $\kappa_x^2+\kappa_z^2\leq k_0^2$, the waves propagate along the radiation direction $\frac{{\bm{\kappa}}}{\lVert{\bm{\kappa}}\rVert}\in{\mathbbmss{R}}^{3\times1}$, with ${{\bm{\kappa}}}=[\kappa_x,{{\gamma}}(\kappa_x,\kappa_z),\kappa_z]^{\mathsf{T}}\in{\mathbbmss{R}}^{3\times1}$, ${\gamma}(\kappa_x,\kappa_z)\triangleq\sqrt{k_0^2-\kappa_x^2-\kappa_z^2}$, and $\lVert{{\bm{\kappa}}}\rVert=k_0$, as shown in {\figurename} {\ref{fig_direction1}}. For $\kappa_x^2+\kappa_z^2> k_0^2$, the waves are evanescent, with their effects confined to regions near the sources (within a few wavelengths) \cite{pizzo2022spatial}. Therefore, in the sequel, we neglect the evanescent-wave components and substitute the simplified \eqref{Green_Function_Sub2} into \eqref{Transmit_Field_Basic}, which yields
\begin{align}\label{Transmit_Field_Basic_Fourier}
e_{\mathsf{t}}({\mathbf{s}})=\iint_{{\mathcal{D}}({\bm\kappa})}\frac{{\rm{d}}\kappa_x}{2\pi}\frac{{\rm{d}}\kappa_z}{2\pi}
{\rm{e}}^{-{\rm{j}}{\bm{\kappa}}^{\mathsf{T}}{\mathbf{s}}}{E}_{\mathsf{t}}(\kappa_x,\kappa_z),
\end{align}  
where ${\mathcal{D}}({\bm\kappa})=\{(\kappa_x,\kappa_y)\in{\mathbbmss{R}}^2|\kappa_x^2+\kappa_y^2\leq k_0^2\}$ and
\begin{align}\label{Transmit_Field_Basic_Fourier_Final}
{E}_{\mathsf{t}}(\kappa_x,\kappa_z)\triangleq\frac{k_0\eta}{2}\int_{{\mathcal{A}}_{\mathsf{t}}}
\frac{{\rm{e}}^{{\rm{j}}{\bm{\kappa}}^{\mathsf{T}}{\mathbf{t}}}}{{\gamma}(\kappa_x,\kappa_z)}x({\mathbf{t}}){\rm{d}}{\mathbf{t}}.
\end{align}
The result in \eqref{Transmit_Field_Basic_Fourier} indicates that the transmitted field $e_{\mathsf{t}}({\mathbf{s}})$ can be viewed as the integral summation of plane waves ${\rm{e}}^{-{\rm{j}}{\bm{\kappa}}^{\mathsf{T}}{\mathbf{s}}}{E}_{\mathsf{t}}(\kappa_x,\kappa_z)$ propagating in the direction $\frac{{\bm{\kappa}}}{\lVert{\bm{\kappa}}\rVert}$, each with a complex-valued amplitude ${E}_{\mathsf{t}}(\kappa_x,\kappa_z)$. 

\subsection{Received Field}
While the \emph{transmitted field} in \eqref{Transmit_Field_Basic} or \eqref{Transmit_Field_Basic_Fourier} is created by the source current, the \emph{received field} measured at $\mathbf{r}$, namely $y(\mathbf{r})$, is generated through the interaction of the transmitted field with scatterers. Similar to \eqref{Green_Function_Sub2}, the received field can be expressed as an integral summation of plane waves from all received directions \cite[Sec. 6.7]{stratton2007electromagnetic}:
\begin{equation}
\begin{split}
y({\mathbf{r}})&=\int_{-\infty}^{+\infty}\int_{-\infty}^{+\infty}\frac{{\rm{d}}\hat{k}_x}{2\pi}\frac{{\rm{d}}\hat{k}_z}{2\pi}
\left({\rm{e}}^{-{\rm{j}}{\hat{\mathbf{k}}}_{+}^{\mathsf{T}}{\mathbf{r}}}\hat{E}_{\mathsf{r}}^{+}(\hat{k}_x,\hat{k}_z)\right.\\
&+\left.{\rm{e}}^{-{\rm{j}}{\hat{\mathbf{k}}}_{-}^{\mathsf{T}}{\mathbf{r}}}\hat{E}_{\mathsf{r}}^{-}(\hat{k}_x,\hat{k}_z)\right),
\end{split}
\end{equation}
where each plane wave has an amplitude $\hat{E}_{\mathsf{r}}^{\pm}(\hat{k}_x,\hat{k}_z)$ for each received direction $\frac{{\hat{\mathbf{k}}}_{\pm}}{\lVert{\hat{\mathbf{k}}}_{\pm}\rVert}=\frac{1}{k_0}[\hat{k}_x,\pm{\hat{\gamma}}(\hat{k}_x,\hat{k}_z),\hat{k}_z]^{\mathsf{T}}\in{\mathbbmss{C}}^{3\times1}$. By neglecting the evanescent-wave components, $y({\mathbf{r}})$ simplifies to the follows:
\begin{align}\label{Received_Field_Simplified}
y({\mathbf{r}})=\iint_{{\mathcal{D}}(\hat{\mathbf{k}})}\frac{{\rm{d}}\hat{k}_x}{2\pi}\frac{{\rm{d}}\hat{k}_z}{2\pi}
{\rm{e}}^{-{\rm{j}}{\hat{\mathbf{k}}}_{\pm}^{\mathsf{T}}{\mathbf{r}}}\hat{E}_{\mathsf{r}}^{\pm}(\hat{k}_x,\hat{k}_z),
\end{align}
where ${\mathcal{D}}(\hat{\mathbf{k}})=\{(\hat{k}_x,\hat{k}_y)\in{\mathbbmss{R}}^2|\hat{k}_x^2+\hat{k}_y^2\leq k_0^2\}$, and ${\rm{e}}^{-{\rm{j}}{\hat{\mathbf{k}}}_{\pm}^{\mathsf{T}}{\mathbf{r}}}\hat{E}_{\mathsf{r}}^{\pm}(\hat{k}_x,\hat{k}_z)$ is shorthand for ${\rm{e}}^{-{\rm{j}}{\hat{\mathbf{k}}}_{+}^{\mathsf{T}}{\mathbf{r}}}\hat{E}_{\mathsf{r}}^{+}(\hat{k}_x,\hat{k}_z)
+{\rm{e}}^{-{\rm{j}}{\hat{\mathbf{k}}}_{-}^{\mathsf{T}}{\mathbf{r}}}\hat{E}_{\mathsf{r}}^{-}(\hat{k}_x,\hat{k}_z)$. The plane waves ${\rm{e}}^{-{\rm{j}}{\hat{\mathbf{k}}}_{+}^{\mathsf{T}}{\mathbf{r}}}$ and ${\rm{e}}^{-{\rm{j}}{\hat{\mathbf{k}}}_{-}^{\mathsf{T}}{\mathbf{r}}}$ correspond to directions within the two sides of the $x$-$z$ plane, as depicted in {\figurename} {\ref{fig_direction1}}. 

To simplify the following analysis, we transform the coordinate system used in \eqref{Received_Field_Simplified} from the $xyz$ system to the $x'y'z'$ system as follows:
\begin{align}
{\hat{\mathbf{k}}}_{\pm}={\mathbf{E}}{{\mathbf{k}}}_{\pm}\Leftrightarrow{\mathbf{E}}^{\mathsf{T}}{\hat{\mathbf{k}}}_{\pm}={{\mathbf{k}}}_{\pm},
\end{align}
where ${{\mathbf{k}}}_{\pm}=[{k}_x,\pm{{\gamma}}({k}_x,{k}_z),{k}_z]^{\mathsf{T}}\in{\mathbbmss{R}}^{3\times1}$. By the definition of ${\mathbf{E}}$ given in Section \ref{Section: System Model}, $\frac{{{\mathbf{k}}}_{\pm}}{\lVert{{\mathbf{k}}}_{\pm}\rVert}$ represents the coordinates of $\frac{{\hat{\mathbf{k}}}_{\pm}}{\lVert{\hat{\mathbf{k}}}_{\pm}\rVert}$ in the $x'y'z'$ system. After applying this invertible variable transformation to the integral in \eqref{Received_Field_Simplified}, it can equivalently be rewritten as follows: 
\begin{align}\label{Received_Field_Simplified_Further}
y({\mathbf{r}})=\iint_{{\mathcal{D}}({\mathbf{k}})}\frac{{\rm{d}}{k}_x}{2\pi}\frac{{\rm{d}}{k}_z}{2\pi}
{\rm{e}}^{-{\rm{j}}{{\mathbf{k}}}_{\pm}^{\mathsf{T}}{\mathbf{E}}^{\mathsf{T}}{\mathbf{r}}}{E}_{\mathsf{r}}^{\pm}({k}_x,{k}_z),
\end{align}
where ${\mathcal{D}}({\mathbf{k}})=\{({k}_x,{k}_y)\in{\mathbbmss{R}}^2|{k}_x^2+{k}_y^2\leq k_0^2\}$. Moreover, ${E}_{\mathsf{r}}^{\pm}({k}_x,{k}_z)$ are the equivalent transformations of $\hat{E}_{\mathsf{r}}^{\pm}({k}_x,{k}_z)$ in the $x'y'z'$ system, representing the amplitude of the plane wave from the received direction $\frac{{{\mathbf{k}}}_{\pm}}{\lVert{{\mathbf{k}}}_{\pm}\rVert}$. 

We comment that deriving a closed-form expression for ${E}_{\mathsf{r}}^{\pm}({k}_x,{k}_z)$ could be a challenging task, as it depends heavily on the transformation matrix $\mathbf{E}$. Fortunately, as will be shown later, a closed-form expression of ${E}_{\mathsf{r}}^{\pm}({k}_x,{k}_z)$ is not necessary for channel modeling, as the focus lies on how ${E}_{\mathsf{t}}(\kappa_x,\kappa_z)$ maps to ${E}_{\mathsf{r}}^{\pm}({k}_x,{k}_z)$.

The plane waves ${\rm{e}}^{-{\rm{j}}{{\mathbf{k}}}_{+}^{\mathsf{T}}{\mathbf{E}}^{\mathsf{T}}{\mathbf{r}}}$ and ${\rm{e}}^{-{\rm{j}}{{\mathbf{k}}}_{-}^{\mathsf{T}}{\mathbf{E}}^{\mathsf{T}}{\mathbf{r}}}$ propagate along the radiation directions within the two sides of the plane where the RX array is placed, as depicted in {\figurename} {\ref{fig_direction2}}. Since we focus only on the received field measured on the left-hand side of the RX array, \eqref{Received_Field_Simplified_Further} can be further simplified as follows:
\begin{align}\label{Received_Field_Simplified_Final}
y({\mathbf{r}})=\iint_{{\mathcal{D}}({\mathbf{k}})}\frac{{\rm{d}}{k}_x}{2\pi}\frac{{\rm{d}}{k}_z}{2\pi}
{\rm{e}}^{-{\rm{j}}{{\mathbf{k}}}^{\mathsf{T}}{\mathbf{E}}^{\mathsf{T}}{\mathbf{r}}}{E}_{\mathsf{r}}({k}_x,{k}_z),
\end{align}
where ${\mathbf{k}}={\mathbf{k}}_{+}$ and ${E}_{\mathsf{r}}({k}_x,{k}_z)={E}_{\mathsf{r}}^{+}({k}_x,{k}_z)$.
\subsection{Multipath Fading Model}
According to \cite{pizzo2022spatial}, the mapping from the \emph{transmit plane-wave spectrum} ${E}_{\mathsf{t}}(\kappa_x,\kappa_z)$ to the \emph{receive plane-wave spectrum} ${E}_{\mathsf{r}}({k}_x,{k}_z)$ follows a linear transformation:
\begin{align}\label{Linear_Mapping}
{E}_{\mathsf{r}}({k}_x,{k}_z)=\iint_{{\mathcal{D}}({\bm\kappa})}{E}_{\mathsf{t}}(\kappa_x,\kappa_z)\hat{\mathsf{H}}_a({\mathbf{k}},{\bm\kappa}){{\rm{d}}\kappa_x}{{\rm{d}}\kappa_z},
\end{align}
where $\hat{\mathsf{H}}_a({\mathbf{k}},{\bm\kappa})\in{\mathbbmss{C}}$ is the wavenumber-domain propagation kernel function that maps transmitted plane waves to received plane waves. Insereing \eqref{Transmit_Field_Basic_Fourier_Final} into \eqref{Linear_Mapping} gives
\begin{align}
{E}_{\mathsf{r}}({k}_x,{k}_z)=\iint_{{\mathcal{D}}({\bm\kappa})}\hat{\mathsf{H}}_a({\mathbf{k}},{\bm\kappa})
\frac{k_0\eta}{2}\int_{{\mathcal{A}}_{\mathsf{t}}}
\frac{{\rm{e}}^{{\rm{j}}{\bm{\kappa}}^{\mathsf{T}}{\mathbf{t}}}x({\mathbf{t}}){\rm{d}}{\mathbf{t}}}{{\gamma}(\kappa_x,\kappa_z)}{{\rm{d}}\kappa_x}{{\rm{d}}\kappa_z},\nonumber
\end{align}
which, together with \eqref{CAPA_Basic_Signal_Model_No_Trans} and \eqref{Received_Field_Simplified_Final}, results in
\begin{equation}\label{4FPWD_Model}
\begin{split}
h({\mathbf{r}},{\mathbf{t}})&=\frac{1}{(2\pi)^2}\frac{k_0\eta}{2}\iiiint_{{\mathcal{D}}({\bm\kappa})\times{\mathcal{D}}({\mathbf{k}})}
{\rm{e}}^{-{\rm{j}}{{\mathbf{k}}}^{\mathsf{T}}{\mathbf{E}}^{\mathsf{T}}{\mathbf{r}}}\\
&\times\hat{\mathsf{H}}_a({\mathbf{k}},{\bm\kappa}){\rm{e}}^{{\rm{j}}{\bm{\kappa}}^{\mathsf{T}}{\mathbf{t}}}{\rm{d}}k_x{\rm{d}}k_z{\rm{d}}\kappa_x{\rm{d}}\kappa_z.
\end{split}
\end{equation}
Substituting \eqref{Transformation_Coordinate} into \eqref{4FPWD_Model} and using the fact that ${\mathbf{E}}^{\mathsf{T}}{\mathbf{E}}={\mathbf{E}}{\mathbf{E}}^{\mathsf{T}}={\mathbf{I}}_3$, we obtain
\begin{equation}\label{4FPWD_Model_Final}
\begin{split}
&h({\mathbf{r}},{\mathbf{t}})=h({\mathbf{o}}_{\mathsf{r}}+{\mathbf{E}}{\mathbf{r}}',{\mathbf{t}})=\frac{1}{(2\pi)^2}\iiiint_{{\mathcal{D}}({\bm\kappa})\times{\mathcal{D}}({\mathbf{k}})}
\\
&\times{\rm{e}}^{-{\rm{j}}{{\mathbf{k}}}^{\mathsf{T}}{\mathbf{r}}'}{\mathsf{H}}_a({\mathbf{k}},{\bm\kappa}){\rm{e}}^{{\rm{j}}{\bm{\kappa}}^{\mathsf{T}}{\mathbf{t}}}{\rm{d}}k_x{\rm{d}}k_z{\rm{d}}\kappa_x{\rm{d}}\kappa_z
= {\mathsf{h}}({\mathbf{r}}',{\mathbf{t}}),
\end{split}
\end{equation}
where ${\mathsf{H}}_a({\mathbf{k}},{\bm\kappa})=\frac{k_0\eta}{2}\hat{\mathsf{H}}_a({\mathbf{k}},{\bm\kappa}){\rm{e}}^{-{\rm{j}}{{\mathbf{k}}}^{\mathsf{T}}{\mathbf{E}}^{\mathsf{T}}{\mathbf{o}}_{\mathsf{r}}}$ is the \emph{wavenumber-domain or angular-domain response} that maps every transmit direction $\frac{\bm\kappa}{\lVert{\bm\kappa}\rVert}$ to each received direction $\frac{\mathbf{k}}{\lVert\mathbf{k}\rVert}$. This response is modeled as a \emph{random process}. Using the fact that ${\mathbf{r}}'=[r_x',0,r_z']^{\mathsf{T}}$ and ${\mathbf{t}}=[t_x,0,t_z]^{\mathsf{T}}$, we rewrite \eqref{4FPWD_Model_Final} as follows:
\begin{equation}\label{4FPWD_Model_Standard}
\begin{split}
{\mathsf{h}}({\mathbf{r}}',{\mathbf{t}})&=\frac{1}{(2\pi)^2}\iiiint_{{\mathcal{D}}({\bm\kappa})\times{\mathcal{D}}({\mathbf{k}})}
{\rm{e}}^{-{\rm{j}}(r_x'k_x+r_z'k_z)}\\
&\times{\mathsf{H}}_a({\mathbf{k}},{\bm\kappa}){\rm{e}}^{{\rm{j}}(t_x\kappa_x+t_z\kappa_z)}{\rm{d}}k_x{\rm{d}}k_z{\rm{d}}\kappa_x{\rm{d}}\kappa_z.
\end{split}
\end{equation}

For analytical tractability, a Rayleigh fading model is considered, where ${\mathsf{H}}_a({\mathbf{k}},{\bm\kappa})$ is modeled as a stationary, circularly symmetric, complex-Gaussian random field. It holds that \cite{pizzo2022spatial}
\begin{align}\label{ZUCG_Gaussian_Random_Field_Origin}
{\mathsf{H}}_a({\mathbf{k}},{\bm\kappa})=S^{\frac{1}{2}}({\mathbf{k}},{\bm\kappa})W({\mathbf{k}},{\bm\kappa}),
\end{align}
where $S({\mathbf{k}},{\bm\kappa})\geq0$ describes the \emph{angular power distribution} of ${\mathsf{h}}({\mathbf{r}}',{\mathbf{t}})$, and $W({\mathbf{k}},{\bm\kappa})$ is modeled as a \emph{zero-mean unit-variance complex-Gaussian (ZUCG)} random field on ${\mathcal{D}}({\mathbf{k}})\times{\mathcal{D}}({\bm\kappa})$. It follows that $W({\mathbf{k}},{\bm\kappa})\sim{\mathcal{CN}}(0,1)$, and the autocorrelation is given by ${\mathbbmss{E}}\{W({\mathbf{k}},{\bm\kappa}){W}^{*}({\mathbf{k}}',{\bm\kappa}')\}=\delta([{\mathbf{k}};{\bm\kappa}]-[{\mathbf{k}}';{\bm\kappa}'])$. We further assume isotropic scattering, which leads to \cite{pizzo2020spatially}
\begin{align}\label{Angular_Domain_Power_Distribution_Isotropic_Scattering}
S({\mathbf{k}},{\bm\kappa})=\frac{A_{s}^2(k_0)}{\gamma(k_x,k_z)\gamma(\kappa_x,\kappa_z)},
\end{align}
where $A_{s}(k_0)$ is a function of $k_0$. The average channel power ${\mathbbmss{E}}\{\lvert {\mathsf{h}}({\mathbf{r}}',{\mathbf{t}}) \rvert^2\}$ is calculated as follows:
\begin{align}
{\mathbbmss{E}}\{\lvert {\mathsf{h}}({\mathbf{r}}',{\mathbf{t}}) \rvert^2\}&=
\iiiint_{{\mathcal{D}}({\bm\kappa})\times{\mathcal{D}}({\mathbf{k}})}\frac{S({\mathbf{k}},{\bm\kappa})}{(2\pi)^4}{\rm{d}}k_x{\rm{d}}k_z{\rm{d}}\kappa_x{\rm{d}}\kappa_z,\nonumber
\end{align}
which, together with the fact that $\iint_{{\mathcal{D}}({\bm\kappa})}\frac{{\rm{d}}\kappa_x{\rm{d}}\kappa_z}{\gamma(\kappa_x,\kappa_z)}=2\pi k_0$, yields ${\mathbbmss{E}}\{\lvert {\mathsf{h}}({\mathbf{r}}',{\mathbf{t}}) \rvert^2\}=\frac{A_{s}^2(k_0)}{(2\pi)^2}k_0^2$. To normalize the channel power, we set $A_{s}(k_0)=\frac{2\pi}{k_0}$.
\section{MISO/SIMO Channels}\label{Section: MISO/SIMO Channels}
After establishing the fading model, we proceed to analyze the performance of CAPA-based MISO/SIMO channels.
\subsection{Channel Capacity}
For MISO/SIMO channels, the transmit signal is expressed as $x({\mathbf{t}})=j({\mathbf{t}})s$, where $s\sim{\mathcal{CN}}(0,1)$ represents the normalized coded data symbol, and $j({\mathbf{t}})$ denotes the source current. Consequently, the signal observed at the RX CAPA can be written as follows:
\begin{align}
{\mathsf{y}}(\mathbf{r}')=s\int_{{\mathcal{A}}_{\mathsf{t}}}{\mathsf{h}}({\mathbf{r}}',{\mathbf{t}})j({\mathbf{t}}){\rm{d}}{\mathbf{t}}+{\mathsf{z}}(\mathbf{r}'),
\end{align}
where $j({\mathbf{t}})$ is subject to the power constraint $\int_{{\mathcal{A}}_{\mathsf{t}}}\lvert j({\mathbf{t}})\rvert^2{\rm{d}}{\mathbf{t}}=P$. The SNR for decoding $s$ is given by \cite{ouyang2024primer,zhao2024continuous}:
\begin{align}\label{SNR_MISO_SIMO_General}
\gamma
=\frac{1}{\sigma^2}\int_{{\mathcal{A}}_{\mathsf{r}}}\left\lvert\int_{{\mathcal{A}}_{\mathsf{t}}}{\mathsf{h}}({\mathbf{r}}',{\mathbf{t}})j({\mathbf{t}}){\rm{d}}{\mathbf{t}}\right\rvert^2{\rm{d}}{\mathbf{r}}'.
\end{align}
\subsubsection{MISO}
We first analyze the MISO case, where the RX aperture size is significantly smaller than both the TX aperture size and the propagation distance $r_y$. Under this condition, the variations in the channel response across the receive aperture are negligible, which yields
\begin{align}
{\mathsf{h}}({\mathbf{r}}',{\mathbf{t}})\approx {\mathsf{h}}([0,0,0]^{\mathsf{T}},{\mathbf{t}})\triangleq h_{\mathsf{r}}({\mathbf{t}}).
\end{align}
As a results, we can simplify \eqref{SNR_MISO_SIMO_General} as follows:
\begin{align}\label{SNR_MISO_General}
\gamma=\frac{\mu({\mathcal{A}}_{\mathsf{r}})}{\sigma^2}\left\lvert\int_{{\mathcal{A}}_{\mathsf{t}}}h_{\mathsf{r}}({\mathbf{t}})
j({\mathbf{t}}){\rm{d}}{\mathbf{t}}\right\rvert^2\triangleq \gamma_{\mathsf{r}},
\end{align}
which reaches its maximum when $j({\mathbf{t}})=\frac{\sqrt{P}h_{\mathsf{r}}^{*}({\mathbf{t}})}{({\int_{{\mathcal{A}}_{\mathsf{t}}}\lvert h_{\mathsf{r}}({\mathbf{t}})\rvert^2{\rm{d}}{\mathbf{t}}})^{1/2}}$. Thus, the SNR for MISO transmission is given by
\begin{align}\label{MISO_SNR_Definition_Standard}
\gamma_{\mathsf{r}}=\frac{\mu({\mathcal{A}}_{\mathsf{r}})P}{\sigma^2}\int_{{\mathcal{A}}_{\mathsf{t}}}\lvert h_{\mathsf{r}}({\mathbf{t}})\rvert^2{\rm{d}}{\mathbf{t}}.
\end{align}
\subsubsection{SIMO}
For SIMO channels, the variations in the channel response across the transmit aperture are negligible, yielding
\begin{subequations}
\begin{align}
{\mathsf{h}}({\mathbf{r}}',{\mathbf{t}})&\approx {\mathsf{h}}({\mathbf{r}}',[0,0,0]^{\mathsf{T}})\triangleq h_{\mathsf{t}}({\mathbf{r}}'),\\
j({\mathbf{t}})&\approx j([0,0,0]^{\mathsf{T}})\approx\sqrt{{P}/{\mu({\mathcal{A}}_{\mathsf{t}})}}.
\end{align}
\end{subequations}
The SNR for SIMO transmission is then given by
\begin{align}
\gamma_{\mathsf{t}}\triangleq\frac{\mu({\mathcal{A}}_{\mathsf{t}})P}{\sigma^2}\int_{{\mathcal{A}}_{\mathsf{r}}}\lvert h_{\mathsf{t}}({\mathbf{r}}')\rvert^2{\rm{d}}{\mathbf{r}}'.
\end{align}

The capacity for MISO/SIMO channels is given by
\begin{align}
{\mathsf{C}}_{\mathsf{c}}=\log_2(1+\gamma_{\mathsf{c}}),~{\mathsf{c}}\in\{{\mathsf{r}},{\mathsf{t}}\}.
\end{align}
Since the capacity of the SIMO channel has a similar form to that of the MISO channel, the subsequent analysis will focus on the MISO case and examine the statistics of $\gamma_{\mathsf{r}}$ and ${\mathsf{C}}_{\mathsf{r}}$.
\subsection{Channel Statistics}
\subsubsection{Autocorrelation}
Inserting ${\mathbf{r}}'=[0,0,0]^{\mathsf{T}}$ into \eqref{4FPWD_Model_Standard} gives
\begin{equation}\label{Channel_Response_MISO_Random_Field}
\begin{split}
h_{\mathsf{r}}({\mathbf{t}})=\frac{1}{(2\pi)^2}\iint_{{\mathcal{D}}({\bm\kappa})}
\hat{\mathsf{H}}_a({\bm\kappa}){\rm{e}}^{{\rm{j}}(t_x\kappa_x+t_z\kappa_z)}{\rm{d}}\kappa_x{\rm{d}}\kappa_z,
\end{split}
\end{equation}
where $\hat{\mathsf{H}}_a({\bm\kappa})\triangleq\iint_{{\mathcal{D}}({\mathbf{k}})}{\mathsf{H}}_a({\mathbf{k}},{\bm\kappa}){\rm{d}}k_x{\rm{d}}k_z$ is a zero-mean Gaussian random field defined over ${\mathcal{D}}({\bm\kappa})$. Consequently, $h_{\mathsf{r}}({\mathbf{t}})$ is a zero-mean Gaussian random field over ${\mathcal{A}}_{\mathsf{t}}$.
\vspace{-5pt}
\begin{lemma}\label{Lemma_Autocorrelation_General}
The autocorrelation function of $h_{\mathsf{r}}({\mathbf{t}})$, denoted as $R_{h_{\mathsf{r}}}({\mathbf{t}},{\mathbf{t}}')\triangleq{\mathbbmss{E}}\{h_{\mathsf{r}}({\mathbf{t}})h_{\mathsf{r}}^{*}({\mathbf{t}}')\}$, is given by
\begin{align}\label{Lemma_Autocorrelation_General_Result}
R_{h_{\mathsf{r}}}({\mathbf{t}},{\mathbf{t}}')=
\iint_{{\mathcal{D}}({\bm\kappa})}
\frac{{\rm{e}}^{{\rm{j}}((t_x-t_x')\kappa_x+(t_z-t_z')\kappa_z)}}{2\pi k_0\gamma(\kappa_x,\kappa_z)}{\rm{d}}\kappa_x{\rm{d}}\kappa_z,
\end{align}
where ${\mathbf{t}}'=[t_x',0,t_z']^{\mathsf{T}}\in{\mathcal{A}}_{\mathsf{t}}$.
\end{lemma}
\vspace{-5pt}
\begin{IEEEproof}
Please refer to Appendix \ref{Proof_Lemma_Autocorrelation_General} for more details.
\end{IEEEproof}
Next, we analyze a special case of \eqref{Channel_Response_MISO_Random_Field}, where the TX is equipped with a linear CAPA. By assuming $L_{{\mathsf{t}},z}\ll L_{{\mathsf{t}},x}$, the planar TX CAPA reduces to a linear configuration along the $x$-axis, which yields ${\mathbf{t}}\approx[t_x, 0, 0]^{\mathsf{T}}$ for ${\mathbf{t}}\in{\mathcal{A}}_{\mathsf{t}}$. As a result, 
\begin{align}\label{SNR_MISO_Linear_Array_Equal}
\gamma_{\mathsf{r}}=\frac{\mu({\mathcal{A}}_{\mathsf{r}})PL_{{\mathsf{t}},z}}{\sigma^2}\int_{-{L_{{\mathsf{t}},x}}/{2}}^{{L_{{\mathsf{t}},x}}/{2}}\lvert h_{{\mathsf{r}}_x}({{t}}_x)\rvert^2{\rm{d}}t_x\triangleq \gamma_{{\mathsf{r}}_x},
\end{align}
where $h_{{\mathsf{r}}_x}({{t}}_x)=h_{{\mathsf{r}}}([t_x, 0, 0]^{\mathsf{T}})$ can be expressed as follows:
\begin{equation}\label{Channel_Response_MISO_Random_Field_Linear}
\begin{split}
h_{{\mathsf{r}}_x}({{t}}_x)=\frac{1}{(2\pi)^2}\int_{-k_0}^{k_0}
\hat{\mathsf{H}}_{a_x}(\kappa_x){\rm{e}}^{{\rm{j}}t_x\kappa_x}{\rm{d}}\kappa_x,
\end{split}
\end{equation}
with $\hat{\mathsf{H}}_{a_x}(\kappa_x)\triangleq\int_{-\sqrt{k_0^2-\kappa_x^2}}^{\sqrt{k_0^2-\kappa_x^2}}\hat{\mathsf{H}}_a({\bm\kappa}){\rm{d}}\kappa_z$ forming a zero-mean Gaussian random field over $[-k_0,k_0]$. Consequently, $h_{{\mathsf{r}}_x}({{t}}_x)$ is a zero-mean Gaussian random field over ${\mathcal{A}}_{{\mathsf{t}}_x}\triangleq[-\frac{L_{{\mathsf{t}},x}}{2},\frac{L_{{\mathsf{t}},x}}{2}]$, whose autocorrelation is characterized as follows.
\vspace{-5pt}
\begin{corollary}\label{Corollary_Autocorrelation_General_Linear}
The autocorrelation function of $h_{{\mathsf{r}}_x}({{t}}_x)$, denoted as $R_{h_{{\mathsf{r}}_x}}(t_x,t_x')\triangleq{\mathbbmss{E}}\{h_{{\mathsf{r}}_x}({{t}}_x)h_{{\mathsf{r}}_x}^{*}({{t}}_x')\}$, is given by
\begin{align}\label{Corollary_Autocorrelation_General_Linear_Result}
R_{h_{{\mathsf{r}}_x}}(t_x,t_x')=\frac{1}{2k_0}
\int_{-k_0}^{k_0}
{{\rm{e}}^{{\rm{j}}(t_x-t_x')\kappa_x}}{\rm{d}}\kappa_x.
\end{align}
\end{corollary}
\vspace{-5pt}
\begin{IEEEproof}
Please refer to Appendix \ref{Proof_Lemma_Autocorrelation_General} for more details.
\end{IEEEproof}
\subsubsection{Linear CAPAs}
In this part, we aim to characterize the statistics of $\int_{{\mathcal{A}}_{{\mathsf{t}}_x}}\lvert h_{{\mathsf{r}}_x}({{t}}_x)\rvert^2{\rm{d}}t_x$. To this end, we denote the eigendecomposition (EVD) of the semipositive definite operator $R_{h_{{\mathsf{r}}_x}}(t_x,t_x')$ as follows: 
\begin{align}\label{EVD_Linear_Random_Operator}
R_{h_{{\mathsf{r}}_x}}(t_x,t_x')=\sum\nolimits_{\ell=1}^{\infty}\sigma_{{\mathsf{r}}_x,\ell}\phi_{{\mathsf{r}}_x,\ell}(t_x)\phi_{{\mathsf{r}}_x,\ell}^{*}(t_x'),
\end{align}
where $\sigma_{{\mathsf{r}}_x,1}\geq\sigma_{{\mathsf{r}}_x,2}\ldots\geq\sigma_{{\mathsf{r}}_x,\infty}\geq0$ are the eigenvalues of $R_{h_{{\mathsf{r}}_x}}(t_x,t_x')$, and $\{\phi_{{\mathsf{r}}_x,\ell}(\cdot)\}_{\ell=1}^{\infty}$ is the set of associated eigenfunctions. Note that $\{\phi_{{\mathsf{r}}_x,\ell}(t_x)\}_{\ell=1}^{\infty}$ forms an orthonormal basis over ${\mathcal{A}}_{{\mathsf{t}}_x}$, which satisfies 
\begin{align}\label{EVD_Linear_Random_Operator_Orthogonality}
\int_{{\mathcal{A}}_{{\mathsf{t}}_x}}\phi_{{\mathsf{r}}_x,\ell}^{*}(t_x)\phi_{{\mathsf{r}}_x,\ell'}(t_x){\rm{d}}t_x=\delta_{\ell,\ell'},
\end{align}
where $\delta_{\ell,\ell'}$ is the Kronecker delta. Based on \eqref{EVD_Linear_Random_Operator}, we conclude the following result.
\vspace{-5pt}
\begin{lemma}\label{Lemma_Linear_Random_Operator_Statistical_Equal}
The Gaussian random field $h_{{\mathsf{r}}_x}({{t}}_x)$ is statistically equivalent to the following representation:
\begin{align}\label{Linear_Random_Operator_Statistical_Equal_Result}
{\overline{h}}_{{\mathsf{r}}_x}({{t}}_x)=\sum_{\ell=1}^{\infty}
\int_{{\mathcal{A}}_{{\mathsf{t}}_x}}\sigma_{{\mathsf{r}}_x,\ell}^{\frac{1}{2}}\phi_{{\mathsf{r}}_x,\ell}(t_x)\phi_{{\mathsf{r}}_x,\ell}^{*}(t_x'){\overline{W}}(t_x'){\rm{d}}t_x',
\end{align}
where ${\overline{W}}(t_x')$ denotes a ZUCG random field over ${\mathcal{A}}_{{\mathsf{t}}_x}$.
\end{lemma}
\vspace{-5pt}
\begin{IEEEproof}
Please refer to Appendix \ref{Proof_Lemma_Linear_Random_Operator_Statistical_Equal} for more details.
\end{IEEEproof}
The results in Lemma \ref{Lemma_Linear_Random_Operator_Statistical_Equal} indicate that 
\begin{align}\label{Linear_Random_Operator_Statistical_Equal_SNR}
\gamma_{{\mathsf{r}}_x}\overset{d}{=}\frac{\mu({\mathcal{A}}_{\mathsf{r}})PL_{{\mathsf{t}},z}}{\sigma^2}\int_{{\mathcal{A}}_{{\mathsf{t}}_x}}\lvert {\overline{h}}_{{\mathsf{r}}_x}({{t}}_x)\rvert^2{\rm{d}}t_x.
\end{align}
In the following analysis, we aim to characterize the statistics of $\int_{{\mathcal{A}}_{{\mathsf{t}}_x}}\lvert {\overline{h}}_{{\mathsf{r}}_x}({{t}}_x)\rvert^2{\rm{d}}t_x$. Since ${\overline{W}}(t_x')$ is a ZUCG random field, $\int_{{\mathcal{A}}_{{\mathsf{t}}_x}}\phi_{{\mathsf{r}}_x,\ell}^{*}(t_x'){\overline{W}}(t_x'){\rm{d}}t_x'\triangleq \Phi_{{\mathsf{r}}_x,\ell}$ is a ZUCG random variable that satisfies the following condition.
\vspace{-5pt}
\begin{lemma}\label{Lemma_Linear_Random_Operator_Statistical_Equal_Cofficient}
The ZUCG random variables $\{\Phi_{{\mathsf{r}}_x,\ell}\}_{\ell=1}^{\infty}$ are independent and identically distributed (i.i.d.), with each following $\Phi_{{\mathsf{r}}_x,\ell}\sim{\mathcal{CN}}(0,1)$ ($\forall \ell$).
\end{lemma}
\vspace{-5pt}
\begin{IEEEproof}
Please refer to Appendix \ref{Proof_Lemma_Linear_Random_Operator_Statistical_Equal_Cofficient} for more details.
\end{IEEEproof}
By substituting $\Phi_{{\mathsf{r}}_x,\ell}=\int_{{\mathcal{A}}_{{\mathsf{t}}_x}}\phi_{{\mathsf{r}}_x,\ell}^{*}(t_x'){\overline{W}}(t_x'){\rm{d}}t_x'$ into \eqref{Linear_Random_Operator_Statistical_Equal_Result} and calculating $\int_{{\mathcal{A}}_{{\mathsf{t}}_x}}\lvert {\overline{h}}_{{\mathsf{r}}_x}({{t}}_x)\rvert^2{\rm{d}}t_x$, we obtain
\begin{equation}\label{Linear_Random_Operator_Equal_Channel_Gain_Initial_Initial} 
\begin{split}
&\int_{{\mathcal{A}}_{{\mathsf{t}}_x}}\lvert {\overline{h}}_{{\mathsf{r}}_x}({{t}}_x)\rvert^2{\rm{d}}t_x=
\sum\nolimits_{\ell_1=1}^{\infty}\sum\nolimits_{\ell_2=1}^{\infty}\sigma_{{\mathsf{r}}_x,\ell_1}^{\frac{1}{2}}\sigma_{{\mathsf{r}}_x,\ell_2}^{\frac{1}{2}}\\
&\times\Phi_{{\mathsf{r}}_x,\ell_1}^{*}\Phi_{{\mathsf{r}}_x,\ell_2}\int_{{\mathcal{A}}_{{\mathsf{t}}_x}}\phi_{{\mathsf{r}}_x,\ell_1}^{*}(t_x)\phi_{{\mathsf{r}}_x,\ell_2}(t_x){\rm{d}}t_x.
\end{split}
\end{equation}
Using \eqref{EVD_Linear_Random_Operator_Orthogonality}, \eqref{Linear_Random_Operator_Equal_Channel_Gain_Initial_Initial} simplifies to the follows:
\begin{align}\label{Linear_Random_Operator_Equal_Channel_Gain_Initial}
\int_{{\mathcal{A}}_{{\mathsf{t}}_x}}\lvert {\overline{h}}_{{\mathsf{r}}_x}({{t}}_x)\rvert^2{\rm{d}}t_x=\sum\nolimits_{\ell=1}^{\infty}
\sigma_{{\mathsf{r}}_x,\ell}\lvert\Phi_{{\mathsf{r}}_x,\ell}\rvert^2.
\end{align}
According to Lemma \ref{Lemma_Linear_Random_Operator_Statistical_Equal_Cofficient}, $\{\Phi_{{\mathsf{r}}_x,\ell}\}_{\ell=1}^{\infty}$ are i.i.d. random variables. Furthermore, since $\Phi_{{\mathsf{r}}_x,\ell}\sim{\mathcal{CN}}(0,1)$, it follows that $\lvert\Phi_{{\mathsf{r}}_x,\ell}\rvert^2\sim{\rm{Exp}}(1)$. Based on this fact and the results in \eqref{SNR_MISO_Linear_Array_Equal} and \eqref{Linear_Random_Operator_Statistical_Equal_SNR}, we derive the following conclusion.
\vspace{-5pt}
\begin{remark}
The SNR $\gamma_{{\mathsf{r}}_x}$ given in \eqref{SNR_MISO_Linear_Array_Equal} is statistically equivalent to a weighted sum of an infinite number of i.i.d. exponentially distributed random variables, each following ${\rm{Exp}}(1)$. The weights are given by $\{\frac{\mu({\mathcal{A}}_{\mathsf{r}})PL_{{\mathsf{t}},z}}{\sigma^2}\sigma_{{\mathsf{r}}_x,\ell}\}_{\ell=1}^{\infty}$. 
\end{remark}
\vspace{-5pt}
Unfortunately, characterizing the statistics of an infinite weighted sum of exponentially distributed random variables is a challenging task. To address this, we next explore the properties of the eigenvalues of the operator $R_{h_{{\mathsf{r}}_x}}(t_x,t_x')$, i.e., $\sigma_{{\mathsf{r}}_x,1}\geq\sigma_{{\mathsf{r}}_x,2}\ldots\geq\sigma_{{\mathsf{r}}_x,\infty}\geq0$. To this end, we define 
\begin{align}\label{Corollary_Autocorrelation_General_Linear_Result_Instead}
K_{h_{{\mathsf{r}}_x}}(t_x,t_x')\triangleq\frac{1}{2\pi}\int_{-k_0}^{k_0}{{\rm{e}}^{{\rm{j}}(t_x-t_x')\kappa_x}}{\rm{d}}\kappa_x,~~ t_x,t_x'\in{\mathcal{A}}_{{\mathsf{t}}_x}. 
\end{align}
Comparing \eqref{Corollary_Autocorrelation_General_Linear_Result_Instead} with \eqref{Corollary_Autocorrelation_General_Linear_Result} leads to $R_{h_{{\mathsf{r}}_x}}(t_x,t_x')=\frac{2\pi}{2k_0}K_{h_{{\mathsf{r}}_x}}(t_x,t_x')$. Let $\varepsilon_{{\mathsf{r}}_x,1}\geq\varepsilon_{{\mathsf{r}}_x,2}\ldots\geq\varepsilon_{{\mathsf{r}}_x,\infty}\geq0$ denote the eigenvalues of $K_{h_{{\mathsf{r}}_x}}(t_x,t_x')$, which yields $\frac{2\pi\varepsilon_{{\mathsf{r}}_x,\ell}}{2k_0}=\sigma_{{\mathsf{r}}_x,\ell}$. For an arbitrary square-integrable function $f_{{\mathsf{r}}_x}(t_x')$ defined on $t_x'\in{\mathcal{A}}_{{\mathsf{t}}_x}$, we define $\overline{f}_{{\mathsf{r}}_x}({{t}}_x)\triangleq\int_{{\mathcal{A}}_{{\mathsf{t}}_x}}K_{h_{{\mathsf{r}}_x}}(t_x,t_x')f_{{\mathsf{r}}_x}(t_x'){\rm{d}}t_x'$ for $t_x\in{\mathcal{A}}_{{\mathsf{t}}_x}$. This can be rewritten as follows:
\begin{align}
\overline{f}_{{\mathsf{r}}_x}({{t}}_x)={\mathbbmss{1}}_{{\mathcal{A}}_{{\mathsf{t}}_x}}(t_x)\!\!\int_{{\mathcal{A}}_{{\mathsf{t}}_x}}\!g_{{{\mathsf{r}}_x}}(t_x\!-\!t_x')
{\mathbbmss{1}}_{{\mathcal{A}}_{{\mathsf{t}}_x}}(t_x')f_{{\mathsf{r}}_x}(t_x'){\rm{d}}t_x',
\end{align}
where ${\mathbbmss{1}}_{{\mathcal{A}}_{{\mathsf{t}}_x}}(\cdot)$ is the indicator function over the set ${\mathcal{A}}_{{\mathsf{t}}_x}$ with ${\mathbbmss{1}}_{{\mathcal{A}}_{{\mathsf{t}}_x}}(t_x')=1$ if $t_x'\in{\mathcal{A}}_{{\mathsf{t}}_x}$ and zero otherwise. The function $g_{{{\mathsf{r}}_x}}(\cdot)$ is defined as follows:
\begin{align}
g_{{{\mathsf{r}}_x}}(x)\triangleq \frac{1}{2\pi}\int_{-k_0}^{k_0}{{\rm{e}}^{{\rm{j}}x\kappa_x}}{\rm{d}}\kappa_x,~x\in{\mathbbmss{R}},
\end{align}
which is the inverse Fourier transform of ${\mathbbmss{1}}_{[-k_0,k_0]}(\kappa_x)$. In other words, the Fourier transform of $g_{{{\mathsf{r}}_x}}(x)$ ($x\in{\mathbbmss{R}}$) is an ideal filter over the range $[-k_0,k_0]$. Using this framework, the eigenvalues $\{\varepsilon_{{\mathsf{r}}_x,\ell}\}_{\ell=1}^{\infty}$ can be characterized by \emph{Landau's eigenvalue theorem} \cite{landau1980eigenvalue,franceschetti2015landau}, which states:
\begin{align}\label{EDoF_Linear_Random_Operator_Statistical_Equal_Result1}
1\geq \varepsilon_{{\mathsf{r}}_x,1}\geq\varepsilon_{{\mathsf{r}}_x,2}\ldots\geq\varepsilon_{{\mathsf{r}}_x,\infty}\geq0,
\end{align}
where $\{\varepsilon_{{\mathsf{r}}_x,\ell}\}_{\ell=1}^{\infty}$ are functions of
\begin{align}
{\mathsf{DOF}}_{{\mathsf{r}}_x}=\frac{\mu([-k_0,k_0])\mu({\mathcal{A}}_{{\mathsf{t}}_x})}{2\pi}=\frac{2k_0L_{{\mathsf{t}},x}}{2\pi}=\frac{2L_{{\mathsf{t}},x}}{\lambda}.
\end{align}
As $L_{{\mathsf{t}},x}\rightarrow\infty$ or ${\mathsf{DOF}}_{{\mathsf{r}}_x}\rightarrow\infty$, the eigenvalues $\{\varepsilon_{{\mathsf{r}}_x,\ell}\}_{\ell=1}^{\infty}$ \emph{polarize} asymptotically \cite{landau1980eigenvalue}: for $\epsilon>0$, it holds that
\begin{equation} \label{EDoF_Linear_Random_Operator_Statistical_Equal_Result2}
\begin{split}
&\lvert\{\ell:\varepsilon_{{\mathsf{r}}_x,\ell}>\epsilon\}\rvert={\mathsf{DOF}}_{{\mathsf{r}}_x}\\
&+\left(\frac{1}{\pi^2}\log\frac{1-\sqrt{\epsilon}}{\sqrt{\epsilon}}\right)\log{{\mathsf{DOF}}_{{\mathsf{r}}_x}}+o(\log{{\mathsf{DOF}}_{{\mathsf{r}}_x}}),
\end{split}
\end{equation}
as $L_{{\mathsf{t}},x}\rightarrow\infty$ or ${\mathsf{DOF}}_{{\mathsf{r}}_x}\rightarrow\infty$.
\vspace{-5pt}
\begin{remark}
The results in \eqref{EDoF_Linear_Random_Operator_Statistical_Equal_Result1} and \eqref{EDoF_Linear_Random_Operator_Statistical_Equal_Result2} imply that as $L_{{\mathsf{t}},x}$ increases, the leading ${\mathsf{DOF}}_{{\mathsf{r}}_x}$ eigenvalues $\{\varepsilon_{{\mathsf{r}}_x,\ell}\}_{\ell=1}^{{\mathsf{DOF}}_{{\mathsf{r}}_x}}$ are near to one, while the rest are near zero. A transition band with a width proportional to $\log{{\mathsf{DOF}}_{{\mathsf{r}}_x}}$ separates these regions, reflecting a behavior similar to that depicted in \cite[{\figurename} 2]{franceschetti2015landau}.
\end{remark}
\vspace{-5pt}
\vspace{-5pt}
\begin{remark}
Landau's eigenvalue theorem was applied in \cite{pizzo2022landau} to analyze the spatial DoFs of CAPA-based LoS channels, with extensions to multipath scenarios. Our work builds upon these findings by accounting for fading randomness and deriving the channel statistics for CAPA-based fading channels. These contributions are elaborated in the subsequent sections.
\end{remark}
\vspace{-5pt}
These observations suggest that for small values of $\ell$, the eigenvalues $\sigma_{{\mathsf{r}}_x,\ell}$ decrease slowly until they reach a critical value at $\ell={\mathsf{DOF}}_{{\mathsf{r}}_x}$, after which they decrease rapidly. This step-like behavior becomes more pronounced as $L_{{\mathsf{t}},x}$ increases \cite{liu2025near}. Since CAPAs are typically electromagnetically large arrays satisfying $L_{{\mathsf{t}},x}\gg \lambda$, the sum $\sum\nolimits_{\ell=1}^{\infty}
\sigma_{{\mathsf{r}}_x,\ell}\lvert\Phi_{{\mathsf{r}}_x,\ell}\rvert^2$ is dominated by the first ${\mathsf{DOF}}_{{\mathsf{r}}_x}$ terms, which yields
\begin{align}\label{EDoF_Linear_Random_Operator_Statistical_Equal_Result_Final}
\int_{{\mathcal{A}}_{{\mathsf{t}}_x}}\lvert {\overline{h}}_{{\mathsf{r}}_x}({{t}}_x)\rvert^2{\rm{d}}t_x\approx\sum\nolimits_{\ell=1}^{{\mathsf{DOF}}_{{\mathsf{r}}_x}}
\sigma_{{\mathsf{r}}_x,\ell}\lvert\Phi_{{\mathsf{r}}_x,\ell}\rvert^2.
\end{align}
\subsubsection{Planar CAPAs}
Next, we consider the case of planar arrays, where the autocorrelation function of the spatial response is given by \eqref{Lemma_Autocorrelation_General_Result}. Following the derivation steps used to obtain \eqref{Linear_Random_Operator_Statistical_Equal_SNR} and \eqref{Linear_Random_Operator_Equal_Channel_Gain_Initial}, we conclude that for planar arrays:
\begin{align}\label{Planar_Random_Operator_Statistical_Equal_SNR}
\gamma_{{\mathsf{r}}}\overset{d}{=}\frac{\mu({\mathcal{A}}_{\mathsf{r}})P}{\sigma^2}\sum\nolimits_{\ell=1}^{\infty}
\sigma_{{\mathsf{r}},\ell}\lvert\Phi_{{\mathsf{r}},\ell}\rvert^2,
\end{align}
where $\{\Phi_{{\mathsf{r}},\ell}\sim{\mathcal{CN}}(0,1)\}_{\ell=1}^{\infty}$ are i.i.d. ZUCG random variables, and $\sigma_{{\mathsf{r}},1}\geq\sigma_{{\mathsf{r}},2}\ldots\geq\sigma_{{\mathsf{r}},\infty}\geq0$ represent the eigenvalues of the semipositive definite operator $R_{h_{\mathsf{r}}}({\mathbf{t}},{\mathbf{t}}')$ as defined in \eqref{Lemma_Autocorrelation_General_Result}. We next employ \emph{Landau's eigenvalue theorem} to characterize the properties of these eigenvalues.

By definition, $\frac{1}{\gamma(\kappa_x,\kappa_z)}$ in \eqref{Lemma_Autocorrelation_General_Result} satisfies $\frac{1}{\gamma(\kappa_x,\kappa_z)}\in[\frac{1}{k_0},\infty)$. On this basis, we bound \eqref{Lemma_Autocorrelation_General_Result} as follows:
\begin{align}
\frac{2\pi}{k_0}\frac{1}{k_0}K_{h_{\mathsf{r}}}({\mathbf{t}},{\mathbf{t}}') \preceq R_{h_{\mathsf{r}}}({\mathbf{t}},{\mathbf{t}}') \preceq \frac{2\pi}{ k_0}M_{h_{\mathsf{r}}}K_{h_{\mathsf{r}}}({\mathbf{t}},{\mathbf{t}}'),
\end{align}
where $M_{h_{\mathsf{r}}}\rightarrow\infty$, and
\begin{align}\label{EDoF_Planar_Random_Operator_Statistical_Equal_Result0}
K_{h_{\mathsf{r}}}({\mathbf{t}},{\mathbf{t}}')\triangleq
\iint_{{\mathcal{D}}({\bm\kappa})}
\frac{{\rm{e}}^{{\rm{j}}((t_x-t_x')\kappa_x+(t_z-t_z')\kappa_z)}}{(2\pi)^2}{\rm{d}}\kappa_x{\rm{d}}\kappa_z.
\end{align}
For ease of explanation, let $\varepsilon_{{\mathsf{r}},1}\geq\varepsilon_{{\mathsf{r}},2}\ldots\geq\varepsilon_{{\mathsf{r}},\infty}\geq0$ denote the eigenvalues of $K_{h_{\mathsf{r}}}({\mathbf{t}},{\mathbf{t}}')$.

For an arbitrary square-integrable function $f_{{\mathsf{r}}}({\mathbf{t}}')$ defined on ${\mathbf{t}}'=[t_x',0,t_z']^{\mathsf{T}}\in{\mathcal{A}}_{\mathsf{t}}$, we rewrite $ \overline{f}_{{\mathsf{r}}}({\mathbf{t}})\triangleq\int_{{\mathcal{A}}_{{\mathsf{t}}}}K_{h_{\mathsf{r}}}({\mathbf{t}},{\mathbf{t}}')f_{{\mathsf{r}}}({\mathbf{t}}'){\rm{d}}{\mathbf{t}}'$ for ${\mathbf{t}}=[t_x,0,t_z]^{\mathsf{T}}\in{\mathcal{A}}_{{\mathsf{t}}}$ as follows:
\begin{align}
\overline{f}_{{\mathsf{r}}}({\mathbf{t}})={\mathbbmss{1}}_{{\mathcal{A}}_{{\mathsf{t}}}}({\mathbf{t}})\int_{{\mathcal{A}}_{{\mathsf{t}}}}g_{{{\mathsf{r}}}}({\mathbf{t}}-{\mathbf{t}}')
{\mathbbmss{1}}_{{\mathcal{A}}_{{\mathsf{t}}}}({\mathbf{t}}')f_{{\mathsf{r}}}({\mathbf{t}}'){\rm{d}}{\mathbf{t}}',
\end{align}
where the function $g_{{{\mathsf{r}}}}(\cdot)$ is defined as follows:
\begin{align}
g_{{{\mathsf{r}}}}({\bm\delta})\triangleq \frac{1}{(2\pi)^2}\iint_{{\mathcal{D}}({\bm\kappa})}{{\rm{e}}^{{\rm{j}}(\delta_x\kappa_x+\delta_z\kappa_z)}}{\rm{d}}\kappa_x{\rm{d}}\kappa_z,
\end{align}
for ${\bm\delta}=[\delta_x,\delta_z]^{\mathsf{T}}\in{\mathbbmss{R}}^{2\times1}$. This function is the inverse Fourier transform of ${\mathbbmss{1}}_{{\mathcal{D}}({\bm\kappa})}([\kappa_x,\kappa_z]^{\mathsf{T}})$. In other words, the Fourier transform of $g_{{{\mathsf{r}}}}({\bm\delta})$ (${\bm\delta}\in{\mathbbmss{R}}^{2\times1}$) is an ideal filter in ${\mathcal{D}}({\bm\kappa})$. Under this setup, the properties of $\{\varepsilon_{{\mathsf{r}},\ell}\}_{\ell=1}^{\infty}$ can be characterized using \emph{Landau's eigenvalue theorem}, which states \cite{landau1975szego,franceschetti2015landau}:
\begin{align}\label{EDoF_Planar_Random_Operator_Statistical_Equal_Result1}
1\geq \varepsilon_{{\mathsf{r}},1}\geq\varepsilon_{{\mathsf{r}},2}\ldots\geq\varepsilon_{{\mathsf{r}},\infty}\geq0,
\end{align}
where $\{\varepsilon_{{\mathsf{r}},\ell}\}_{\ell=1}^{\infty}$ are functions of
\begin{align}
{\mathsf{DOF}}_{{\mathsf{r}}}=\frac{\mu({\mathcal{D}}({\bm\kappa}))\mu({\mathcal{A}}_{{\mathsf{t}}})}{(2\pi)^2}
=\frac{\pi k_0^2L_{{\mathsf{t}},x}L_{{\mathsf{t}},z}}{(2\pi)^2}=\frac{\pi L_{{\mathsf{t}},x}L_{{\mathsf{t}},z}}{\lambda^2}.
\end{align}
As $L_{{\mathsf{t}},x},L_{{\mathsf{t}},z}\rightarrow\infty$ or ${\mathsf{DOF}}_{{\mathsf{r}}}\rightarrow\infty$, the eigenvalues $\{\varepsilon_{{\mathsf{r}},\ell}\}_{\ell=1}^{\infty}$ \emph{polarize} asymptotically \cite{landau1975szego,franceschetti2015landau}: for $\epsilon>0$, 
\begin{equation} \label{EDoF_Planar_Random_Operator_Statistical_Equal_Result2}
\begin{split}
&\lvert\{\ell:\varepsilon_{{\mathsf{r}},\ell}>\epsilon\}\rvert={\mathsf{DOF}}_{{\mathsf{r}}}+o({\mathsf{DOF}}_{{\mathsf{r}}}),
\end{split}
\end{equation}
as $L_{{\mathsf{t}},x},L_{{\mathsf{t}},z}\rightarrow\infty$ or ${\mathsf{DOF}}_{{\mathsf{r}}}\rightarrow\infty$.
\vspace{-5pt}
\begin{remark}
The results in \eqref{EDoF_Planar_Random_Operator_Statistical_Equal_Result1} and \eqref{EDoF_Planar_Random_Operator_Statistical_Equal_Result2} imply that there are roughly ${\mathsf{DOF}}_{{\mathsf{r}}}$ significant eigenvalues near one, the rest becoming negligible. This result is less precise than its counterpart \eqref{EDoF_Linear_Random_Operator_Statistical_Equal_Result2} in that the transition band is not specified.
\end{remark}
\vspace{-5pt}
The above arguments imply that for electromagnetically large arrays such as CAPAs, which satisfy $L_{{\mathsf{t}},x},L_{{\mathsf{t}},z}\gg \lambda$, the eigenvalues of $K_{h_{\mathsf{r}}}({\mathbf{t}},{\mathbf{t}}')$ are dominated by the first ${\mathsf{DOF}}_{{\mathsf{r}}}$ terms, $\{\varepsilon_{{\mathsf{r}},\ell}\}_{\ell=1}^{{\mathsf{DOF}}_{{\mathsf{r}}}}$. Based on the sandwich relationship in \eqref{EDoF_Planar_Random_Operator_Statistical_Equal_Result0}, it is reasonable to deduce that the eigenvalues of $R_{h_{\mathsf{r}}}({\mathbf{t}},{\mathbf{t}}')$ are also dominated by the leading ${\mathsf{DOF}}_{{\mathsf{r}}}$ terms, which yields 
\begin{align}
\sum\nolimits_{\ell=1}^{\infty}
\sigma_{{\mathsf{r}},\ell}\lvert\Phi_{{\mathsf{r}},\ell}\rvert^2\approx\sum\nolimits_{\ell=1}^{{\mathsf{DOF}}_{{\mathsf{r}}}}
\sigma_{{\mathsf{r}},\ell}\lvert\Phi_{{\mathsf{r}},\ell}\rvert^2.
\end{align}
\subsubsection{Extension to SIMO}
Owing to the reciprocity between MISO and SIMO channels, the above discussions extend naturally to the SIMO case. For simplicity, we first consider the case of linear arrays with $L_{{\mathsf{r}},z}\ll L_{{\mathsf{r}},x}$, which yields
\begin{align}\label{SNR_SIMO_Linear_Array_Equal}
\gamma_{\mathsf{t}}=\frac{\mu({\mathcal{A}}_{\mathsf{t}})PL_{{\mathsf{r}},z}}{\sigma^2}\int_{-{L_{{\mathsf{r}},x}}/{2}}^{{L_{{\mathsf{r}},x}}/{2}}\lvert h_{{\mathsf{t}}_x}({{r}}_x')\rvert^2{\rm{d}}r_x'\triangleq \gamma_{{\mathsf{t}}_x},
\end{align}
with $h_{{\mathsf{t}}_x}({{r}}_x')=h_{\mathsf{t}}([r_x', 0, 0]^{\mathsf{T}})$. Following the steps used to derive \eqref{Corollary_Autocorrelation_General_Linear_Result}, the autocorrelation of $h_{{\mathsf{t}}_x}({{r}}_x')$ is given by
\begin{subequations}
\begin{align}
R_{h_{{\mathsf{t}}_x}}(r_x',r_x'')&=\frac{1}{2k_0}
\int_{-k_0}^{k_0}
{{\rm{e}}^{-{\rm{j}}(r_x'-r_x'')k_x}}{\rm{d}}k_x\\
&=\frac{1}{2k_0}
\int_{-k_0}^{k_0}
{{\rm{e}}^{{\rm{j}}(r_x'-r_x'')k_x}}{\rm{d}}k_x,
\end{align}
\end{subequations}
which has the same form as \eqref{Corollary_Autocorrelation_General_Linear_Result}. Therefore, \emph{Landau's eigenvalue theorem} can be applied to characterize the eigenvalues of $R_{h_{{\mathsf{t}}_x}}(r_x',r_x'')$, denote by $\sigma_{{\mathsf{t}}_x,1}\geq\sigma_{{\mathsf{t}}_x,2}\ldots\geq\sigma_{{\mathsf{t}}_x,\infty}\geq0$. Similar to \eqref{EDoF_Linear_Random_Operator_Statistical_Equal_Result_Final}, we obtain
\begin{subequations}\label{Linear_Random_Operator_Statistical_Equal_SNR_SIMO}
\begin{align}
\gamma_{{\mathsf{t}}_x}&\overset{d}{=}\frac{1}{\sigma^2}\mu({\mathcal{A}}_{\mathsf{t}})PL_{{\mathsf{r}},z}\sum\nolimits_{\ell=1}^{\infty}
\sigma_{{\mathsf{t}}_x,\ell}\lvert\Phi_{{\mathsf{t}}_x,\ell}\rvert^2\\
&\approx\frac{1}{\sigma^2}\mu({\mathcal{A}}_{\mathsf{t}})PL_{{\mathsf{r}},z}\sum\nolimits_{\ell=1}^{{\mathsf{DOF}}_{{\mathsf{t}}_x}}
\sigma_{{\mathsf{t}}_x,\ell}\lvert\Phi_{{\mathsf{t}}_x,\ell}\rvert^2,
\end{align}
\end{subequations}
where ${\mathsf{DOF}}_{{\mathsf{t}}_x}=\frac{2L_{{\mathsf{r}},x}}{\lambda}$, and $\{\Phi_{{\mathsf{t}}_x,\ell}\sim{\mathcal{CN}}(0,1)\}_{\ell=1}^{\infty}$ are i.i.d. ZUCG random variables. For planar arrays, we obtain
\begin{subequations}\label{Planar_Random_Operator_Statistical_Equal_SNR_SIMO}
\begin{align}
\gamma_{{\mathsf{t}}}&\overset{d}{=}\frac{1}{\sigma^2}\mu({\mathcal{A}}_{\mathsf{t}})P\sum\nolimits_{\ell=1}^{\infty}
\sigma_{{\mathsf{t}},\ell}\lvert\Phi_{{\mathsf{t}},\ell}\rvert^2\\
&\approx\frac{1}{\sigma^2}\mu({\mathcal{A}}_{\mathsf{t}})P\sum\nolimits_{\ell=1}^{{\mathsf{DOF}}_{{\mathsf{t}}}}
\sigma_{{\mathsf{t}},\ell}\lvert\Phi_{{\mathsf{t}},\ell}\rvert^2,
\end{align}
\end{subequations}
where ${\mathsf{DOF}}_{{\mathsf{t}}}=\frac{\pi L_{{\mathsf{r}},x}L_{{\mathsf{r}},z}}{\lambda^2}$, $\{\Phi_{{\mathsf{t}},\ell}\sim{\mathcal{CN}}(0,1)\}_{\ell=1}^{\infty}$ are i.i.d. ZUCG random variables, and $\sigma_{{\mathsf{t}},1}\geq\sigma_{{\mathsf{t}},2}\ldots\geq\sigma_{{\mathsf{t}},\infty}\geq0$ represent the eigenvalues of the following operator:
\begin{align}
R_{h_{\mathsf{t}}}({\mathbf{r}}',{\mathbf{r}}'')=
\iint_{{\mathcal{D}}({\mathbf{k}})}
\frac{{\rm{e}}^{{\rm{j}}((r_x'-r_x'')k_x+(r_z'-r_z'')k_z)}}{2\pi k_0\gamma(k_x,k_z)}{\rm{d}}k_x{\rm{d}}k_z,
\end{align}
where ${\mathbf{r}}=[r_x,0,r_z]^{\mathsf{T}}$ and ${\mathbf{r}}'=[r_x',0,r_z']^{\mathsf{T}}$.

In summary, the SNRs for MISO and SIMO channels employing linear or planar CAPAs can be asymptotically approximated as a finite weighted sum of exponentially distributed random variables. The weights are determined by the eigenvalues of the autocorrelation function. Detailed numerical methods for calculating these eigenvalues are given in \cite{atkinson2007solving}.
\subsection{Performance Analysis}\label{Section:MISO:Performance Analysis}
We now analyze the performance of CAPA-based MISO/SIMO channels. For brevity, we focus on the capacity of MISO channels utilizing planar arrays. The derived results, however, can be readily extended to other configurations. 
\subsubsection{Average SNR}
We begin with the average SNR:
\begin{subequations}
\begin{align}
{\mathbbmss{E}}\{\gamma_{{\mathsf{r}}}\}&=\frac{\mu({\mathcal{A}}_{\mathsf{r}})P}{\sigma^2}{\mathbbmss{E}}\left\{\int_{{\mathcal{A}}_{\mathsf{t}}}\lvert h_{\mathsf{r}}({\mathbf{t}})\rvert^2{\rm{d}}{\mathbf{t}}\right\}\label{Average_SNR_Definition}\\
&=\frac{\mu({\mathcal{A}}_{\mathsf{r}})P}{\sigma^2}\int_{{\mathcal{A}}_{\mathsf{t}}}{\mathbbmss{E}}\{\lvert h_{\mathsf{r}}({\mathbf{t}})\rvert^2\}{\rm{d}}{\mathbf{t}},
\end{align}
\end{subequations}
where \eqref{Average_SNR_Definition} is based on \eqref{MISO_SNR_Definition_Standard}. Referring to \eqref{Lemma_Autocorrelation_General_Result}, we have
\begin{align}
{\mathbbmss{E}}\{\lvert h_{\mathsf{r}}({\mathbf{t}})\rvert^2\}=R_{h_{\mathsf{r}}}({\mathbf{t}},{\mathbf{t}})
=\iint_{{\mathcal{D}}({\bm\kappa})}
\frac{\frac{1}{2\pi k_0}{\rm{d}}\kappa_x{\rm{d}}\kappa_z}{\gamma(\kappa_x,\kappa_z)}=1.
\end{align}
Substituting this result back, we have
\begin{align}
{\mathbbmss{E}}\{\gamma_{{\mathsf{r}}}\}=\frac{\mu({\mathcal{A}}_{\mathsf{r}})P}{\sigma^2}\int_{{\mathcal{A}}_{\mathsf{t}}}{\rm{d}}{\mathbf{t}}=
\frac{\mu({\mathcal{A}}_{\mathsf{r}})\mu({\mathcal{A}}_{\mathsf{t}})P}{\sigma^2}.
\end{align}
These calculations demonstrate that the average received SNR scales linearly with the product of the transmit and receive aperture sizes, as intuitively expected.  
\subsubsection{Diversity Gain}
We next analyze the diversity gain by studying the capacity of MISO channels as follows:
\begin{align}\label{MISO_Rate_CAPA_Basic}
{\mathsf{C}}_{\mathsf{r}}=\log_2(1+\gamma_{\mathsf{r}})\overset{d}{=}\log_2\left(1+\overline{\gamma}a_{\mathsf{r}}\right),
\end{align}
where $\overline{\gamma}=\frac{\mu({\mathcal{A}}_{\mathsf{r}})P}{\sigma^2}$ and $a_{\mathsf{r}}=\sum_{\ell=1}^{{\mathsf{DOF}}_{{\mathsf{r}}}}
\sigma_{{\mathsf{r}},\ell}\lvert\Phi_{{\mathsf{r}},\ell}\rvert^2$. 

The \emph{maximal (full) diversity gain} is revealed by analyzing the OP, defined as follows:
\begin{align}
{\mathcal{P}}_{\mathsf{r}}=\Pr({\mathsf{C}}_{\mathsf{r}}<R)=\Pr\left(a_{\mathsf{r}}<{\hat{a}}_{\mathsf{r}}\right)=F_{a_{\mathsf{r}}}({\hat{a}}_{\mathsf{r}}),
\end{align}
where $R>0$ denotes the target rate, ${\hat{a}}_{\mathsf{r}}=\frac{2^R-1}
{{\overline{\gamma}}}$, and $F_{a_{\mathsf{r}}}(\cdot)$ denotes the cumulative distribution function (CDF) of $a_{\mathsf{r}}$. The full diversity gain is defined as follows \cite{zheng2003diversity}:
\begin{align}
d_{\mathsf{r}}^{\star}\triangleq\lim_{\overline{\gamma}\rightarrow\infty}-\frac{\log{{\mathcal{P}}_{\mathsf{r}}}}{\log{\overline{\gamma}}}=
\lim_{\overline{\gamma}\rightarrow\infty}-\frac{\log(F_{a_{\mathsf{r}}}({\hat{a}}_{\mathsf{r}}))}{\log{\overline{\gamma}}},
\end{align}
which measures the SNR exponent of the OP in the high-SNR regime. The next task is to characterize the statistics of $a_{\mathsf{r}}$, and the main results are summarized as follows.
\vspace{-5pt}
\begin{lemma}
The probability density function (PDF) and CDF of $a_{\mathsf{r}}$ are given by
\begin{align}
&f_{a_{\mathsf{r}}}(x)=
\frac{\sigma_{{\mathsf{r}},\min}^{{\mathsf{DOF}}_{{\mathsf{r}}}}}{\prod_{\ell=1}^{{\mathsf{DOF}}_{{\mathsf{r}}}}\sigma_{{\mathsf{r}},\ell}}\sum_{q=0}^{\infty}\frac{\psi_q x^{{\mathsf{DOF}}_{{\mathsf{r}}}+q-1}{\rm e}^{-\frac{x}{\sigma_{{\mathsf{r}},\min}}}}{\sigma_{{\mathsf{r}},\min}^{{\mathsf{DOF}}_{{\mathsf{r}}}+q}\Gamma({\mathsf{DOF}}_{{\mathsf{r}}}+q)},
\label{PDF_Channel_Gain_MISO}\\
&F_{a_{\mathsf{r}}}({x})=
\frac{\sigma_{{\mathsf{r}},\min}^{{\mathsf{DOF}}_{{\mathsf{r}}}}}{\prod_{\ell=1}^{{\mathsf{DOF}}_{{\mathsf{r}}}}\sigma_{{\mathsf{r}},\ell}}\sum_{q=0}^{\infty}\frac{\psi_q \Upsilon({\mathsf{DOF}}_{{\mathsf{r}}}+q,{x}/\sigma_{{\mathsf{r}},\min})}{\Gamma({\mathsf{DOF}}_{{\mathsf{r}}}+q)},\label{CDF_Channel_Gain_MISO}
\end{align}
respectively, where $\Gamma\left(z\right)=\int_{0}^{\infty}{\rm e}^{-t}t^{z-1}{\rm d}t$ is the Gamma function \cite[Eq. (8.310.1)]{gradshteyn2014table}, $\Upsilon\left(s,x\right)=\int_{0}^{x}t^{s-1}{\rm e}^{-t}{\rm d}t$ is the lower incomplete Gamma function \cite[Eq. (8.350.1)]{gradshteyn2014table}, $\sigma_{{\mathsf{r}},\min}=\min_{\ell\in\{1,\ldots,{\mathsf{DOF}}_{{\mathsf{r}}}\}}\sigma_{{\mathsf{r}},\ell}=\sigma_{{\mathsf{r}},{\mathsf{DOF}}_{{\mathsf{r}}}}$, ${\psi _0} = 1$, and $\psi_q$ ($q\geq1$) can be calculated recursively by
\begin{align}
{\psi _{q}} = \sum\nolimits_{k = 1}^{q} {\left[ {\sum\nolimits_{\ell = 1}^{{\mathsf{DOF}}_{{\mathsf{r}}}}{{{\left( {1 - \sigma_{{\mathsf{r}},\min}/\sigma_{{\mathsf{r}},\ell}} \right)}^k}} } \right]} \frac{\psi _{q - k}}{{q}}.
\end{align}
\end{lemma}
\vspace{-5pt}
\begin{IEEEproof}
Please refer to \cite[Eq. (2.9)]{moschopoulos1985distribution} for more details.
\end{IEEEproof}
Employing \eqref{CDF_Channel_Gain_MISO} yields the following theorem.
\vspace{-5pt}
\begin{theorem}\label{Theorem_MISO_OP_High_SNR_Asymptotic}
The OP for the MISO case is given by ${\mathcal{P}}_{\mathsf{r}}=F_{a_{\mathsf{r}}}({\hat{a}}_{\mathsf{r}})$. As $\overline{\gamma}\rightarrow\infty$, the OP satisfies
\begin{align}\label{MISO_OP_High_SNR_Asymptotic}
\lim_{\overline{\gamma}\rightarrow\infty}{\mathcal{P}}_{\mathsf{r}}\simeq
\frac{{\hat{a}}_{\mathsf{r}}^{{\mathsf{DOF}}_{{\mathsf{r}}}}}{{\mathsf{DOF}}_{{\mathsf{r}}}!\prod_{\ell=1}^{{\mathsf{DOF}}_{{\mathsf{r}}}}\sigma_{{\mathsf{r}},\ell}}
=\frac{\left({2^R-1}\right)^{{\mathsf{DOF}}_{{\mathsf{r}}}}\frac{1}{{\overline{\gamma}}^{{\mathsf{DOF}}_{{\mathsf{r}}}}}}{{\mathsf{DOF}}_{{\mathsf{r}}}!\prod_{\ell=1}^{{\mathsf{DOF}}_{{\mathsf{r}}}}\sigma_{{\mathsf{r}},\ell}}.
\end{align}
The maximal diversity gain and the associated array gain are given by $d_{\mathsf{r}}^{\star}={{\mathsf{DOF}}_{{\mathsf{r}}}}$ and $\frac{({\mathsf{DOF}}_{{\mathsf{r}}}!\prod_{\ell=1}^{{\mathsf{DOF}}_{{\mathsf{r}}}}\sigma_{{\mathsf{r}},\ell})^{{1}/{\mathsf{DOF}}_{{\mathsf{r}}}}}{2^R-1}$, respectively.
\end{theorem}
\vspace{-5pt}
\begin{IEEEproof}
Pleases refer to Appendix \ref{Proof_Theorem_MISO_OP_High_SNR_Asymptotic} for more details.
\end{IEEEproof}
\vspace{-5pt}
\begin{remark}
The above arguments imply that the maximal diversity gain achieved by a CAPA is given by ${\mathsf{DOF}}_{{\mathsf{r}}}=\frac{\pi L_{{\mathsf{t}},x}L_{{\mathsf{t}},z}}{\lambda^2}=\pi N_{{\mathsf{t}},x}N_{{\mathsf{t}},z}$, which is determined by both the aperture size and the wavelength.
\end{remark}
\vspace{-5pt}
\subsubsection{Multiplexing Gain}
The \emph{spatial multiplexing gain} describes how the data rate scales with the SNR at high SNRs, in contrast to that for single-antenna channels. The maximal multiplexing gain is defined as follows:
\begin{align}
r_{\mathsf{r}}^{\star}\triangleq\lim_{\overline{\gamma}\rightarrow\infty}\frac{{\mathcal{R}}_{\mathsf{r}}}{\log_2(1+{\overline{\gamma}})}=
\lim_{\overline{\gamma}\rightarrow\infty}\frac{{\mathbbmss{E}}\{{\mathsf{C}}_{\mathsf{r}}\}}{\log_2(1+{\overline{\gamma}})},
\end{align}
where ${\mathcal{R}}_{\mathsf{r}}\triangleq{\mathbbmss{E}}\{{\mathsf{C}}_{\mathsf{r}}\}$ is the ECC, and $\log_2(1+{\overline{\gamma}})$ measures the data rate of a single-antenna channel. 
\vspace{-5pt}
\begin{theorem}\label{Theorem_MISO_ADR_High_SNR_Asymptotic}
The ECC for the MISO case is given by 
\begin{equation}\label{MISO_ADR_Explicit}
\begin{split}
{\mathcal{R}}_{\mathsf{r}}&=\frac{\sigma_{{\mathsf{r}},\min}^{{\mathsf{DOF}}_{{\mathsf{r}}}}}{\prod_{\ell=1}^{{\mathsf{DOF}}_{{\mathsf{r}}}}\sigma_{{\mathsf{r}},\ell}}
\sum_{q=0}^{\infty}\sum_{v=0}^{{\mathsf{DOF}}_{{\mathsf{r}}}+q-1}
\frac{\psi_q/\log{2}}{({\mathsf{DOF}}_{{\mathsf{r}}}+q-1-v)!}\\
&\times\left[\frac{(-1)^{{\mathsf{DOF}}_{{\mathsf{r}}}+q-v}{\rm e}^{\frac{1}{{\overline{\gamma}}{\sigma_{{\mathsf{r}},\min}}}}}{({\overline{\gamma}}{\sigma_{{\mathsf{r}},\min}})^{{\mathsf{DOF}}_{{\mathsf{r}}}+q-1-v}}{\rm{Ei}}
\left({\frac{-1}{{\overline{\gamma}}{\sigma_{{\mathsf{r}},\min}}}}\right)\right.\\
&+\left.\sum_{u=1}^{{\mathsf{DOF}}_{{\mathsf{r}}}+q-1-v}\Gamma(u)\left(\frac{-1}{{\overline{\gamma}}{\sigma_{{\mathsf{r}},\min}}}\right)^{{\mathsf{DOF}}_{{\mathsf{r}}}+q-1-v-u}\right],
\end{split}
\end{equation}
where ${\rm{Ei}}\left(x\right)=-\int_{-x}^{\infty}{\rm{e}}^{-t}t^{-1}{\rm{d}}t$ denotes the exponential integral function \cite[Eq. (8.211.1)]{gradshteyn2014table}. The high-SNR ECC satisfies
\begin{align}\label{MISO_ADR_High_SNR_Asymptotic}
\lim\nolimits_{\overline{\gamma}\rightarrow\infty}{\mathcal{R}}_{\mathsf{r}}\simeq
\log_2{\overline{\gamma}}-{\mathcal{L}}_{\mathsf{r}},
\end{align}
where 
\begin{equation}
\begin{split}
{\mathcal{L}}_{\mathsf{r}}=\frac{-\sigma_{{\mathsf{r}},\min}^{{\mathsf{DOF}}_{{\mathsf{r}}}}}{\prod_{\ell=1}^{{\mathsf{DOF}}_{{\mathsf{r}}}}\sigma_{{\mathsf{r}},\ell}}
\sum_{q=0}^{\infty}\frac{\psi_q
(\psi({\mathsf{DOF}}_{{\mathsf{r}}}+q)+\log{\sigma_{{\mathsf{r}},\min}})}{\log{2}},
\end{split}
\end{equation}
and $\psi\left(x\right)=\frac{{\rm d}}{{\rm d}x}\ln{\Gamma\left(x\right)}$ is Euler's digamma function.
\end{theorem}
\vspace{-5pt}
\begin{IEEEproof}
Please refer to Appendix \ref{Proof_Theorem_MISO_OP_High_SNR_Asymptotic} for more details.
\end{IEEEproof}
\vspace{-5pt}
\begin{remark}
Theorem \ref{Theorem_MISO_ADR_High_SNR_Asymptotic} indicates that the maximal multiplexing gain in the CAPA-based MISO channel is $r_{\mathsf{r}}^{\star}=1$, and ${\mathcal{L}}_{\mathsf{r}}$ is the associated high-SNR power offset in 3-dB units.
\end{remark}
\vspace{-5pt}
\subsubsection{Diversity-Multiplexing Trade-off}
When analyzing $d_{\mathsf{r}}^{\star}$, a fixed target data rate $R$ is considered. However, in practice, it is more meaningful to consider a target rate proportional to the SNR, i.e., $R(\overline{\gamma})=r_{\mathsf{r}}\log_2(1+\overline{\gamma})$ for $r_{\mathsf{r}}\in(0,1)$. Under this condition, the high-SNR OP satisfies
\begin{align}
\lim_{\overline{\gamma}\rightarrow\infty}\Pr({\mathsf{C}}_{\mathsf{r}}<r_{\mathsf{r}}\log_2(1+\overline{\gamma}))\simeq
(\rho_{\mathsf{r}}(r_{\mathsf{r}})\overline{\gamma})^{-d_{\mathsf{r}}(r_{\mathsf{r}})},
\end{align}
where the corresponding diversity order is given as follows:
\begin{align}
d_{\mathsf{r}}(r_{\mathsf{r}})=\lim_{\overline{\gamma}\rightarrow\infty}-\frac{\log(\Pr({\mathsf{C}}_{\mathsf{r}}<r_{\mathsf{r}}\log_2(1+\overline{\gamma})))}{\log{\overline{\gamma}}},
\end{align}
which measures the SNR exponent of the OP in the high-SNR regime when achieving a target rate $r_{\mathsf{r}}\log_2(1+\overline{\gamma})$, or equivalently, a spatial multiplexing gain $r_{\mathsf{r}}$. Furthermore, the array gain $\rho_{\mathsf{r}}(r_{\mathsf{r}})$ associated with $d_{\mathsf{r}}(r_{\mathsf{r}})$ satisfies \cite{ordonez2012array}
\begin{align}
(\rho_{\mathsf{r}}(r_{\mathsf{r}}))^{-d_{\mathsf{r}}(r_{\mathsf{r}})}=\lim_{\overline{\gamma}\rightarrow\infty}
\frac{\Pr({\mathsf{C}}_{\mathsf{r}}<r_{\mathsf{r}}\log_2(1+\overline{\gamma}))}{\overline{\gamma}^{-d_{\mathsf{r}}(r_{\mathsf{r}})}}.
\end{align}
Intuitively, a larger target multiplexing gain may result in a smaller SNR exponent of the OP, indicating a trade-off between multiplexing gain and diversity gain. The detailed relationship between $d_{\mathsf{r}}$ and $r_{\mathsf{r}}$ is given as follows.
\vspace{-5pt}
\begin{theorem}\label{Theorem_MISO_DMT_High_SNR_Asymptotic}
In a CAPA-based MISO channel, the DMT is given by $d_{\mathsf{r}}(r_{\mathsf{r}})={\mathsf{DOF}}_{{\mathsf{r}}}(1-r_{\mathsf{r}})$, and the array gain in the DMT framework is given by $\rho_{\mathsf{r}}(r_{\mathsf{r}})=({\mathsf{DOF}}_{{\mathsf{r}}}!\prod_{\ell=1}^{{\mathsf{DOF}}_{{\mathsf{r}}}}\sigma_{{\mathsf{r}},\ell})^{\frac{1}{{\mathsf{DOF}}_{{\mathsf{r}}}(1-r_{\mathsf{r}})}}$. The maximal multiplexing gain satisfies $\sup\{r_{\mathsf{r}}:d_{\mathsf{r}}(r_{\mathsf{r}})>0\}=1=r_{\mathsf{r}}^{\star}$, and the maximal diversity gain is given by $d_{\mathsf{r}}(0)={\mathsf{DOF}}_{{\mathsf{r}}}=d_{\mathsf{r}}^{\star}$.
\end{theorem}
\vspace{-5pt}
\begin{IEEEproof}
Please refer to Appendix \ref{Proof_Theorem_MISO_OP_High_SNR_Asymptotic} for more details.
\end{IEEEproof}
\vspace{-5pt}
\begin{remark}
The above discussions suggest that the DMT achieved by a CAPA is influenced by both the aperture size and the wavelength. The DMT can be improved by either increasing the aperture size or utilizing higher frequency bands, which aligns with the two key features of CAPAs.
\end{remark}
\vspace{-5pt}
For reference, Table \ref{Table_MISO_Results} summarizes the results for the MISO case, where the array gains are expressed with respect to $\frac{P}{\sigma^2}$.

\begin{table}[!t]
\centering
\setlength{\abovecaptionskip}{0pt}
\resizebox{0.48\textwidth}{!}{
\begin{tabular}{|c|c|c|c|}
\hline
CAPA & \multicolumn{1}{c|}{Array Gain (corresponding to $d_{\mathsf{r}}^{\star}$)} & \multicolumn{1}{c|}{DMT} & \multicolumn{1}{c|}{Array Gain in the DMT Framework} \\ \hline
Linear             & $\frac{\mu({\mathcal{A}}_{\mathsf{r}})L_{{\mathsf{t}},z}({\mathsf{DOF}}_{{\mathsf{r}}_x}!\prod_{\ell=1}^{{\mathsf{DOF}}_{{\mathsf{r}}_x}}\sigma_{{\mathsf{r}}_x,\ell})^{{1}/{\mathsf{DOF}}_{{\mathsf{r}}_x}}}{2^R-1}$                               & ${\mathsf{DOF}}_{{\mathsf{r}}_x}(1-r_{\mathsf{r}})$                        & ${\mu({\mathcal{A}}_{\mathsf{r}})L_{{\mathsf{t}},z}({\mathsf{DOF}}_{{\mathsf{r}}_x}!\prod_{\ell=1}^{{\mathsf{DOF}}_{{\mathsf{r}}_x}}\sigma_{{\mathsf{r}}_x,\ell})^{\frac{1}{{\mathsf{DOF}}_{{\mathsf{r}}_x}(1-r_{\mathsf{r}})}}}$                                          \\ \hline
Planar             & $\frac{\mu({\mathcal{A}}_{\mathsf{r}})({\mathsf{DOF}}_{{\mathsf{r}}}!\prod_{\ell=1}^{{\mathsf{DOF}}_{{\mathsf{r}}}}\sigma_{{\mathsf{r}},\ell})^{{1}/{\mathsf{DOF}}_{{\mathsf{r}}}}}{2^R-1}$                               & ${\mathsf{DOF}}_{{\mathsf{r}}}(1-r_{\mathsf{r}})$                        & $\mu({\mathcal{A}}_{\mathsf{r}})({\mathsf{DOF}}_{{\mathsf{r}}}!\prod_{\ell=1}^{{\mathsf{DOF}}_{{\mathsf{r}}}}\sigma_{{\mathsf{r}},\ell})^{\frac{1}{{\mathsf{DOF}}_{{\mathsf{r}}}(1-r_{\mathsf{r}})}}$                                          \\ \hline
\end{tabular}}
\caption{Theoretical results for MISO}
\label{Table_MISO_Results}
\vspace{-15pt}
\end{table}
\section{MIMO Channels}\label{Section: MIMO Channels}
In the following section, we focus on the MIMO case.
\subsection{Channel Capacity}
The singular value decomposition (SVD) of the spatial response given in \eqref{4FPWD_Model_Standard} can be written as follows:
\begin{align}\label{SVD_MIMO}
{\mathsf{h}}({\mathbf{r}}',{\mathbf{t}})=\sum\nolimits_{\ell=1}^{\infty}\varphi_{\ell}({\mathbf{r}}')\xi_{\ell}({\mathbf{t}})\sqrt{\sigma_{\ell}},
\end{align}
where $\sqrt{\sigma_{1}}\geq\sqrt{\sigma_{2}}\ldots\geq\sqrt{\sigma_{\infty}}\geq0$ are the singular values of ${\mathsf{h}}({\mathbf{r}}',{\mathbf{t}})$, and $\{\varphi_{\ell}(\cdot)\}_{\ell=1}^{\infty}$ and $\{\xi_{\ell}(\cdot)\}_{\ell=1}^{\infty}$ are the associated left-singular and right-singular functions, respectively. The sets $\{\varphi_{\ell}(\cdot)\}_{\ell=1}^{\infty}$ and $\{\xi_{\ell}(\cdot)\}_{\ell=1}^{\infty}$ form orthonormal bases over ${\mathcal{A}}_{{\mathsf{r}}}$ and ${\mathcal{A}}_{{\mathsf{t}}}$, respectively, which satisfy properties similar to those in \eqref{EVD_Linear_Random_Operator_Orthogonality}. The SVD decomposes the spatial channel into parallel, non-interfering sub-channels, enabling optimal power allocation via a water-filling strategy to achieve the channel capacity. However, for CAPA-based fading channels, the properties of $\{\sqrt{\sigma_{\ell}}\}_{\ell=1}^{\infty}$ remain unknown. In the sequel, we investigate the statistics of ${\mathsf{h}}({\mathbf{r}}',{\mathbf{t}})$ and analyze the achievable rate.
\subsection{Channel Statistics}
According to \eqref{SVD_MIMO}, we have
\begin{align}
{R}_{\heartsuit}({\mathbf{t}},{\mathbf{t}}')\triangleq\int_{{{\mathcal{A}}_{{\mathsf{r}}}}}{\mathsf{h}}({\mathbf{r}}',{\mathbf{t}}){\mathsf{h}}^{*}({\mathbf{r}}',{\mathbf{t}}'){\rm{d}}{{\mathbf{r}}'}
=\sum_{\ell=1}^{\infty}\xi_{\ell}({\mathbf{t}})\xi_{\ell}^{*}({\mathbf{t}}'){\sigma_{\ell}}
,\nonumber
\end{align}
where $\sum_{\ell=1}^{\infty}\xi_{\ell}({\mathbf{t}})\xi_{\ell}^{*}({\mathbf{t}}'){\sigma_{\ell}}$ represents the EVD of ${R}_{\heartsuit}({\mathbf{t}},{\mathbf{t}}')$. Define $\hat{R}_{\heartsuit}({\mathbf{t}},{\mathbf{t}}')\triangleq{\mathbbmss{E}}\left\{{R}_{\heartsuit}({\mathbf{t}},{\mathbf{t}}')\right\}$ as the TX-side autocorrelation function. Following the steps to obtain \eqref{Lemma_Autocorrelation_General_Result}, we have
\begin{align}
\hat{R}_{\heartsuit}({\mathbf{t}},{\mathbf{t}}')=
\iint_{{\mathcal{D}}({\bm\kappa})}
\frac{\mu({{\mathcal{A}}_{{\mathsf{r}}}}){\rm{e}}^{{\rm{j}}((t_x-t_x')\kappa_x+(t_z-t_z')\kappa_z)}}{2\pi k_0\gamma(\kappa_x,\kappa_z)}{\rm{d}}\kappa_x{\rm{d}}\kappa_z,\nonumber
\end{align}
which takes a form analogous to \eqref{Lemma_Autocorrelation_General_Result}. Thus, \emph{Landau's eigenvalue theorem} can be used to analyze the eigenvalue behavior of $\hat{R}_{\heartsuit}({\mathbf{t}},{\mathbf{t}}')$. Similarly, the RX-side autocorrelation function of ${\mathsf{h}}({\mathbf{r}}',{\mathbf{t}})$ is defined as $\hat{R}_{\spadesuit}({\mathbf{r}}',{\mathbf{r}}'')\triangleq{\mathbbmss{E}}\left\{{R}_{\spadesuit}({\mathbf{r}}',{\mathbf{r}}'')\right\}$, where
\begin{align}
{R}_{\spadesuit}({\mathbf{r}}',{\mathbf{r}}'')\triangleq\int_{{{\mathcal{A}}_{{\mathsf{t}}}}}{\mathsf{h}}({\mathbf{r}}',{\mathbf{t}}){\mathsf{h}}^{*}({\mathbf{r}}'',{\mathbf{t}}){\rm{d}}{{\mathbf{t}}}
=\sum_{\ell=1}^{\infty}\varphi_{\ell}({\mathbf{r}}')\varphi_{\ell}^{*}({\mathbf{r}}''){\sigma_{\ell}}.\nonumber
\end{align}
Following similar derivation steps to derive \eqref{Lemma_Autocorrelation_General_Result}, we have
\begin{align}
\hat{R}_{\spadesuit}({\mathbf{r}}',{\mathbf{r}}'')=
\iint_{{\mathcal{D}}({\mathbf{k}})}
\frac{\mu({{\mathcal{A}}_{{\mathsf{t}}}}){\rm{e}}^{-{\rm{j}}((r_x'-r_x'')k_x+(r_z'-r_z'')k_z)}}{2\pi k_0\gamma(k_x,k_z)}{\rm{d}}k_x{\rm{d}}k_z.\nonumber
\end{align}
Note that \emph{Landau's eigenvalue theorem} can also analyze the eigenvalues of $\hat{R}_{\spadesuit}({\mathbf{r}}',{\mathbf{r}}'')$. The eigenvalues of the TX-side and RX-side autocorrelation functions are characterized as follows.
\subsubsection{Linear CAPAs}
For a MIMO fading channel employing linear CAPAs along the $x$/$x'$-axis at both the TX and RX, the RX-side and TX-side autocorrelation functions exhibit $\frac{2L_{{\mathsf{t}},x}}{\lambda}=2N_{{\mathsf{t}},x}$ and $\frac{2L_{{\mathsf{r}},x}}{\lambda}=2N_{{\mathsf{r}},x}$ leading eigenvalues, respectively. 
\subsubsection{Planar CAPAs}
For a MIMO fading channel using planar CAPAs at both the TX and RX, the RX-side and TX-side autocorrelation functions exhibit $\frac{\pi L_{{\mathsf{t}},x}L_{{\mathsf{t}},z}}{\lambda^2}=\pi N_{{\mathsf{t}},x}N_{{\mathsf{t}},z}$ and $\frac{\pi L_{{\mathsf{r}},x}L_{{\mathsf{r}},z}}{\lambda^2}=\pi N_{{\mathsf{r}},x}N_{{\mathsf{r}},z}$ leading eigenvalues, respectively.  
\vspace{-5pt}
\begin{remark}\label{remark_MIMO_DoF}
Since ${R}_{\heartsuit}({\mathbf{t}},{\mathbf{t}}')$ shares the same eigenvalues as ${R}_{\spadesuit}({\mathbf{r}}',{\mathbf{r}}'')$, we conclude that, statistically, the considered MIMO channel has $\min\{\pi N_{{\mathsf{r}},x}N_{{\mathsf{r}},z},\pi N_{{\mathsf{t}},x}N_{{\mathsf{t}},z}\}$ significant singular values or dominant sub-channels when planar CAPAs are employed. For linear CAPAs along the $x$/$x'$-axis, the number of significant singular values is $\min\{2N_{{\mathsf{r}},x},2N_{{\mathsf{t}},x}\}$.
\end{remark}
\vspace{-5pt}
\subsection{A Wavenumber-Domain Transmission Framework}
This section implements the wavenumber-domain transmission framework from \cite{sanguinetti2022wavenumber,pizzo2022fourier} to enable CAPA-based MIMO communications that leverage these dominant sub-channels. Before implementing this framework, we first simplify the spatial channel model using the techniques described in \cite{pizzo2022fourier}.
\subsubsection{Linear CAPAs}
We begin with the scenario where linear CAPAs along the $x$/$x'$-axis are employed at both the TX and RX. In this case, the spatial response in \eqref{4FPWD_Model_Standard} reduces to
\begin{equation}\label{4FPWD_Model_Standard_MIMO_Simplified}
\begin{split}
{\mathsf{h}}({\mathbf{r}}',{\mathbf{t}})=\int_{-k_0}^{k_0}\int_{-k_0}^{k_0}
{\rm{e}}^{-{\rm{j}}r_x'k_x}{\tilde{\mathsf{H}}_a(k_x,\kappa_x)}{\rm{e}}^{{\rm{j}}t_x\kappa_x}{\rm{d}}k_x{\rm{d}}\kappa_x,
\end{split}
\end{equation}
where the function $\tilde{\mathsf{H}}_a(k_x,\kappa_x)$ is defined as follows:
\begin{align}
\tilde{\mathsf{H}}_a(k_x,\kappa_x)=\int_{-\sqrt{k_0^2-k_x^2}}^{\sqrt{k_0^2-k_x^2}}\int_{-\sqrt{k_0^2-\kappa_x^2}}^{\sqrt{k_0^2-\kappa_x^2}}\frac{{\mathsf{H}}_a({\mathbf{k}},{\bm\kappa})}{(2\pi)^2}
{\rm{d}}k_z{\rm{d}}\kappa_z.
\end{align}
Based on \eqref{ZUCG_Gaussian_Random_Field_Origin} and \eqref{Angular_Domain_Power_Distribution_Isotropic_Scattering}, $\tilde{\mathsf{H}}_a(k_x,\kappa_x)$ is a zero-mean Gaussian random field over $[-k_0,k_0]\times[-k_0,k_0]$, which satisfies
\begin{align}\label{Wavenumber_Domain_Random_Field_Linear}
{\mathbbmss{E}}\{\tilde{\mathsf{H}}_a(k_x,\kappa_x)\tilde{\mathsf{H}}_a^{*}(k_x',\kappa_x')\}=\frac{\delta(k_x-k_x')\delta(\kappa_x-\kappa_x')}{4k_0^2}.
\end{align}

As discussed earlier, when using linear arrays, the number of dominant sub-channels for the CAPA-based MIMO channel equals $\min\{\frac{2L_{{\mathsf{t}},x}}{\lambda},\frac{2L_{{\mathsf{r}},x}}{\lambda}\}$. To utilize these sub-channels, we uniformly partition the integration region $[-k_0,k_0]$ for $k_x$ and $\kappa_x$ into $\frac{2L_{{\mathsf{r}},x}}{\lambda}=2N_{{\mathsf{r}},x}$ and $\frac{2L_{{\mathsf{t}},x}}{\lambda}=2N_{{\mathsf{t}},x}$ angular sets with spacing interval $\frac{2k_0}{2N_{{\mathsf{r}},x}}=\frac{2\pi}{L_{{\mathsf{r}},x}}$ and $\frac{2k_0}{2N_{{\mathsf{t}},x}}=\frac{2\pi}{L_{{\mathsf{t}},x}}$, respectively:
\begin{align}
&\{[{2\pi\ell}/{L_{{\mathsf{r}},x}},{2\pi(\ell+1)}/{L_{{\mathsf{r}},x}}]|\ell=-{N_{{\mathsf{r}},x}},\ldots,{N_{{\mathsf{r}},x}}-1\},\nonumber\\
&\{[{2\pi m}/{L_{{\mathsf{t}},x}},{2\pi(m+1)}/{L_{{\mathsf{t}},x}}]|m=-{N_{{\mathsf{t}},x}},\ldots,{N_{{\mathsf{t}},x}}-1\}.\nonumber
\end{align}
Applying the mean-value theorem, we approximate the MIMO model for linear CAPAs in \eqref{4FPWD_Model_Standard_MIMO_Simplified} as follows:
\begin{align}\label{4FPWD_SIMO_Model_Linear_Final_Samling_Theorem}
{\mathsf{h}}({\mathbf{r}}',{\mathbf{t}})
\approx\sum_{\ell=-{N_{{\mathsf{r}},x}}}^{{N_{{\mathsf{r}},x}}-1}\sum_{m=-{N_{{\mathsf{t}},x}}}^{{N_{{\mathsf{t}},x}}-1}
{\rm{e}}^{-{\rm{j}}\frac{2\pi\ell r_x'}{L_{{\mathsf{r}},x}}}{h}_{a,\ell,m}
{\rm{e}}^{{\rm{j}}\frac{2\pi m t_x}{L_{{\mathsf{t}},x}}},
\end{align}
where ${h}_{a,\ell,m}
\triangleq{\rm{e}}^{{\rm{j}}\frac{\pi t_x}{L_{{\mathsf{t}},x}}-{\rm{j}}\frac{\pi r_x}{L_{{\mathsf{r}},x}}}
\int_{\frac{2\pi m}{L_{{\mathsf{t}},x}}}^{\frac{2\pi(m+1)}{L_{{\mathsf{t}},x}}}
\int_{\frac{2\pi\ell}{L_{{\mathsf{r}},x}}}^{\frac{2\pi(\ell+1)}{L_{{\mathsf{r}},x}}}
\tilde{\mathsf{H}}_a(k_x,\kappa_x)\times{\rm{d}}k_x{\rm{d}}\kappa_x$ denotes the wavenumber-domain coefficient. The above partitioning becomes asymptotically accurate when $\min\{2N_{{\mathsf{r}},x},2N_{{\mathsf{t}},x}\}\gg1$ or $\min\{L_{{\mathsf{r}},x},L_{{\mathsf{t}},x}\}\gg\lambda$. Based on \eqref{Wavenumber_Domain_Random_Field_Linear}, $\{{h}_{a,\ell,m}\sim{\mathcal{CN}}(0,\sigma_{a,\ell,m}^2)\}_{\ell,m}$ are i.i.d. zero-mean Gaussian random variables, where
\begin{align}
\sigma_{a,\ell,m}^2=\int_{\frac{2\pi m}{L_{{\mathsf{t}},x}}}^{\frac{2\pi(m+1)}{L_{{\mathsf{t}},x}}}
\int_{\frac{2\pi\ell}{L_{{\mathsf{r}},x}}}^{\frac{2\pi(\ell+1)}{L_{{\mathsf{r}},x}}}
\frac{{\rm{d}}k_x{\rm{d}}\kappa_x}{4k_0^2}=\frac{1}{4N_{{\mathsf{t}},x}N_{{\mathsf{r}},x}}.
\end{align}
The sets $\{{\rm{e}}^{-{\rm{j}}\frac{2\pi\ell r_x'}{L_{{\mathsf{r}},x}}}\}_{\ell=-{L_{{\mathsf{r}},x}}/{\lambda}}^{{L_{{\mathsf{r}},x}}/{\lambda}-1}$ and $\{{\rm{e}}^{{\rm{j}}\frac{2\pi m t_x}{L_{{\mathsf{t}},x}}}\}_{m=-{L_{{\mathsf{t}},x}}/{\lambda}}^{{L_{{\mathsf{t}},x}}/{\lambda}-1}$ form orthogonal Fourier bases on $r_x'\in[-\frac{L_{{\mathsf{r}},x}}{2},\frac{L_{{\mathsf{r}},x}}{2}]$ and $t_x\in[-\frac{L_{{\mathsf{t}},x}}{2},\frac{L_{{\mathsf{t}},x}}{2}]$, respectively.
\subsubsection{Planar CAPAs}
For planar arrays, the number of dominant sub-channels equals $\min\{\frac{\pi L_{{\mathsf{t}},x}L_{{\mathsf{t}},z}}{\lambda^2},\frac{\pi L_{{\mathsf{r}},x}L_{{\mathsf{r}},z}}{\lambda^2}\}$. To effectively utilize these sub-channels, we partition $(k_x,k_z)\in{\mathcal{D}}({\mathbf{k}})$ and $(\kappa_x,\kappa_z)\in{\mathcal{D}}({\bm{\kappa}})$ into the following angular sets:
\begin{align}
&{\mathcal{W}}_{\mathsf{r}}(\ell_x,\ell_z)=\left\{\!\begin{smallmatrix}(k_x,k_z)\in\left[\frac{2\pi\ell_x}{L_{{\mathsf{r}},x}},\frac{2\pi(\ell_x+1)}{L_{{\mathsf{r}},x}}\right]\times
\left[\frac{2\pi\ell_z}{L_{{\mathsf{r}},z}},\frac{2\pi(\ell_z+1)}{L_{{\mathsf{r}},z}}\right]\end{smallmatrix}\!\right\}\nonumber,\\
&{\mathcal{W}}_{\mathsf{t}}(m_x,m_z)=\left\{\!\begin{smallmatrix}(\kappa_x,\kappa_z)\in\left[\frac{2\pi m_x}{L_{{\mathsf{t}},x}},\frac{2\pi(m_x+1)}{L_{{\mathsf{t}},x}}\right]\times\left[\frac{2\pi m_z}{L_{{\mathsf{t}},z}},\frac{2\pi(m_z+1)}{L_{{\mathsf{t}},z}}\right]\end{smallmatrix}\!\right\}\nonumber,
\end{align}
for $(\ell_x,\ell_z)\in{\mathcal{E}}_{\mathsf{r}}$ and $(m_x,m_z)\in{\mathcal{E}}_{\mathsf{t}}$, respectively, where
\begin{align}
&{\mathcal{E}}_{\mathsf{r}}=\{(\ell_x,\ell_z)\in{\mathbbmss{Z}}^2|({\ell_x\lambda}/{L_{{\mathsf{r}},x}})^2+({\ell_z\lambda}/{L_{{\mathsf{r}},z}})^2\leq1\}\nonumber,\\
&{\mathcal{E}}_{\mathsf{t}}=\{(m_x,m_z)\in{\mathbbmss{Z}}^2|({m_x\lambda}/{L_{{\mathsf{t}},x}})^2+({m_z\lambda}/{L_{{\mathsf{t}},z}})^2\leq1\}.\nonumber
\end{align}
The cardinalities of these sets satisfy \cite{pizzo2022fourier}
\begin{subequations}\label{4FPWD_Cardinalities}
\begin{align}
\lvert{\mathcal{E}}_{\mathsf{r}}\rvert&=\pi{L_{{\mathsf{r}},x}L_{{\mathsf{r}},z}}/{\lambda^2}+o(\pi{L_{{\mathsf{r}},x}L_{{\mathsf{r}},z}}/{\lambda^2}),\\
\lvert{\mathcal{E}}_{\mathsf{t}}\rvert&=\pi{L_{{\mathsf{t}},x}L_{{\mathsf{t}},z}}/{\lambda^2}+o(\pi{L_{{\mathsf{t}},x}L_{{\mathsf{t}},z}}/{\lambda^2}).
\end{align}
\end{subequations}
Applying the mean-value theorem yields the following planar-array model approximation:
\begin{equation}\label{4FPWD_SIMO_Model_Planar_Final_Samling_Theorem}
\begin{split}
{\mathsf{h}}({\mathbf{r}}',{\mathbf{t}})&\approx \sum\nolimits_{(\ell_x,\ell_z)\in{\mathcal{E}}_{\mathsf{r}}}\sum\nolimits_{(m_x,m_z)\in{\mathcal{E}}_{\mathsf{t}}}
{\rm{e}}^{{\rm{j}}\frac{2\pi m_xt_x}{L_{{\mathsf{t}},x}}}\\
&\times{\rm{e}}^{{\rm{j}}\frac{2\pi m_zt_z}{L_{{\mathsf{t}},z}}}{h}_{a,\ell_x,\ell_z,m_x,m_z}{\rm{e}}^{-{\rm{j}}\frac{2\pi\ell_xr_x'}{L_{{\mathsf{r}},x}}}{\rm{e}}^{-{\rm{j}}\frac{2\pi\ell_zr_z'}{L_{{\mathsf{r}},z}}},
\end{split}
\end{equation}
where ${h}_{a,\ell_x,\ell_z,m_x,m_z}\sim{\mathcal{CN}}(0,\sigma_{a,{\mathsf{r}},\ell_x,\ell_z}^2\sigma_{a,{\mathsf{t}},m_x,m_z}^2)$, with\footnote{A detailed calculation method for the integrals in \eqref{Fading_Coefficients_MIMO} is provided in \cite[Appendix IV-C]{pizzo2020spatially}. The associated code is accessible at: \url{https://github.com/lucasanguinetti/Holographic-MIMO-Small-Scale-Fading}.}
\begin{subequations}\label{Fading_Coefficients_MIMO}
\begin{align}
\sigma_{a,{\mathsf{r}},\ell_x,\ell_z}^2&=\iint_{{\mathcal{W}}_{\mathsf{r}}(\ell_x,\ell_z)\cap{\mathcal{D}}({\mathbf{k}})}\frac{A_{s}(k_0)}{(2\pi)^2} \frac{{\rm{d}}k_x{\rm{d}}k_z}{\gamma(k_x,k_z)},\\
\sigma_{a,{\mathsf{t}},m_x,m_z}^2&=\iint_{{\mathcal{W}}_{\mathsf{t}}(m_x,m_z)\cap{\mathcal{D}}({\bm{\kappa}})}\frac{{\rm{d}}\kappa_x{\rm{d}}\kappa_z}{(2\pi)^2} \frac{A_{s}(k_0)}{\gamma(\kappa_x,\kappa_z)}.
\end{align}
\end{subequations}
The sets $\{{\rm{e}}^{-{\rm{j}}\frac{2\pi\ell_xr_x'}{L_{{\mathsf{r}},x}}}{\rm{e}}^{-{\rm{j}}\frac{2\pi\ell_zr_z'}{L_{{\mathsf{r}},z}}}\}_{(\ell_x,\ell_z)\in{\mathcal{E}}_{\mathsf{r}}}$ and $\{{\rm{e}}^{{\rm{j}}\frac{2\pi m_xt_x}{L_{{\mathsf{t}},x}}}{\rm{e}}^{{\rm{j}}\frac{2\pi m_zt_z}{L_{{\mathsf{t}},z}}}\}_{(m_x,m_z)\in{\mathcal{E}}_{\mathsf{t}}}$ form orthogonal Fourier bases on $(r_x',r_z')\in{\mathcal{A}}_{\mathsf{r}}$ and $(t_x,t_z)\in{\mathcal{A}}_{\mathsf{t}}$, respectively.

The expansion in \eqref{4FPWD_SIMO_Model_Planar_Final_Samling_Theorem} becomes asymptotically tight when $\min\{\frac{\pi L_{{\mathsf{t}},x}L_{{\mathsf{t}},z}}{\lambda^2},\frac{\pi L_{{\mathsf{r}},x}L_{{\mathsf{r}},z}}{\lambda^2}\}\gg1$. Combining this with Remark \ref{remark_MIMO_DoF}, we conclude that for electromagnetically large CAPAs where $\min\{L_{{\mathsf{r}},x},L_{{\mathsf{t}},x},L_{{\mathsf{r}},z},L_{{\mathsf{t}},z}\}\gg\lambda$, the number of significant singular values in the spatial channel equals the number of angular sets used for spatial model approximation. This relationship yields a uniform MIMO channel representation:
\begin{align}\label{MIMO_CAPA_General_Statistical_Model}
{\mathsf{h}}({\mathbf{r}}',{\mathbf{t}})\approx\sum\nolimits_{i=1}^{{\mathsf{D}}_{\mathsf{r}}}\sum\nolimits_{j=1}^{{\mathsf{D}}_{\mathsf{t}}}
\psi_{i}({\mathbf{r}}')h_{i,j}\phi_{j}({\mathbf{t}}),
\end{align}
where the set $\{h_{i,j}\sim{\mathcal{CN}}(0,\varrho_{{\mathsf{r}},i}^2\varrho_{{\mathsf{t}},j}^2)\}_{i=1,j=1}^{i={\mathsf{D}}_{\mathsf{r}},j={\mathsf{D}}_{\mathsf{t}}}$ contains the wavenumber-domain channel coefficients, and the orthogonal Fourier bases $\{\phi_{j}({\mathbf{t}})\}_{j=1}^{{\mathsf{D}}_{\mathsf{t}}}$ and $\{\psi_{i}({\mathbf{r}})\}_{i=1}^{{\mathsf{D}}_{\mathsf{r}}}$ satisfy \cite{ouyang2024primer}
\begin{subequations}\label{MIMO_CAPA_General_Statistical_Model_Orthogonal}
\begin{align}
&\int_{\mathcal{A}_{\mathsf{t}}}\phi_{j}({\mathbf{t}})\phi_{j'}^{*}({\mathbf{t}}){\rm{d}}{\mathbf{t}}=\mu(\mathcal{A}_{\mathsf{t}})\delta_{j,j'}=L_{{\mathsf{t}},x}L_{{\mathsf{t}},z}\delta_{j,j'},
\label{MISO_CAPA_General_Statistical_Model_Orthogonal}\\
&\int_{\mathcal{A}_{\mathsf{t}}}\psi_{i}({\mathbf{r}}')\psi_{i'}^{*}({\mathbf{r}}'){\rm{d}}{\mathbf{r}}'=\mu(\mathcal{A}_{\mathsf{r}})\delta_{i,i'}=L_{{\mathsf{r}},x}L_{{\mathsf{r}},z}\delta_{i,i'}.
\label{SIMO_CAPA_General_Statistical_Model_Orthogonal}
\end{align}
\end{subequations}
Table \ref{Table_MISO_SIMO_Parameters} summarizes the parameter settings for the system.
\vspace{-5pt}
\begin{remark}
Equation \eqref{MIMO_CAPA_General_Statistical_Model} approximates the SVD in \eqref{SVD_MIMO}. This approximation achieves asymptotic accuracy for electromagnetically large CAPAs where $\min\{L_{{\mathsf{r}},x},L_{{\mathsf{t}},x},L_{{\mathsf{r}},z},L_{{\mathsf{t}},z}\}\gg\lambda$. The authors in \cite{pizzo2022fourier} interpreted \eqref{MIMO_CAPA_General_Statistical_Model} through the Fourier relationship between spatial response and wavenumber-domain (or angular-domain) response. Their analysis confirmed that ${\mathsf{h}}({\mathbf{r}}',{\mathbf{t}})$ can be approximated without information loss by \eqref{MIMO_CAPA_General_Statistical_Model} when $\min\{L_{{\mathsf{r}},x},L_{{\mathsf{t}},x},L_{{\mathsf{r}},z},L_{{\mathsf{t}},z}\}\gg\lambda$. 
\end{remark}
\vspace{-5pt}

\begin{table}[!t]
\centering
\setlength{\abovecaptionskip}{0pt}
\resizebox{0.45\textwidth}{!}{
\begin{tabular}{|c|c|c|c|c|c|}
\hline
\textbf{}                & \textbf{CAPA}   & ${\mathsf{D}}_{\mathsf{t}}$ or ${\mathsf{D}}_{\mathsf{r}}$ & $\{\phi_{j}({\mathbf{t}})\}$ or $\{\psi_{i}({\mathbf{r}}')\}$ & $\{\varrho_{{\mathsf{t}},j}^2\}$ or $\{\varrho_{{\mathsf{r}},i}^2\}$ & $\mu(\mathcal{A}_{\mathsf{t}})$ or $\mu(\mathcal{A}_{\mathsf{r}})$ \\ \hline
\multirow{2}{*}{\textbf{TX}} & \textbf{Linear} & $2N_{{\mathsf{t}},x}$ & $\{{\rm{e}}^{{\rm{j}}\frac{2\pi\ell t_x}{L_{{\mathsf{t}},x}}}\}_{\ell=-\frac{L_{{\mathsf{t}},x}}{\lambda}}^{\frac{L_{{\mathsf{t}},x}}{\lambda}-1}$     & $\{\frac{1}{2\sqrt{N_{{\mathsf{t}},x}}}\}$ & $L_{{\mathsf{t}},x}L_{{\mathsf{t}},z}$      \\ \cline{2-6} 
                      & \textbf{Planar} & $\pi N_{{\mathsf{t}},x}N_{{\mathsf{t}},z}$ & $\{{\rm{e}}^{{\rm{j}}\frac{2\pi m_xt_x}{L_{{\mathsf{t}},x}}}{\rm{e}}^{{\rm{j}}\frac{2\pi m_zt_z}{L_{{\mathsf{t}},z}}}\}_{(m_x,m_z)\in{\mathcal{E}}_{\mathsf{t}}}$     & $\{\sigma_{a,{\mathsf{t}},m_x,m_z}^2\}$ & $L_{{\mathsf{t}},x}L_{{\mathsf{t}},z}$      \\ \hline
\multirow{2}{*}{\textbf{RX}} & \textbf{Linear} & $2N_{{\mathsf{r}},x}$ & $\{{\rm{e}}^{-{\rm{j}}\frac{2\pi\ell r_x}{L_{{\mathsf{r}},x}}}\}_{\ell=-\frac{L_{{\mathsf{r}},x}}{\lambda}}^{\frac{L_{{\mathsf{r}},x}}{\lambda}-1}$     & $\{\frac{1}{2\sqrt{N_{{\mathsf{r}},x}}}\}$ & $L_{{\mathsf{r}},x}L_{{\mathsf{r}},z}$      \\ \cline{2-6} 
                      & \textbf{Planar} & $\pi N_{{\mathsf{r}},x}N_{{\mathsf{r}},z}$ & $\{{\rm{e}}^{-{\rm{j}}\frac{2\pi\ell_xr_x}{L_{{\mathsf{r}},x}}}{\rm{e}}^{-{\rm{j}}\frac{2\pi\ell_zr_z}{L_{{\mathsf{r}},z}}}\}_{(\ell_x,\ell_z)\in{\mathcal{E}}_{\mathsf{r}}}$     & $\{\sigma_{a,{\mathsf{r}},\ell_x,\ell_z}^2\}$ & $L_{{\mathsf{r}},x}L_{{\mathsf{r}},z}$      \\ \hline
\end{tabular}}
\caption{Parameter settings for MIMO channels}
\label{Table_MISO_SIMO_Parameters}
\vspace{-15pt}
\end{table}

\subsubsection{Transmission Framework}
By defining ${\bm\Psi}({\mathbf{r}}')\triangleq [\psi_{1}({\mathbf{r}}'),\ldots,\psi_{{\mathsf{D}}_{\mathsf{r}}}({\mathbf{r}}')]$ and ${\bm\Phi}({\mathbf{t}})\triangleq [\phi_{1}({\mathbf{t}}),\ldots,\phi_{{\mathsf{D}}_{\mathsf{t}}}({\mathbf{t}})]^{\mathsf{T}}$, we rewrite \eqref{MIMO_CAPA_General_Statistical_Model} as follows:
\begin{align}\label{MIMO_CAPA_General_Statistical_Model_Matrix}
{\mathsf{h}}({\mathbf{r}}',{\mathbf{t}})\approx {\bm\Psi}({\mathbf{r}}'){\mathbf{H}}{\bm\Phi}({\mathbf{t}}),
\end{align}
where ${\mathbf{H}}\triangleq[h_{i,j}]\in{\mathbbmss{C}}^{{\mathsf{D}}_{\mathsf{r}}\times {\mathsf{D}}_{\mathsf{t}}}$. It follows from \eqref{MIMO_CAPA_General_Statistical_Model_Orthogonal} that
\begin{align}
\int_{\mathcal{A}_{\mathsf{t}}}\frac{{\bm\Phi}({\mathbf{t}}){\bm\Phi}^{\mathsf{H}}({\mathbf{t}})}{\mu(\mathcal{A}_{\mathsf{t}})}{\rm{d}}{\mathbf{t}}={\mathbf{I}}_{{\mathsf{D}}_{\mathsf{t}}},~
\int_{\mathcal{A}_{\mathsf{r}}}\frac{{\bm\Psi}^{\mathsf{H}}({\mathbf{r}}'){\bm\Psi}({\mathbf{r}}')}{\mu(\mathcal{A}_{\mathsf{r}})}{\rm{d}}{\mathbf{r}}={\mathbf{I}}_{{\mathsf{D}}_{\mathsf{r}}}.
\end{align}
Furthermore, the rank of ${\mathbf{H}}$ is $\min\{{\mathsf{D}}_{\mathsf{r}},{\mathsf{D}}_{\mathsf{t}}\}$, which equals the number of significant sub-channels in the spatial response ${\mathsf{h}}({\mathbf{r}}',{\mathbf{t}})$, as noted in Remark \ref{remark_MIMO_DoF}.

\begin{figure*}[!t]
 \centering
\setlength{\abovecaptionskip}{0pt}
\includegraphics[height=0.12\textwidth]{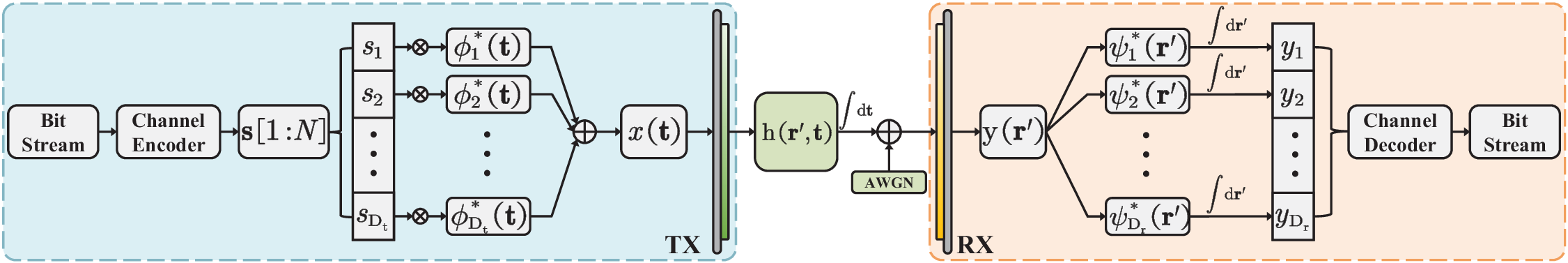}
\caption{Illustration of the wavenumber-domain transmission framework for CAPAs.}
\label{Figure: MIMO}
\vspace{-15pt}
\end{figure*}

Building on the model in \eqref{MIMO_CAPA_General_Statistical_Model} or \eqref{MIMO_CAPA_General_Statistical_Model_Matrix} and utilizing the orthogonality of Fourier bases, we propose the transmission framework in {\figurename} {\ref{Figure: MIMO}} to achieve capacity for CAPA-based MIMO channels. The bit stream is first encoded into a codeword ${\mathbf{s}}[1:N]\in{\mathbbmss{C}}^{{{\mathsf{D}}_{\mathsf{t}}}\times N}$ of length $N$. For each channel use, the coded data vector $\mathbf{s}=[s_1;\ldots;s_{{\mathsf{D}}_{\mathsf{t}}}]\in{\mathbbmss{C}}^{{\mathsf{D}}_{\mathsf{t}}\times1}$ is mapped to the spatial signal using the Fourier basis $\{\phi_{j}({\mathbf{t}})\}_{j=1}^{{\mathsf{D}}_{\mathsf{t}}}$ as follows:
\begin{align}\label{MIMO_Transmit}
x(\mathbf{t})=\sum\nolimits_{j=1}^{{\mathsf{D}}_{\mathsf{t}}}s_j\phi_{j}^{*}({\mathbf{t}})={\bm\Phi}^{\mathsf{H}}({\mathbf{t}}){\mathbf{s}}.
\end{align}
This signal propagates through the spatial channel ${\mathsf{h}}({\mathbf{r}}',{\mathbf{t}})$. Substituting \eqref{MIMO_CAPA_General_Statistical_Model_Matrix} and \eqref{MIMO_Transmit} into \eqref{CAPA_Basic_Signal_Model} gives
\begin{subequations}
\begin{align}
{\mathsf{y}}(\mathbf{r}')&={\bm\Psi}({\mathbf{r}}'){\mathbf{H}}\int_{{\mathcal{A}}_{\mathsf{t}}}{\bm\Phi}({\mathbf{t}}){\bm\Phi}^{\mathsf{H}}({\mathbf{t}})
{\rm{d}}{\mathbf{t}}\times{\mathbf{s}}+{\mathsf{z}}(\mathbf{r}')\\
&=\mu(\mathcal{A}_{\mathsf{t}}){\bm\Psi}({\mathbf{r}}'){\mathbf{H}}{\mathbf{s}}+{\mathsf{z}}(\mathbf{r}').
\end{align}
\end{subequations}
At the receiver, ${\mathsf{D}}_{\mathsf{r}}$ equalizers are used, each computing $y_i=\int_{{\mathcal{A}}_{\mathsf{r}}}\psi_{i}^{*}({\mathbf{r}}'){\mathsf{y}}(\mathbf{r}'){\rm{d}}{\mathbf{r}}'$ for $i=1,\ldots,{\mathsf{D}}_{\mathsf{r}}$. The output ${\mathbf{y}}=[y_1;\ldots;y_{{\mathsf{D}}_{\mathsf{r}}}]\in{\mathbbmss{C}}^{{\mathsf{D}}_{\mathsf{r}}\times1}$ is fed to a maximum-likelihood decoder to recover the bit stream. This results in
\begin{subequations}\label{MIMO_Filtering_Sigal_Model}
\begin{align}
{\mathbf{y}}&=\int_{{\mathcal{A}}_{\mathsf{r}}}{\bm\Psi}^{\mathsf{H}}({\mathbf{r}}')\left(\mu(\mathcal{A}_{\mathsf{t}}){\bm\Psi}({\mathbf{r}}'){\mathbf{H}}{\mathbf{s}}\right)
{\rm{d}}{\mathbf{r}}'
+{\mathbf{z}}\\
&=\mu(\mathcal{A}_{\mathsf{t}})\mu(\mathcal{A}_{\mathsf{r}}){\mathbf{H}}{\mathbf{s}}+{\mathbf{z}},
\end{align}
\end{subequations}
where ${\mathbf{z}}=\int_{{\mathcal{A}}_{\mathsf{r}}}{\bm\Psi}^{\mathsf{H}}({\mathbf{r}}'){\mathsf{z}}(\mathbf{r}'){\rm{d}}{\mathbf{r}}'\sim{\mathcal{CN}}({\mathbf{0}},\sigma^2\mu({\mathcal{A}}_{\mathsf{r}}){\mathbf{I}}_{{\mathsf{D}}_{\mathsf{r}}})$. \vspace{-5pt}
\begin{remark}
The results in \eqref{MIMO_Filtering_Sigal_Model} show that data is transmitted through the channel characterized by ${\mathbf{H}}$. Since $\mathbf{H}$ is constructed using angular/wavenumber-domain coefficients, the framework in {\figurename} {\ref{Figure: MIMO}} is called the wavenumber-domain transmission framework.
\end{remark}
\vspace{-5pt}
\subsection{Performance Analysis}\label{Section:MIMO:Performance Analysis}
We assume the coded data vectors are i.i.d. Gaussian variables that satisfy the power constraint ${\mathbbmss{E}}\{\int_{\mathcal{A}_{\mathsf{t}}}\lvert{x}(\mathbf{t})\rvert^2{\rm{d}}{\mathbf{t}}\}=\mu(\mathcal{A}_{\mathsf{t}}){\mathsf{tr}}(\mathbbmss{E}\{{\mathbf{s}}{\mathbf{s}}^{\mathsf{H}}\})=P$ with ${\mathbf{s}}\sim{\mathcal{CN}}({\mathbf{0}},\frac{P}{{\mathsf{D}}_{\mathsf{t}}\mu(\mathcal{A}_{\mathsf{t}})}{\mathbf{I}}_{{\mathsf{D}}_{\mathsf{t}}})$\footnote{We use an equal-power allocation-based input since this strategy achieves capacity in the high-SNR regime.}. The channel capacity is then expressed as follows:
\begin{align}\label{MIMO_Rate_CAPA_Basic}
{\mathsf{C}}_{\mathsf{mm}}=\log_2\det({\mathbf{I}}_{{\mathsf{D}}_{\mathsf{r}}}+{\overline{\gamma}}{\mathbf{H}}{\mathbf{H}}^{\mathsf{H}}),
\end{align}
with $\overline{\gamma}=\frac{\mu(\mathcal{A}_{\mathsf{r}})\mu(\mathcal{A}_{\mathsf{t}})P}{{{\mathsf{D}}_{\mathsf{t}}}\sigma^2}$. Recalling that $[{\mathbf{H}}]_{i,j}\sim{\mathcal{CN}}(0,\varrho_{{\mathsf{r}},i}^2\varrho_{{\mathsf{t}},j}^2)$ gives ${\mathbf{H}}\overset{d}{=}{\mathbf{R}}^{\frac{1}{2}}{\tilde{\mathbf{H}}}
{\mathbf{T}}^{\frac{1}{2}}$, where ${\mathbf{R}}={\mathsf{diag}}\{[\varrho_{{\mathsf{r}},1}^2;\ldots;\varrho_{{\mathsf{r}},{\mathsf{D}}_{\mathsf{r}}}^2]\}$, ${\mathbf{T}}={\mathsf{diag}}\{[\varrho_{{\mathsf{t}},1}^2;\ldots;\varrho_{{\mathsf{t}},{\mathsf{D}}_{\mathsf{t}}}^2]\}$, and ${\tilde{\mathbf{H}}}\in{\mathbbmss{C}}^{{\mathsf{D}}_{\mathsf{r}}\times {\mathsf{D}}_{\mathsf{t}}}$ is a complex Gaussian matrix with i.i.d. zero-mean unit-variance entries. 
\subsubsection{Multiplexing Gain}
The ECC of the MIMO channel can be expressed as follows:
\begin{subequations}
\begin{align}
{\mathbbmss{E}}\{{\mathsf{C}}_{\mathsf{mm}}\}&={\mathbbmss{E}}_{\tilde{\mathbf{H}}}\{\log_2\det({\mathbf{I}}_{{\mathsf{D}}_{\mathsf{r}}}+{\overline{\gamma}}
{\mathbf{R}}^{\frac{1}{2}}{\tilde{\mathbf{H}}}
{\mathbf{T}}{\tilde{\mathbf{H}}}^{\mathsf{H}}{\mathbf{R}}^{\frac{1}{2}})\}\label{MIMO_Capacity_Transform0}\\
&={\mathbbmss{E}}_{\tilde{\mathbf{H}}}\{\log_2\det({\mathbf{I}}_{{\mathsf{D}}_{\mathsf{t}}}+{\overline{\gamma}}
{\mathbf{T}}^{\frac{1}{2}}{\tilde{\mathbf{H}}}^{\mathsf{H}}{\mathbf{R}}{\tilde{\mathbf{H}}}{\mathbf{T}}^{\frac{1}{2}})\},\label{MIMO_Capacity_Transform1}
\end{align}
\end{subequations}
where \eqref{MIMO_Capacity_Transform1} follows from Sylvester's determinant identity. When $\mathbf{T}$ and $\mathbf{R}$ have distinct eigenvalues $\varrho_{{\mathsf{t}},j}^2$ and $\varrho_{{\mathsf{r}},i}^2$, respectively, an analytically tractable closed-form solution for ${\mathbbmss{E}}\{{\mathsf{C}}_{\mathsf{mm}}\}$ is available by utilizing the moment generating function (MGF) of ${\mathsf{C}}_{\mathsf{mm}}$; see \cite[Eq. (63)]{ghaderipoor2012application}. However, in our system model, $\mathbf{T}$ and $\mathbf{R}$ may contain repeated eigenvalues due to the \emph{angular power distribution} $S({\mathbf{k}},{\bm\kappa})$, making it challenging to derive a closed-form ${\mathbbmss{E}}\{{\mathsf{C}}_{\mathsf{mm}}\}$. We therefore focus on the high-SNR ECC.
\vspace{-5pt}
\begin{theorem}\label{Theorem_MIMO_CAPA_Multiplexing}
When ${\mathsf{D}}_{\mathsf{r}}={\mathsf{D}}_{\mathsf{t}}$, we have
\begin{align}\label{Result1_Theorem_MIMO_CAPA_Multiplexing}
\lim_{{\overline{\gamma}}\rightarrow\infty}{\mathbbmss{E}}\{{\mathsf{C}}_{\mathsf{mm}}\}\simeq {\mathsf{D}}_{\mathsf{t}}\log_2{\overline{\gamma}}+\log_2\det({\mathbf{R}}{\mathbf{T}})+\epsilon_0,
\end{align}
where $\epsilon_0 =\frac{1}{\log{2}}\sum_{\epsilon=0}^{{\mathsf{D}}_{\mathsf{t}}-1}\psi({\mathsf{D}}_{\mathsf{t}}-\epsilon)$. When ${\mathsf{D}}_{\mathsf{r}}\ne {\mathsf{D}}_{\mathsf{t}}$, we have
\begin{align}\label{Result2_Theorem_MIMO_CAPA_Multiplexing}
\lim_{{\overline{\gamma}}\rightarrow\infty}{\mathbbmss{E}}\{{\mathsf{C}}_{\mathsf{mm}}\}\simeq {\mathsf{n}}\log_2{\overline{\gamma}}+\log_2\det({\mathbf{A}})+\epsilon_1,
\end{align}
where $\epsilon_1={\mathbbmss{E}}\{\log_2\det({\mathbf{G}}{\mathbf{B}}{\mathbf{G}}^{\mathsf{H}})\}\in[{\sigma_{\diamond}^2}\epsilon_2,{\sigma_{\circ}^2}\epsilon_2]$, ${\mathbf{G}}\in{\mathbbmss{C}}^{{\mathsf{n}}\times {\mathsf{m}}}$ is a complex Gaussian matrix with zero-mean unit-variance entries, and $\epsilon_2 =\frac{1}{\log{2}}\sum_{\epsilon=0}^{{\mathsf{n}}-1}\psi({\mathsf{m}}-\epsilon)$. Furthermore, we have ${\mathsf{n}}=\min\{{\mathsf{D}}_{\mathsf{t}},{\mathsf{D}}_{\mathsf{r}}\}$, ${\mathsf{m}}=\max\{{\mathsf{D}}_{\mathsf{t}},{\mathsf{D}}_{\mathsf{r}}\}$, and
\begin{align}
\left\{\!\!\!\begin{array}{ll}
{\mathbf{A}}={\mathbf{T}},{\mathbf{B}}={\mathbf{R}},\sigma_{\diamond}^2=\min_{i}\varrho_{{\mathsf{r}},i}^2,\sigma_{\circ}^2=\max_{i}\varrho_{{\mathsf{r}},i}^2&\!\!{\mathsf{n}}={\mathsf{D}}_{\mathsf{t}}\\
{\mathbf{A}}={\mathbf{R}},{\mathbf{B}}={\mathbf{T}},\sigma_{\diamond}^2=\min_{j}\varrho_{{\mathsf{t}},j}^2,\sigma_{\circ}^2=\max_{j}\varrho_{{\mathsf{t}},j}^2&\!\!{\mathsf{n}}={\mathsf{D}}_{\mathsf{r}}
\end{array}
\right..\nonumber
\end{align}
\end{theorem}
\vspace{-5pt}
\begin{IEEEproof}
Please refer to Appendix \ref{Proof_MIMO_Theorem} for more details.
\end{IEEEproof}
\vspace{-5pt}
\begin{remark}
The results in Theorem \ref{Theorem_MIMO_CAPA_Multiplexing} suggest that the maximal multiplexing gain in our considered CAPA-based MIMO channel is $r_{\mathsf{mm}}^{\star}=\min\{{\mathsf{D}}_{\mathsf{t}},{\mathsf{D}}_{\mathsf{r}}\}$, i.e., the rank of $\mathbf{H}$.
\end{remark}
\vspace{-5pt}
\subsubsection{Diversity Gain}
Obtaining a tractable expression for the OP ${\mathcal{P}}_{\mathsf{mm}}=\Pr({\mathsf{C}}_{\mathsf{mm}}<R)$ is more challenging than analyzing the ECC. We thus focus on its high-SNR characteristics.
\vspace{-5pt}
\begin{theorem}\label{Theorem_MIMO_CAPA_Diversity}
Given $R>0$, it holds that
\begin{align}\label{Result_Theorem_MIMO_CAPA_Diversity}
\lim_{{\overline{\gamma}}\rightarrow\infty}{\mathcal{P}}_{\mathsf{mm}}\simeq \frac{\epsilon_{{\mathsf{t}},{\mathsf{r}}}{\overline{\gamma}}^{-{\mathsf{D}}_{\mathsf{r}}{\mathsf{D}}_{\mathsf{t}}}}
{(\prod_{j=1}^{{\mathsf{D}}_{\mathsf{t}}}\varrho_{{\mathsf{t}},j}^2)^{{\mathsf{D}}_{\mathsf{r}}}(\prod_{i=1}^{{\mathsf{D}}_{\mathsf{r}}}\varrho_{{\mathsf{r}},i}^2)^{{\mathsf{D}}_{\mathsf{t}}}},
\end{align}
where $\epsilon_{{\mathsf{t}},{\mathsf{r}}}=G_{{\mathsf{n}}+1,{\mathsf{n}}+1}^{0,{\mathsf{n}}+1}\left(\left._{0,1,\ldots,{\mathsf{n}}}^{1,1+{\mathsf{m}},\ldots,{\mathsf{n}}+{\mathsf{m}}}\right|2^{R}\right)$ with $G_{\cdot,\cdot}^{\cdot,\cdot}(\cdot)$ denoting the Meijer G-function \cite[Eq. (9.301)]{gradshteyn2014table}.
\end{theorem}
\vspace{-5pt}
\begin{IEEEproof}
Please refer to Appendix \ref{Proof_MIMO_Theorem} for more details.
\end{IEEEproof}
\vspace{-5pt}
\begin{remark}
Theorem \ref{Theorem_MIMO_CAPA_Diversity} indicates that the maximal diversity gain in CAPA-based MIMO channels equals ${\mathsf{D}}_{\mathsf{t}}{\mathsf{D}}_{\mathsf{r}}$, and the array gain is given by ${(\prod_{j=1}^{{\mathsf{D}}_{\mathsf{t}}}\varrho_{{\mathsf{t}},j}^2)^{\frac{1}{{\mathsf{D}}_{\mathsf{t}}}}
(\prod_{i=1}^{{\mathsf{D}}_{\mathsf{r}}}\varrho_{{\mathsf{r}},i}^2)^{\frac{1}{{\mathsf{D}}_{\mathsf{r}}}}}
{\epsilon_{{\mathsf{t}},{\mathsf{r}}}^{-\frac{1}{{\mathsf{D}}_{\mathsf{r}}{\mathsf{D}}_{\mathsf{t}}}}}$.
\end{remark}
\vspace{-5pt}
\subsubsection{Diversity-Multiplexing Trade-off}
The DMT for CAPA-based MIMO systems is characterized as follows:
\begin{align}
d_{\mathsf{mm}}(r_{\mathsf{mm}})=\lim_{\overline{\gamma}\rightarrow\infty}\frac{\log(\Pr({\mathsf{C}}_{\mathsf{mm}}<r_{\mathsf{mm}}\log_2(1+\overline{\gamma})))}{-\log{\overline{\gamma}}},
\end{align}
where $d_{\mathsf{mm}}(r_{\mathsf{mm}})$ represents the achievable diversity gain for a target multiplexing gain $r_{\mathsf{mm}}$. The array gain in the DMT framework satisfies
\begin{align}
a_{\mathsf{mm}}^{-d_{\mathsf{mm}}(r_{\mathsf{mm}})}(r_{\mathsf{mm}})=\lim_{\overline{\gamma}\rightarrow\infty}
\frac{\Pr({\mathsf{C}}_{\mathsf{mm}}<r_{\mathsf{mm}}\log_2(1+\overline{\gamma}))}{\overline{\gamma}^{-d_{\mathsf{mm}}(r_{\mathsf{mm}})}}.\nonumber
\end{align}
The corresponding results are summarized below.
\vspace{-5pt}
\begin{theorem}\label{Theorem_MIMO_CAPA_DMT}
Given the target multiplexing gain $r_{\mathsf{mm}}$, it has
\begin{align}
d_{\mathsf{mm}}(r_{\mathsf{mm}})=G_d(\lfloor r_{\mathsf{mm}}\rfloor)-G_r(\lfloor r_{\mathsf{mm}}\rfloor)r_{\mathsf{mm}}
\end{align}
for $r_{\mathsf{mm}}\in(0,\min\{{\mathsf{D}}_{\mathsf{t}},{\mathsf{D}}_{\mathsf{r}}\})$, where $G_d(x)={\mathsf{D}}_{\mathsf{r}}{\mathsf{D}}_{\mathsf{t}}-x(x+1)$ and $G_r(x)={\mathsf{D}}_{\mathsf{r}}+{\mathsf{D}}_{\mathsf{t}}-(2x+1)$.
The array gain associated with $d_{\mathsf{mm}}(r_{\mathsf{mm}})$ satisfies 
\begin{align}
a_{\mathsf{mm}}(r_{\mathsf{mm}})\in[f_{r_{\mathsf{mm}}}(\min\nolimits_{i,j}\varrho_{{\mathsf{t}},j}^2\varrho_{{\mathsf{r}},i}^2),
f_{r_{\mathsf{mm}}}(\max\nolimits_{i,j}\varrho_{{\mathsf{t}},j}^2\varrho_{{\mathsf{r}},i}^2)], \nonumber
\end{align}
where $f_{r_{\mathsf{mm}}}(x)=(K_{{\mathsf{m}},{\mathsf{n}}}\det({\mathbf{K}}_{{\mathsf{m}},{\mathsf{n}}}(\lfloor r_{\mathsf{mm}}\rfloor))(\prod\nolimits_{t=1}^{\lfloor r_{\mathsf{mm}}\rfloor}\Gamma(t)t!\Big)\times(\frac{x^{-G_d(\lfloor r_{\mathsf{mm}}\rfloor)}}{G_r(\lfloor r_{\mathsf{mm}}\rfloor)}))^{-\frac{1}{d_{\mathsf{mm}}(r_{\mathsf{mm}})}}$. Moreover, ${\mathsf{n}}=\min\{{\mathsf{D}}_{\mathsf{t}},{\mathsf{D}}_{\mathsf{r}}\}$, ${\mathsf{m}}=\max\{{\mathsf{D}}_{\mathsf{t}},{\mathsf{D}}_{\mathsf{r}}\}$, $K_{{\mathsf{m}},{\mathsf{n}}}=\prod_{t=1}^{{\mathsf{n}}}\frac{1}{({\mathsf{n}}-t)!({\mathsf{m}}-t)!}$, and matrix ${\mathbf{K}}_{{\mathsf{m}},{\mathsf{n}}}(x)$ is defined as $[{\mathbf{K}}_{{\mathsf{m}},{\mathsf{n}}}(x)]_{u,v}=\sum_{i=0}^{{\mathsf{m}}-{\mathsf{n}}}\binom{{\mathsf{m}}-{\mathsf{n}}}{i}\frac{(-1)^i}{u+v+i}$ for $u,v=1,\ldots,{\mathsf{n}}-x-1$.
\end{theorem}
\vspace{-5pt}
\begin{IEEEproof}
Please refer to Appendix \ref{Proof_MIMO_Theorem} for more details.
\end{IEEEproof}
\vspace{-5pt}
\begin{corollary}\label{Corollary_MIMO_IID_Rayleigh_Array_Gain}
When $\varrho_{{\mathsf{r}},i}=\varrho_{{\mathsf{r}}}$ for $i=1,\ldots, {\mathsf{D}}_{\mathsf{r}}$ and $\varrho_{{\mathsf{t}},j}=\varrho_{{\mathsf{t}}}$ for $j=1,\ldots, {\mathsf{D}}_{\mathsf{t}}$, the array gain in the DMT framework is given by $a_{\mathsf{mm}}(r_{\mathsf{mm}})=f_{r_{\mathsf{mm}}}({\varrho_{{\mathsf{r}}}^{2}\varrho_{{\mathsf{t}}}^{2}})$.
\end{corollary}
\vspace{-5pt}
\begin{IEEEproof}
Similar to the proof of Theorem \ref{Theorem_MIMO_CAPA_DMT}.
\end{IEEEproof}
\vspace{-5pt}
\begin{remark}
The above results show the wavenumber-domain transmission framework achieves a maximal diversity gain of ${\mathsf{D}}_{\mathsf{t}}{\mathsf{D}}_{\mathsf{r}}$ and a multiplexing gain of $\min\{{\mathsf{D}}_{\mathsf{t}},{\mathsf{D}}_{\mathsf{r}}\}$. Since ${\mathsf{D}}_{\mathsf{t}}$ and ${\mathsf{D}}_{\mathsf{r}}$ denote the numbers of significant eigenvalues of the TX-side and RX-side autocorrelation functions of ${\mathsf{h}}({\mathbf{r}}',{\mathbf{t}})$, respectively, we conclude that the wavenumber-domain framework effectively utilizes the spatial DoFs offered by deploying CAPAs.
\end{remark}
\vspace{-5pt}
\section{Spatially-Discrete Arrays}\label{Section_SPDA}
\begin{figure}[!t]
 \centering
\setlength{\abovecaptionskip}{0pt}
\includegraphics[height=0.22\textwidth]{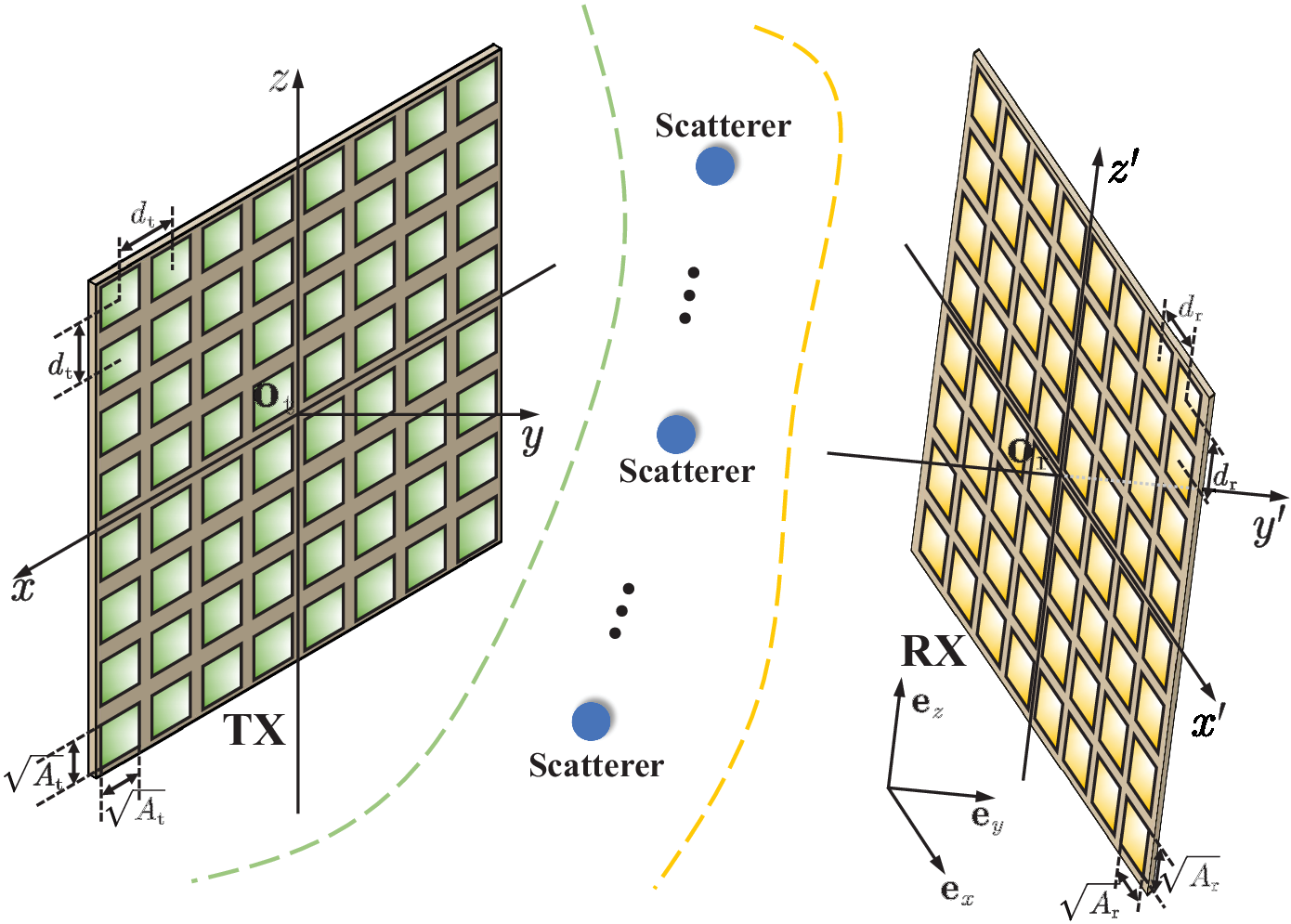}
\caption{Illustration of an SPDA-based channel.}
\label{Figure: System_SPD_Model}
\vspace{-15pt}
\end{figure}

Next, we compare the performance of CAPAs with SPDAs, as shown in {\figurename} {\ref{Figure: System_SPD_Model}}. The TX SPDA comprises $M_{\mathsf{t}}=M_{{\mathsf{t}},x}M_{{\mathsf{t}},z}$ antenna elements, while the RX SPDA consists of $M_{\mathsf{r}}=M_{{\mathsf{r}},x}M_{{\mathsf{r}},z}$ antenna elements. The quantities $M_{{\mathsf{t}},x}=\lfloor {L_{{\mathsf{t}},x}}/{d_{\mathsf{t}}}\rfloor$ and $M_{{\mathsf{r}},x}=\lfloor {L_{{\mathsf{r}},x}}/{d_{\mathsf{r}}}\rfloor$ indicate the number of elements along the $x$/$x'$-axis, whereas $M_{{\mathsf{t}},z}=\lfloor {L_{{\mathsf{t}},z}}/{d_{\mathsf{t}}}\rfloor$ and $M_{{\mathsf{r}},z}=\lfloor {L_{{\mathsf{r}},z}}/{d_{\mathsf{r}}}\rfloor$ represent the number of elements along the $z$/$z'$-axis. Both TX and RX elements have physical dimensions of $\sqrt{A_{\mathsf{t}}}$ and $\sqrt{A_{\mathsf{r}}}$ along the $x$/$x'$- and $z$/$z'$-axes, respectively. The inter-element distances $d_{\mathsf{t}}$ and $d_{\mathsf{r}}$ satisfy $d_{\mathsf{t}}\geq \sqrt{A_{\mathsf{t}}}$ and $d_{\mathsf{r}}\geq \sqrt{A_{\mathsf{r}}}$. For simplicity, we assume $M_{{\mathsf{t}},x}$, $M_{{\mathsf{t}},z}$, $M_{{\mathsf{r}},x}$, and $M_{{\mathsf{r}},z}$ are even numbers unless otherwise specified.

The central locations of the $(m_{{\mathsf{t}},x},m_{{\mathsf{t}},z})$th TX element and $(m_{{\mathsf{r}},x},m_{{\mathsf{r}},z})$th RX element are respectively given by ${\mathbf{t}}_{m_{{\mathsf{t}},x},m_{{\mathsf{t}},z}}=[d_{\mathsf{t}}(m_{{\mathsf{t}},x}+\frac{1}{2}),0,d_{\mathsf{t}}(m_{{\mathsf{t}},z}+\frac{1}{2})]^{\mathsf{T}}$ and ${\mathbf{r}}_{m_{{\mathsf{r}},x},m_{{\mathsf{r}},z}}'=[d_{\mathsf{r}}(m_{{\mathsf{r}},x}+\frac{1}{2}),0,d_{\mathsf{r}}(m_{{\mathsf{r}},z}+\frac{1}{2})]^{\mathsf{T}}$, where $m_{{\mathsf{s}},i}\in{\mathcal{M}}_{{\mathsf{s}},i}\triangleq\{-\frac{{M}_{{\mathsf{s}},i}}{2},\ldots,\frac{{M}_{{\mathsf{s}},i}}{2}-1\}$ for ${\mathsf{s}}\in\{{\mathsf{t}},{\mathsf{r}}\}$ and $i\in\{x,z\}$. The channel coefficient between them is approximated by
\begin{align}
H_{m_{{\mathsf{r}},x},m_{{\mathsf{r}},z},m_{{\mathsf{t}},x},m_{{\mathsf{t}},z}}\approx\sqrt{A_{\mathsf{r}}A_{\mathsf{t}}}
{\mathsf{h}}({\mathbf{r}}_{m_{{\mathsf{r}},x},m_{{\mathsf{r}},z}}'\!,\!{\mathbf{t}}_{m_{{\mathsf{t}},x},m_{{\mathsf{t}},z}}),
\end{align}
where $\sqrt{A_{\mathsf{r}}A_{\mathsf{t}}}$ accounts for the aperture size effects of both TX and RX antenna elements. For conciseness, we will limit our subsequent analysis to the case of linear TX/RX SPDAs aligned along $x$/$x'$-axes.
\subsection{MISO/SIMO Channels}\label{Section_SPDA: Linear Arrays}
We first examine MISO/SIMO channels. Because MISO and SIMO channels exhibit reciprocity, we concentrate our analysis on the MISO case with a linear TX SPDA and a single RX antenna element centered at ${\mathbf{o}}_{\mathsf{r}}$, where $M_{{\mathsf{t}},z}=M_{{\mathsf{r}},x}=M_{{\mathsf{r}},z}=1$. The $m_{{\mathsf{t}},x}$th TX element's central location is ${\mathbf{t}}_{m_{{\mathsf{t}},x}}=[d_{\mathsf{t}}(m_{{\mathsf{t}},x}+\frac{1}{2}),0,0]^{\mathsf{T}}$ for $m_{{\mathsf{t}},x}\in{\mathcal{M}}_{{\mathsf{t}},x}$. Following the derivation of \eqref{MISO_SNR_Definition_Standard}, the received SNR for the SPDA-based MISO channel is expressed as follows:
\begin{align}\label{MISO_SNR_Definition_Standard_SPDA}
\dot{\gamma}_{\mathsf{r}}=\frac{{{A}}_{\mathsf{r}}P}{\sigma^2}\sum\nolimits_{m_{{\mathsf{t}},x}\in{\mathcal{M}}_{{\mathsf{t}},x}}\int_{{\mathcal{A}}_{m_{{\mathsf{t}},x}}}\lvert h_{\mathsf{r}}({\mathbf{t}})\rvert^2{\rm{d}}{\mathbf{t}},
\end{align}
where ${\mathcal{A}}_{m_{{\mathsf{t}},x}}$ is the aperture of the $m_{{\mathsf{t}},x}$th TX element.

Since $A_{\mathsf{t}}=\lvert{\mathcal{A}}_{m_{{\mathsf{t}},x}}\rvert$ is negligible relative to the propagation distance, we can ignore signal strength variations within the aperture of each antenna element. This reduces \eqref{MISO_SNR_Definition_Standard_SPDA} to
\begin{align}\label{SIMO_Linear_SPDA}
\dot{\gamma}_{\mathsf{r}}\approx\frac{A_{\mathsf{t}}{{A}}_{\mathsf{r}}P}{\sigma^2}\sum_{m_{{\mathsf{t}},x}\in{\mathcal{M}}_{{\mathsf{t}},x}}
\lvert h_{\mathsf{r}}([d_{\mathsf{t}}(m_{{\mathsf{t}},x}+{1}/{2}),0,0]^{\mathsf{T}})\rvert^2.
\end{align}
Define ${\mathbf{h}}_{\mathsf{r}}\triangleq[h_{\mathsf{r}}([d_{\mathsf{t}}(m_{{\mathsf{t}},x}+{1}/{2}),0,0]^{\mathsf{T}})]_{{m_{{\mathsf{t}},x}\in{\mathcal{M}}_{{\mathsf{t}},x}}}
\in{\mathbbmss{C}}^{{M}_{{\mathsf{t}},x}\times1}$. From \eqref{Corollary_Autocorrelation_General_Linear_Result}, the $(m,m')$th element of the correlation matrix ${\mathbf{R}}_{\mathsf{r}}\triangleq{\mathbbmss{E}}\{{\mathbf{h}}_{\mathsf{r}}{\mathbf{h}}_{\mathsf{r}}^{\mathsf{H}}\}\in{\mathbbmss{C}}^{{M}_{{\mathsf{t}},x}\times{M}_{{\mathsf{t}},x}}$ is given by
\begin{align}\label{SPDA_MISO_Correlation_Used}
[{\mathbf{R}}_{\mathsf{r}}]_{m,m'}=\frac{1}{2k_0}\int_{-k_0}^{k_0}
{{\rm{e}}^{{\rm{j}}(m-m')d_{\mathsf{t}}\kappa_x}}{\rm{d}}\kappa_x.
\end{align}
Let $\{\sigma_{{\mathbf{h}}_{\mathsf{r}},m_{{\mathsf{t}},x}}\}_{m_{{\mathsf{t}},x}=1}^{{M}_{{\mathsf{t}},x}}$ be the eigenvalues of ${\mathbf{R}}_{\mathsf{r}}$. It has \cite{liu2023near-field}
\begin{align}\label{SPDA_MISO_SNR_Used}
\lVert{\mathbf{h}}_{\mathsf{r}}\rVert^2\overset{d}{=}\tilde{\mathbf{h}}_{\mathsf{r}}^{\mathsf{H}}{\mathbf{R}}_{\mathsf{r}}\tilde{\mathbf{h}}_{\mathsf{r}}
\overset{d}{=}\sum\nolimits_{m_{{\mathsf{t}},x}=1}^{{M}_{{\mathsf{t}},x}}\sigma_{{\mathbf{h}}_{\mathsf{r}},m_{{\mathsf{t}},x}}\lvert\Psi_{{\mathbf{h}}_{\mathsf{r}},m_{{\mathsf{t}},x}}\rvert^2,
\end{align}
where $\tilde{\mathbf{h}}_{\mathsf{r}}\sim{\mathcal{CN}}({\mathbf{0}},{\mathbf{I}}_{M_{{\mathsf{t}},x}})$ and $\{\Psi_{{\mathbf{h}}_{\mathsf{r}},m_{{\mathsf{t}},x}}\}_{m_{{\mathsf{t}},x}=1}^{{M}_{{\mathsf{t}},x}}$ are i.i.d. ZUCG random variables. This formulation makes the DMT analysis of SPDAs analogous to the MISO case in Section \ref{Section: MISO/SIMO Channels}, with further details provided in \cite[Section IV-B2]{liu2023near-field}.

When $L_{{\mathsf{t}},x}\gg d_{\mathsf{t}}$ or $M_{{\mathsf{t}},x}\gg1$, the eigenvalues of ${\mathbf{R}}_{\mathsf{r}}$ exhibit nearly the same behavior as those of $R_{h_{{\mathsf{r}}_x}}(t_x,t_x')$ in \eqref{Corollary_Autocorrelation_General_Linear_Result} but each scaled by a factor of $\frac{M_{{\mathsf{t}},x}}{L_{{\mathsf{t}},x}}\approx\frac{1}{d_{\mathsf{t}}}$; see \cite{zhu2017eigenvalue,do2023parabolic} and related references. This property extends to planar arrays, where the scaling factor is $\frac{M_{{\mathsf{t}},x}M_{{\mathsf{t}},z}}{L_{{\mathsf{t}},x}L_{{\mathsf{t}},z}}\approx\frac{1}{d_{\mathsf{t}}^2}$. This means that when setting $M_{{\mathsf{t}},x}\geq{\mathsf{DOF}}_{{\mathsf{r}}_x}=\frac{2L_{{\mathsf{t}},x}}{\lambda}=2N_{{\mathsf{t}},x}$ or $d_{\mathsf{t}}\leq\frac{\lambda}{2}$, the eigenvalues of ${\mathbf{R}}_{\mathsf{r}}$ also exhibit a step-like pattern with approximately $2N_{{\mathsf{t}},x}$ dominant values for $L_{{\mathsf{t}},x}\gg\lambda$, i.e., $\sigma_{{\mathbf{h}}_{\mathsf{r}},\ell}=\frac{M_{{\mathsf{t}},x}}{L_{{\mathsf{t}},x}}\sigma_{{\mathsf{r}}_x,\ell}$ for $\ell\in[1,{\mathsf{DOF}}_{{\mathsf{r}}_x}]$ and nearly zero otherwise. Table \ref{Table_SPDA_Results} summarizes the results for linear SPDAs with $d_{\mathsf{t}}=\frac{\lambda}{2}$, where $\overline{\gamma}=\frac{P}{\sigma^2}$, $d(r)$ is the achievable diversity order for a given multiplexing gain $r$ as $\overline{\gamma}\rightarrow\infty$, ${\mathsf{A}}_{\mathsf{SPDA}}$ is the array gain corresponding to the largest diversity gain, and $\eta_{\mathsf{t}},\eta_{\mathsf{r}}\in[0,1]$ are the array occupation ratios (AORs) for the TX and RX SPDAs, respectively.
\subsection{MIMO Channels}
We now analyze the MIMO scenario with linear SPDAs deployed along the $x$/$x'$-axis. The signal model is given by
\begin{align}\label{MIMO_Linear_SPDA}
{\mathbf{y}}_{\mathsf{mm}}=\sqrt{A_{\mathsf{r}}A_{\mathsf{t}}}{\mathbf{H}}_{\mathsf{mm}}{\mathbf{s}}_{\mathsf{mm}}+{\mathbf{n}}_{\mathsf{mm}},
\end{align}
where ${\mathbf{n}}_{\mathsf{mm}}\sim{\mathcal{CN}}({\mathbf{0}},\sigma^2{\mathbf{I}}_{M_{{\mathsf{r}},x}})$, ${\mathbf{s}}_{\mathsf{mm}}\sim{\mathcal{CN}}({\mathbf{0}},\frac{P}{M_{{\mathsf{t}},x}}{\mathbf{I}}_{M_{{\mathsf{t}},x}})$, and ${\mathbf{H}}_{\mathsf{mm}}={\mathbf{H}}_{{\mathsf{r}},x}{\mathbf{H}}_{a}{\mathbf{H}}_{{\mathsf{t}},x}^{\mathsf{H}}\in{\mathbbmss{C}}^{M_{{\mathsf{r}},x}\times M_{{\mathsf{t}},x}}$. Here, ${\mathbf{H}}_{{\mathsf{t}},x}=[{\mathbf{h}}_{{\mathsf{t}},x,m}]_{m=-{L_{{\mathsf{t}},x}}/{\lambda}}^{{L_{{\mathsf{t}},x}}/{\lambda}-1}
\in{\mathbbmss{C}}^{M_{{\mathsf{t}},x}\times2N_{{\mathsf{t}},x}}$, ${\mathbf{H}}_{{\mathsf{r}},x}=[{\mathbf{h}}_{{\mathsf{r}},x,\ell}]_{\ell=-{L_{{\mathsf{r}},x}}/{\lambda}}^{{L_{{\mathsf{r}},x}}/{\lambda}-1}
\in{\mathbbmss{C}}^{M_{{\mathsf{r}},x}\times2N_{{\mathsf{r}},x}}$, and ${\mathbf{H}}_{a}=[{h}_{a,\ell,m}]_{\forall \ell,m}\in{\mathbbmss{C}}^{2N_{{\mathsf{r}},x}\times2N_{{\mathsf{t}},x}}$. Moreover, ${\mathbf{h}}_{{\mathsf{t}},x,m}=[{\rm{e}}^{-{\rm{j}}\frac{2\pi m d_{\mathsf{t}}}{L_{{\mathsf{t}},x}}(m_{{\mathsf{t}},x}+\frac{1}{2})}]_{m_{{\mathsf{t}},x}=-{{M}_{{\mathsf{t}},x}}/{2}}^{{{M}_{{\mathsf{t}},x}}/{2}-1}$ and ${\mathbf{h}}_{{\mathsf{r}},x,\ell}=[{\rm{e}}^{-{\rm{j}}\frac{2\pi\ell d_{\mathsf{r}}}{L_{{\mathsf{r}},x}}(m_{{\mathsf{r}},x}+\frac{1}{2})}]_{m_{{\mathsf{r}},x}=-{{M}_{{\mathsf{r}},x}}/{2}}^{{{M}_{{\mathsf{r}},x}}/{2}-1}$. The channel capacity satisfies
\begin{equation}\label{MIMO_Rate_SPDA_Basic}
\begin{split}
\hat{\mathsf{C}}_{\mathsf{mm}}&=\log_2\det({\mathbf{I}}_{M_{{\mathsf{r}},x}}+{\hat{\gamma}}{\mathbf{H}}_{\mathsf{mm}}{\mathbf{H}}_{\mathsf{mm}}^{\mathsf{H}})\\
&\overset{d}{=}\log_2\det({\mathbf{I}}_{2N_{{\mathsf{r}},x}}+{\hat{\gamma}}{\mathbf{R}}_{a}^{1/2}{\mathbf{H}}_{a}{\mathbf{T}}_{a}{\mathbf{H}}_{a}^{\mathsf{H}}{\mathbf{R}}_{a}^{1/2}),
\end{split}
\end{equation}
with $\hat{\gamma}=\frac{A_{\mathsf{r}}A_{\mathsf{t}}P}{M_{{\mathsf{t}},x}\sigma^2}$, ${\mathbf{T}}_{a}={\mathbf{H}}_{{\mathsf{t}},x}^{\mathsf{H}}{\mathbf{H}}_{{\mathsf{t}},x}\in{\mathbbmss{C}}^{2N_{{\mathsf{t}},x}\times2N_{{\mathsf{t}},x}}$, and ${\mathbf{R}}_{a}={\mathbf{H}}_{{\mathsf{r}},x}^{\mathsf{H}}{\mathbf{H}}_{{\mathsf{r}},x}\in{\mathbbmss{C}}^{2N_{{\mathsf{r}},x}\times2N_{{\mathsf{r}},x}}$. We note that $\hat{\mathsf{C}}_{\mathsf{mm}}$ can be analyzed in the same manner as the channel capacity ${\mathsf{C}}_{\mathsf{mm}}$ given in \eqref{MIMO_Rate_CAPA_Basic}.

To fully capture the angular-domain information in ${\mathbf{H}}_{a}$, the conditions $M_{{\mathsf{r}},x}\geq 2N_{{\mathsf{r}},x}$ and $M_{{\mathsf{t}},x}\geq 2N_{{\mathsf{t}},x}$ must hold. Given that $M_{{\mathsf{r}},x}=\lfloor {L_{{\mathsf{r}},x}}/{d_{\mathsf{r}}}\rfloor$, $M_{{\mathsf{t}},x}=\lfloor {L_{{\mathsf{t}},x}}/{d_{\mathsf{t}}}\rfloor$, $N_{{\mathsf{r}},x}= \frac{L_{{\mathsf{r}},x}}{\lambda}$, and $N_{{\mathsf{t}},x}= \frac{L_{{\mathsf{t}},x}}{\lambda}$, we conclude that $d_{\mathsf{t}}\leq \frac{\lambda}{2}$ and $d_{\mathsf{r}}\leq \frac{\lambda}{2}$ should be satisfied. For brevity, we examine the case of $d_{\mathsf{t}}=d_{\mathsf{r}}= \frac{\lambda}{2}$, which yields $M_{{\mathsf{r}},x}= 2N_{{\mathsf{r}},x}$, $M_{{\mathsf{t}},x}= 2N_{{\mathsf{t}},x}$, ${\mathbf{h}}_{{\mathsf{r}},x,\ell}=[{\rm{e}}^{-{\rm{j}}\frac{2\pi\ell }{2N_{{\mathsf{r}},x}}(m_{{\mathsf{r}},x}+\frac{1}{2})}]_{m_{{\mathsf{r}},x}=-N_{{\mathsf{r}},x}}^{N_{{\mathsf{r}},x}-1}$, and ${\mathbf{h}}_{{\mathsf{t}},x,m}=[{\rm{e}}^{-{\rm{j}}\frac{2\pi{m} }{2N_{{\mathsf{t}},x}}(m_{{\mathsf{t}},x}+\frac{1}{2})}]_{m_{{\mathsf{t}},x}=-N_{{\mathsf{t}},x}}^{N_{{\mathsf{t}},x}-1}$. The resulting ${\mathbf{H}}_{{\mathsf{r}},x}$ and ${\mathbf{H}}_{{\mathsf{t}},x}$ become non-normalized discrete Fourier transform matrices with ${\mathbf{H}}_{{\mathsf{r}},x}{\mathbf{H}}_{{\mathsf{r}},x}^{\mathsf{H}}=2N_{{\mathsf{r}},x}{\mathbf{I}}_{2N_{{\mathsf{r}},x}}$ and ${\mathbf{H}}_{{\mathsf{t}},x}{\mathbf{H}}_{{\mathsf{t}},x}^{\mathsf{H}}=2N_{{\mathsf{t}},x}{\mathbf{I}}_{2N_{{\mathsf{t}},x}}$. Using the approaches in previous sections, we derive the DMT and the array gain, as shown in Table \ref{Table_SPDA_Results}.

\begin{table*}[!t]
\centering
\setlength{\abovecaptionskip}{0pt}
\resizebox{0.9\textwidth}{!}{
\begin{tabular}{|c|c|c|c|c|c|}
\hline
\textbf{System} & \textbf{Capacity} & \textbf{DMT} $d(r)$ & \textbf{Array Gain} ${\mathsf{A}}_{\mathsf{SPDA}}$ (with respect to $\overline{\gamma}=\frac{P}{\sigma^2}$) & \textbf{TX AOR} $\eta_{\mathsf{t}}$ & \textbf{RX AOR} $\eta_{\mathsf{r}}$\\ \hline
\textbf{MISO}   & $\log_2(1+\overline{\gamma}A_{\mathsf{r}}A_{\mathsf{t}}\lVert{\mathbf{h}}_{\mathsf{r}}\rVert^2)$       & $2N_{{\mathsf{t}},x}(1-r)$ & $\frac{1}{2^R-1}\frac{2N_{{\mathsf{t}},x}A_{\mathsf{t}}A_{\mathsf{r}}}{L_{{\mathsf{t}},x}}((2N_{{\mathsf{t}},x})!\prod_{\ell=1}^{2N_{{\mathsf{t}},x}}\sigma_{{\mathsf{r}}_x,\ell})^{\frac{1}{2N_{{\mathsf{t}},x}}}$  & $\frac{2N_{{\mathsf{t}},x}A_{\mathsf{t}}}{\mu(\mathcal{A}_{\mathsf{t}})}$  & $\frac{A_{\mathsf{r}}}{\mu(\mathcal{A}_{\mathsf{r}})}$    \\ \hline
\textbf{MIMO}   & $\log_2\det({\mathbf{I}}_{2N_{{\mathsf{r}},x}}+\overline{\gamma}(2N_{{\mathsf{r}},x}A_{\mathsf{r}})A_{\mathsf{t}}{\mathbf{H}}_{a}{\mathbf{H}}_{a}^{\mathsf{H}})$       & $(2N_{{\mathsf{t}},x}-r)(2N_{{\mathsf{r}},x}-r)$ & $(2N_{{\mathsf{r}},x}A_{\mathsf{r}})A_{\mathsf{t}}{(\prod_{j=1}^{2N_{{\mathsf{t}},x}}\varrho_{{\mathsf{t}},j}^2)^{\frac{1}{2N_{{\mathsf{t}},x}}}
(\prod_{i=1}^{2N_{{\mathsf{r}},x}}\varrho_{{\mathsf{r}},i}^2)^{\frac{1}{2N_{{\mathsf{r}},x}}}}
{\epsilon_{{\mathsf{t}},{\mathsf{r}}}^{-\frac{1}{4N_{{\mathsf{r}},x}N_{{\mathsf{t}},x}}}}$   & $\frac{2N_{{\mathsf{t}},x}A_{\mathsf{t}}}{\mu(\mathcal{A}_{\mathsf{t}})}$ & $\frac{2N_{{\mathsf{r}},x}A_{\mathsf{r}}}{\mu(\mathcal{A}_{\mathsf{r}})}$     \\ \hline
\end{tabular}}
\caption{Channel capacity for linear SPDAs.}
\label{Table_SPDA_Results}
\vspace{-15pt}
\end{table*}

Table \ref{Table_SPDA_Results} shows that setting $d_{\mathsf{t}}=d_{\mathsf{r}}= \frac{\lambda}{2}$ enables SPDAs to achieve the same DMT as CAPAs. Let ${\mathsf{A}}_{\mathsf{CAPA}}$ denote the array gain of CAPAs. By comparing the results in Table \ref{Table_SPDA_Results} with those in Sections \ref{Section:MISO:Performance Analysis} and \ref{Section:MIMO:Performance Analysis}, we conclude that $\frac{{\mathsf{A}}_{\mathsf{SPDA}}}{{\mathsf{A}}_{\mathsf{CAPA}}}=\eta_{\mathsf{t}}\eta_{\mathsf{r}}\leq1$, where equality holds when $\eta_{\mathsf{t}}=\eta_{\mathsf{r}}=1$.

These findings indicate that when antenna elements are spaced at half-wavelength intervals ($d_{\mathsf{t}}=d_{\mathsf{r}}=\frac{\lambda}{2}$), an SPDA achieves the same DMT as a CAPA but with a lower array gain. If the SPDA elements are spaced at sub-half-wavelength intervals, the rank of the channel matrix or correlation matrix reaches $2N_{{\mathsf{r}},x}$ or $2N_{{\mathsf{t}},x}$, which ensures that the SPDA maintains the same DMT as the CAPA. However, when the spacing exceeds half a wavelength, the rank of the channel or correlation matrix reduces to $M_{{\mathsf{r}},x}<2N_{{\mathsf{r}},x}$ or $M_{{\mathsf{t}},x}<2N_{{\mathsf{t}},x}$, which leads to a lower DMT than that of the CAPA. These observations lead to the following conclusion.
\vspace{-5pt}
\begin{remark}\label{remark_compare_s_c}
When an SPDA employs half-wavelength or sub-half-wavelength spacing, it achieves the same DMT as a CAPA but with a lower array gain. However, if the spacing exceeds half a wavelength, the DMT of the SPDA falls below that of the CAPA.
\end{remark}
\vspace{-5pt}
\section{Numerical Results}\label{Section: Numerical Results}
In this section, computer simulation results will be used to demonstrate the performance of CAPAs and validate the accuracy of the developed analytical results. Unless explicitly stated otherwise, we set $u_{x}=u_z=0$ m, $u_y=10$ m, ${\mathbf{E}}=\mathbf{I}_3$, $N_{{\mathsf{t}},x}=N_{{\mathsf{t}},z}=N_{{\mathsf{r}},x}=N_{{\mathsf{r}},z}=10$, $L_{{\mathsf{t}},x}=L_{{\mathsf{t}},z}=L_{{\mathsf{r}},x}=L_{{\mathsf{r}},z}=10\lambda$, $\lambda= 0.0107$ m, $A_{\mathsf{t}}=A_{\mathsf{r}}=\frac{\lambda^2}{32}$, and $d_{\mathsf{t}}=d_{\mathsf{r}}=d$. The noise strength is $\sigma^2= N_0B$ with bandwidth $B = 1$ MHz and effective density $N_0 = -174$ dBm/Hz.
\subsection{MISO/SIMO Channels}
For MISO/SIMO channels, we set $\mu({\mathcal{A}}_{\mathsf{r}})=A_{\mathsf{r}}$ for MISO and $\mu({\mathcal{A}}_{\mathsf{t}})=A_{\mathsf{t}}$ for SIMO to ensure a fair comparison with SPDAs. In this case, MISO capacity equals SIMO capacity.
\subsubsection{Linear Arrays}
\begin{figure}[!t]
    \centering
    \subfigbottomskip=0pt
	\subfigcapskip=-5pt
\setlength{\abovecaptionskip}{0pt}
    \subfigure[OP. $R=5$ bps/Hz.]
    {
        \includegraphics[height=0.175\textwidth]{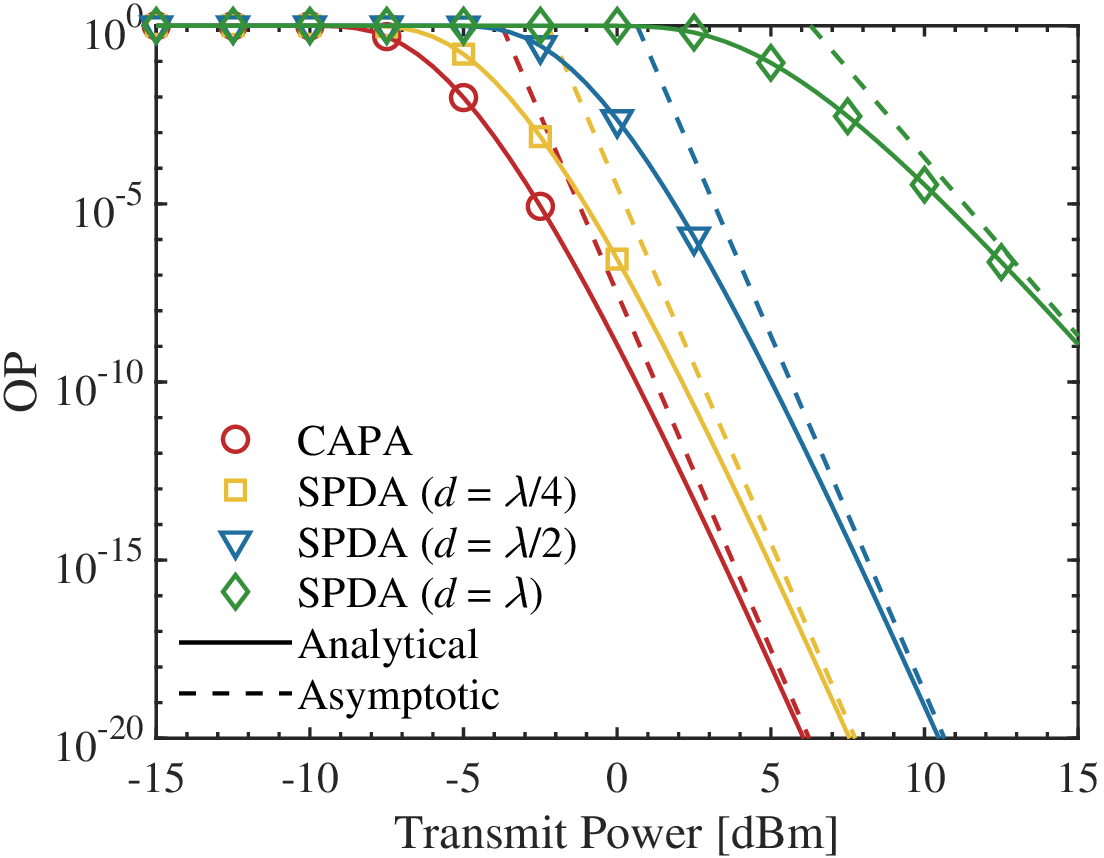}
	   \label{fig6a}	
    }
   \subfigure[ECC.]
    {
        \includegraphics[height=0.173\textwidth]{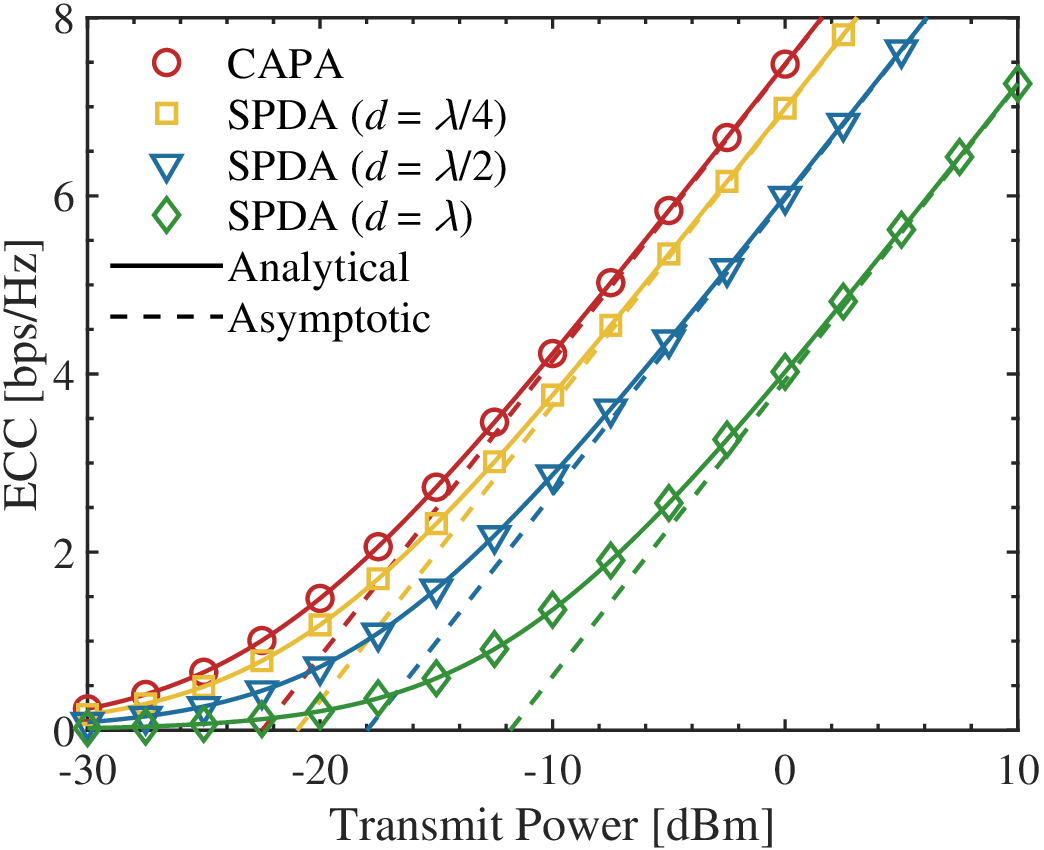}
	   \label{fig6b}	
    }
\caption{The OP and ECC of MISO/SIMO using linear arrays.}
    \label{Figure6}
    \vspace{-10pt}
\end{figure}

For linear arrays, we set $L_{{\mathsf{t}},z}=\sqrt{A_{\mathsf{t}}}$ and $L_{{\mathsf{r}},z}=\sqrt{A_{\mathsf{r}}}$. {\figurename} {\ref{fig6a}} and {\figurename} {\ref{fig6b}} illustrate OP and ECC as functions of transmit power, where symbols denote simulated results. We observe that the analytical results closely match the simulations, which confirms the accuracy of our closed-form expressions. For comparison, we include the OP and ECC of SPDAs with antenna spacings of $d=\frac{\lambda}{4}$, $d=\frac{\lambda}{2}$, and $d=\lambda$, where analytical results are derived from \eqref{SPDA_MISO_Correlation_Used} and \eqref{SPDA_MISO_SNR_Used}. The results indicate that CAPA achieves the lowest OP and highest ECC in all scenarios, which demonstrates its superior SE. Additionally, SPDAs demonstrate improved SE performance with smaller antenna spacing (i.e., denser deployments), as the increased array gain enhances performance.

\begin{figure}[!t]
\centering
    \subfigbottomskip=0pt
	\subfigcapskip=-5pt
\setlength{\abovecaptionskip}{0pt}
    \subfigure[Diversity-multiplexing trade-off.]
    {
        \includegraphics[height=0.175\textwidth]{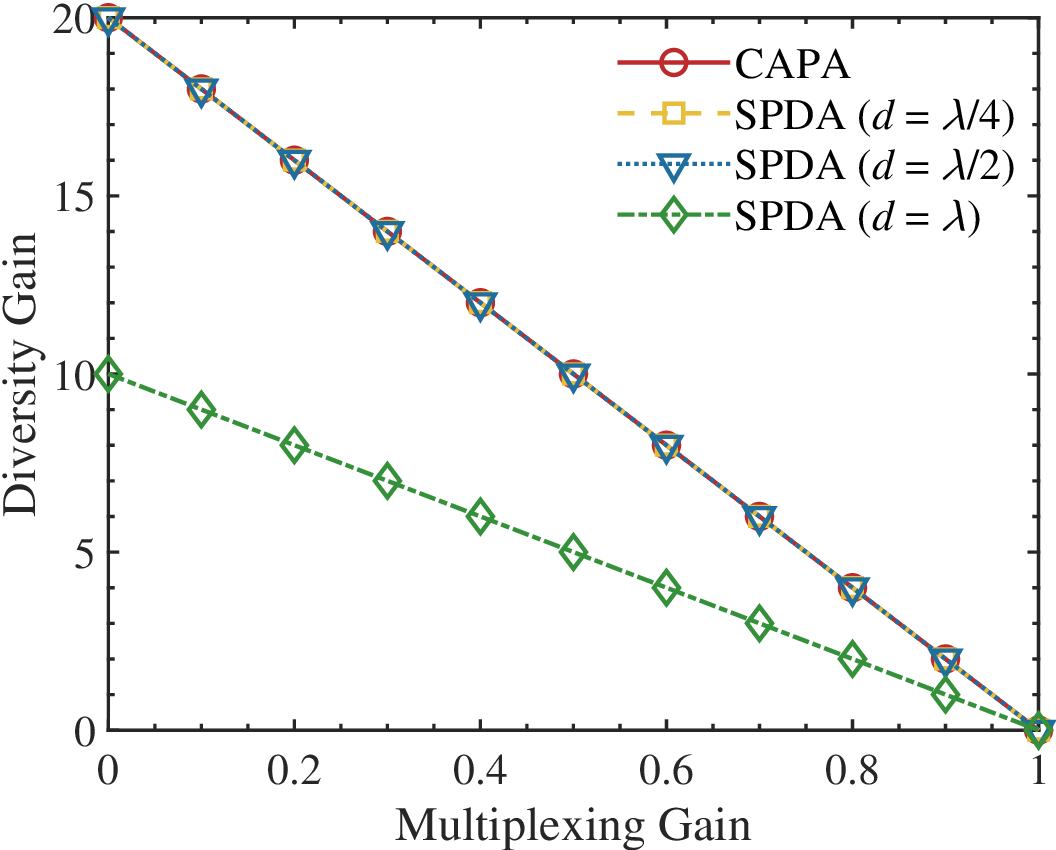}
	   \label{fig7a}	
    }
   \subfigure[Array gain vs. multiplexing gain.]
    {
        \includegraphics[height=0.175\textwidth]{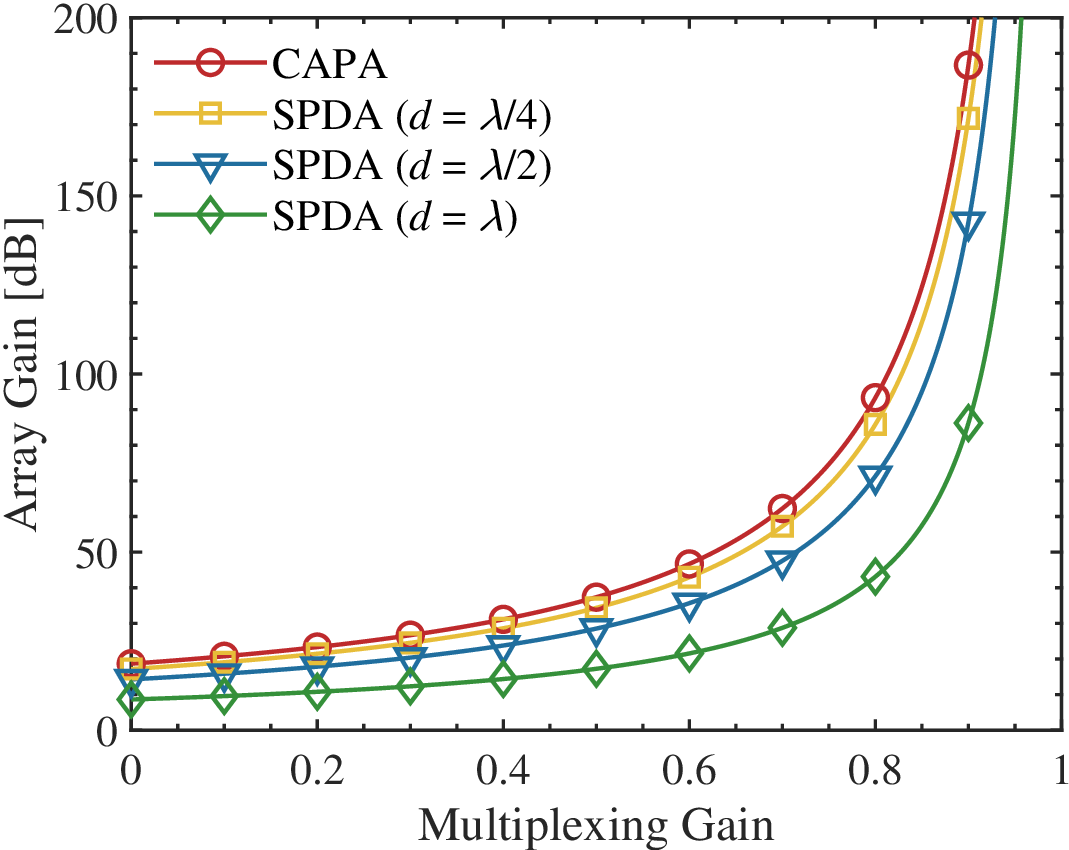}
	   \label{fig7b}	
    }
\caption{The diversity, multiplexing, and array gain trade-off in MISO/SIMO channels using linear arrays.}
\label{Figure7}
\vspace{-10pt}
\end{figure}

\begin{figure*}[!t]
\centering
    \subfigbottomskip=0pt
	\subfigcapskip=-5pt
\setlength{\abovecaptionskip}{0pt}
    \subfigure[ECC.]
    {
        \includegraphics[height=0.21\textwidth]{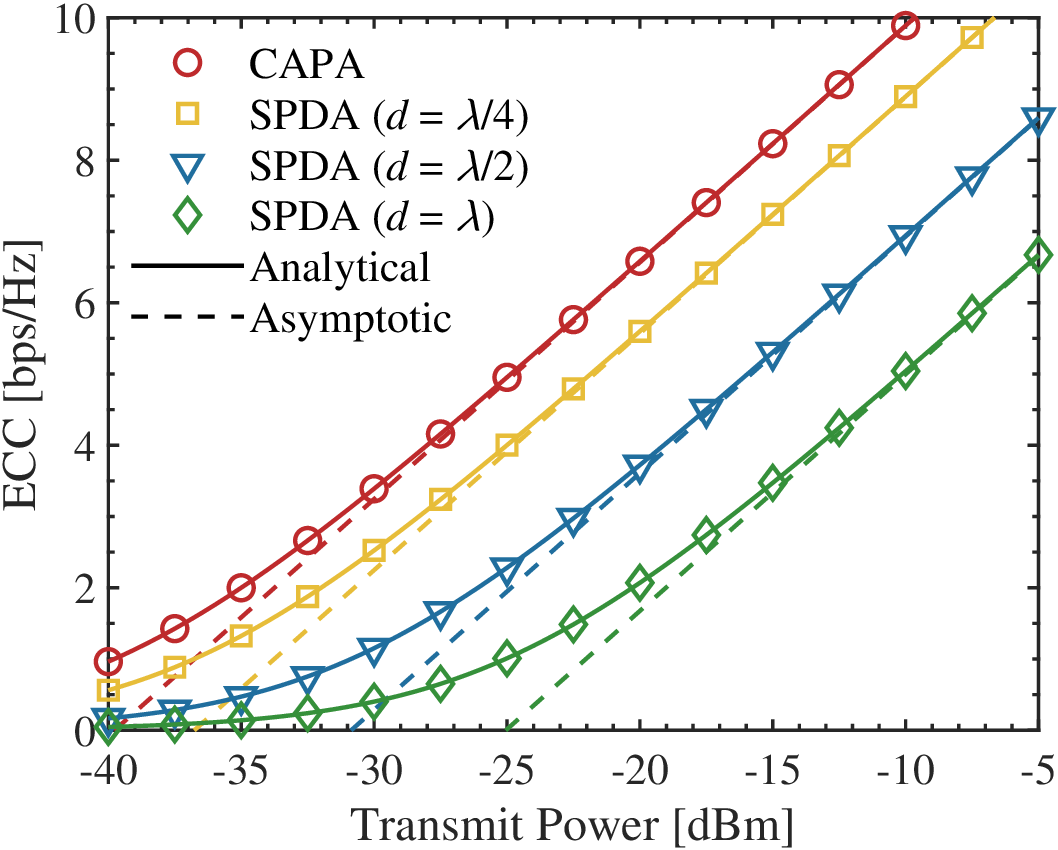}
	   \label{fig8a}	
    }
    \subfigure[Diversity-multiplexing trade-off.]
    {
        \includegraphics[height=0.21\textwidth]{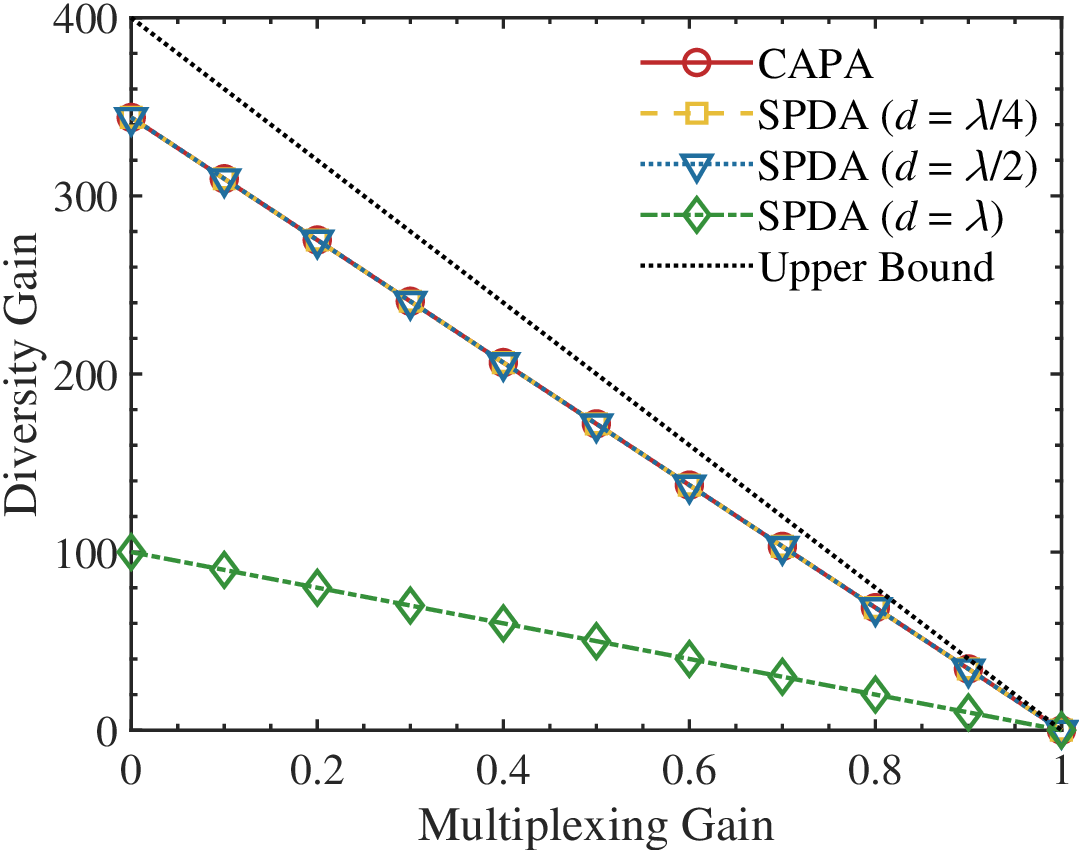}
	   \label{fig8b}	
    }
   \subfigure[Array gain vs. multiplexing gain.]
    {
        \includegraphics[height=0.21\textwidth]{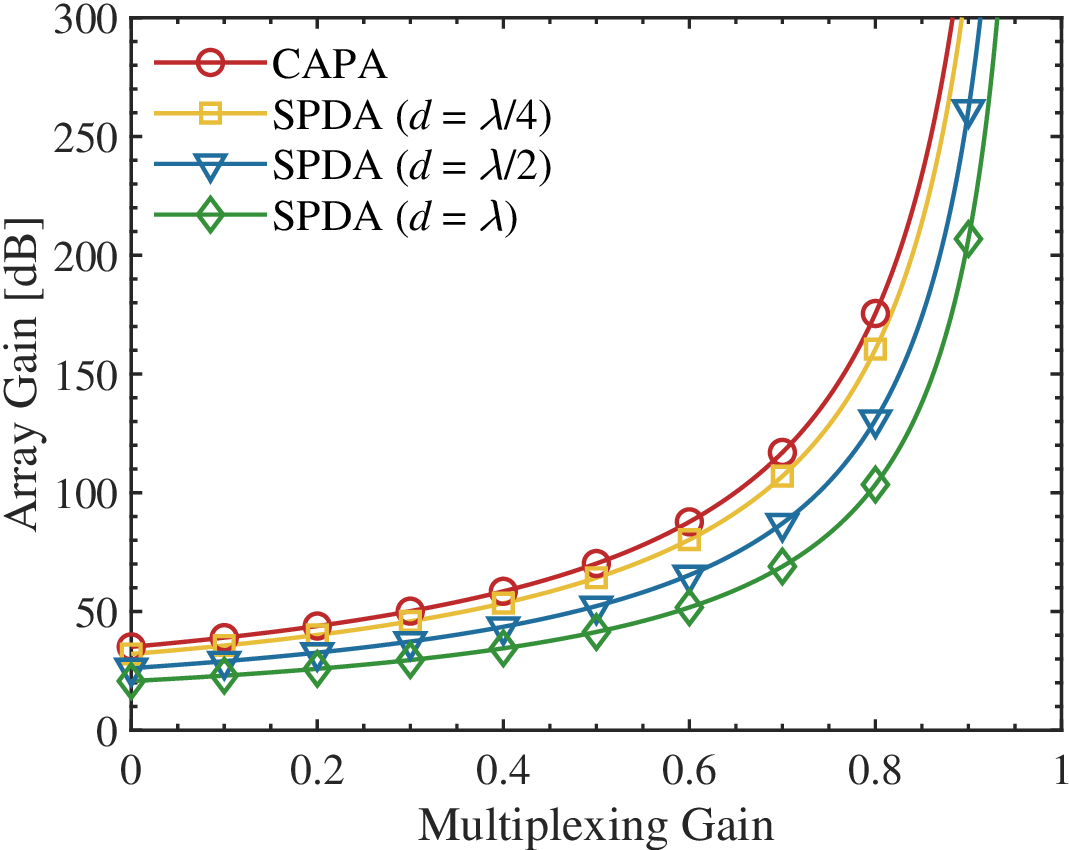}
	   \label{fig8c}	
    }
\caption{The ECC and diversity, multiplexing, and array gain trade-off in MISO/SIMO channels using planar arrays.}
\label{Figure8}
\vspace{-10pt}
\end{figure*}

\begin{figure*}[!t]
\centering
    \subfigbottomskip=0pt
	\subfigcapskip=-5pt
\setlength{\abovecaptionskip}{0pt}
    \subfigure[ECC.]
    {
        \includegraphics[height=0.21\textwidth]{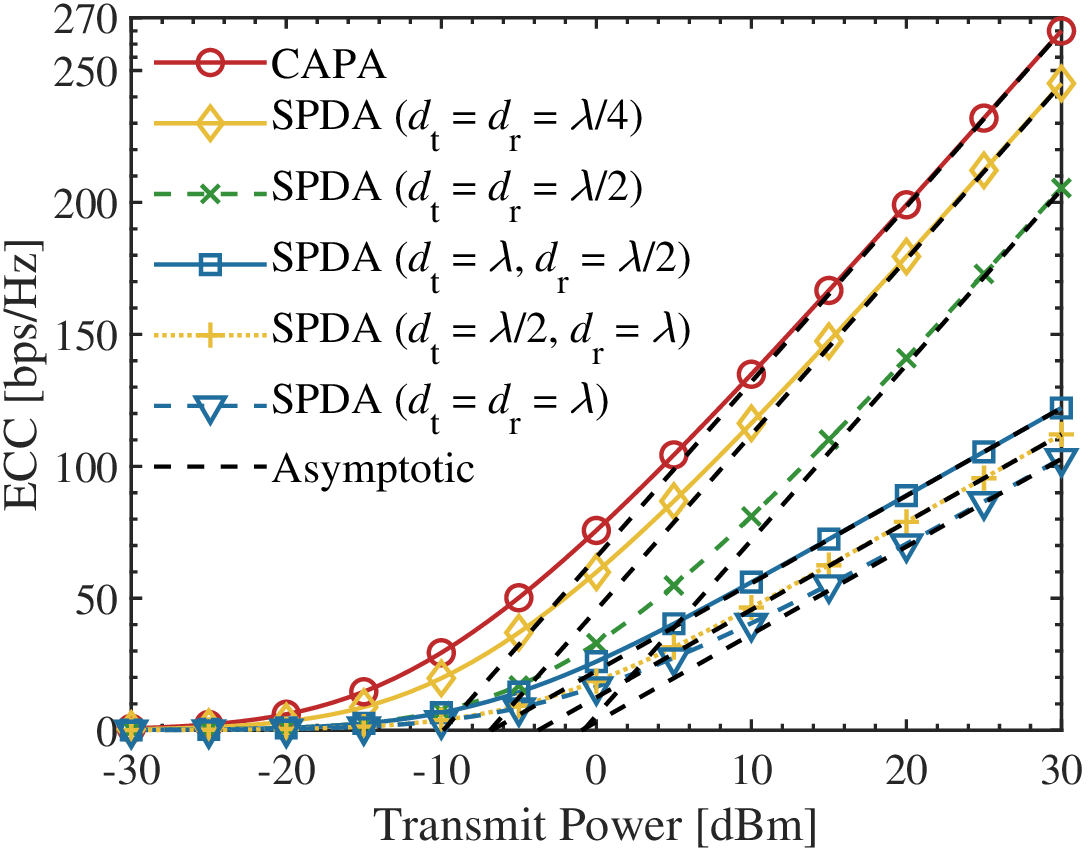}
	   \label{fig9a}	
    }
    \subfigure[Diversity-multiplexing trade-off.]
    {
        \includegraphics[height=0.21\textwidth]{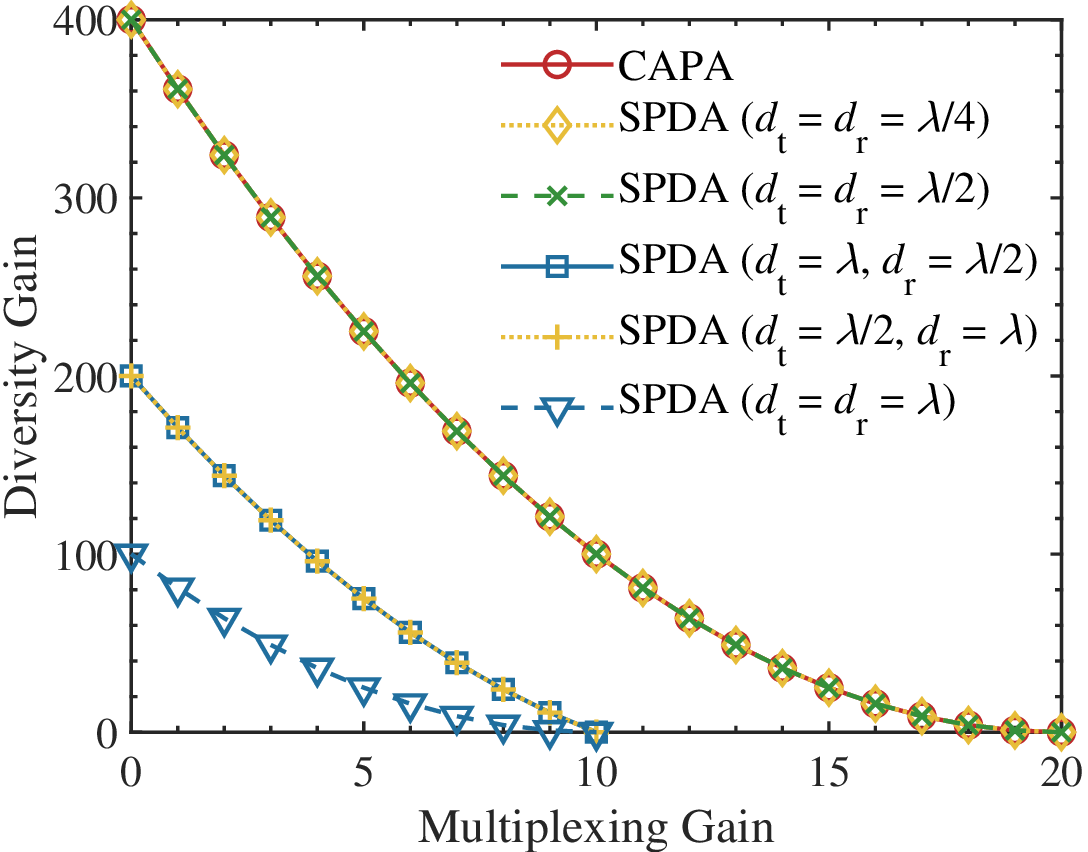}
	   \label{fig9b}	
    }
   \subfigure[Array gain vs. multiplexing gain.]
    {
        \includegraphics[height=0.21\textwidth]{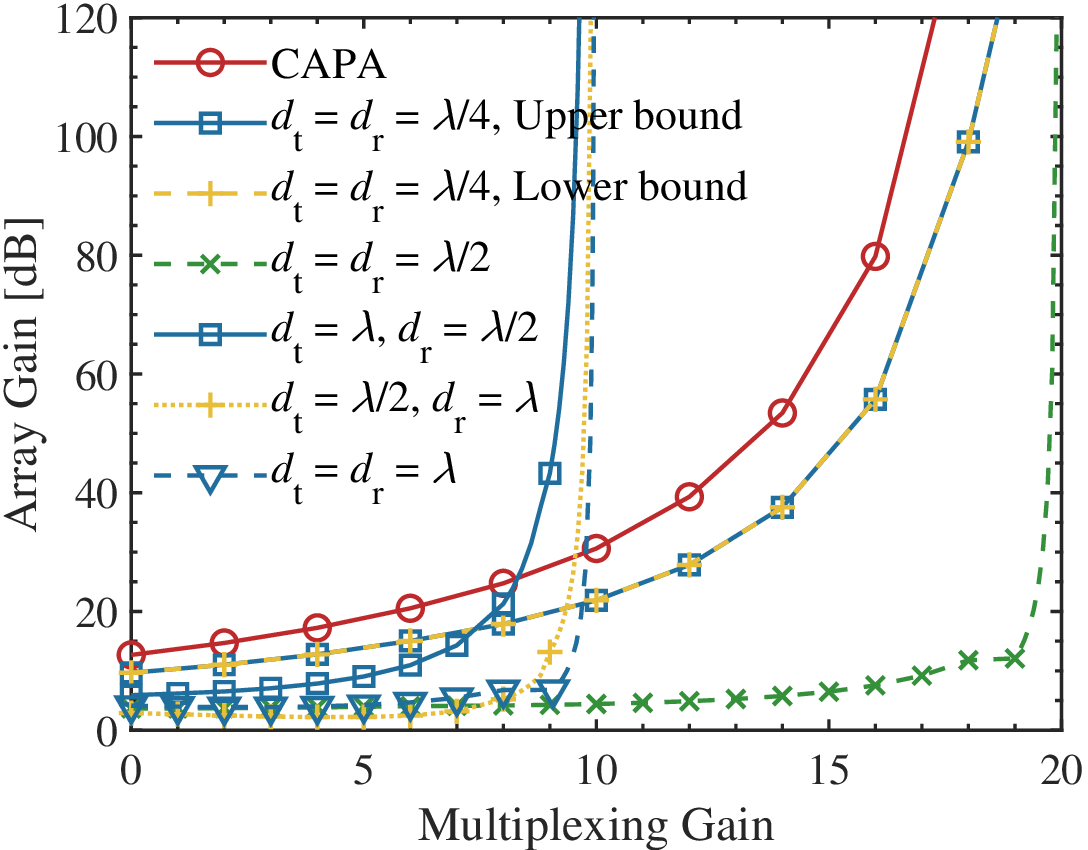}
	   \label{fig9c}	
    }
\caption{The diversity, multiplexing, and array gain trade-off in MIMO channels using linear arrays.}
\label{Figure9}
\vspace{-10pt}
\end{figure*}

The asymptotic results for OP and ECC are also plotted in {\figurename} {\ref{Figure6}}, which almost perfectly match with the numerical results in the high-SNR regime. This agreement validates the accuracy of the derived diversity order and high-SNR slope. {\figurename} {\ref{fig6b}} demonstrates that while CAPA achieves the same high-SNR slope (multiplexing gain) as SPDAs, its superior array gain enhances ECC by reducing the high-SNR power offset. Regarding OP performance in {\figurename} {\ref{fig6a}}, $\frac{\lambda}{4}$-SPDA ($d=\frac{\lambda}{4}$) and $\frac{\lambda}{2}$-SPDA ($d=\frac{\lambda}{2}$) exhibit the same diversity order as CAPA, all of which exceed that of ${\lambda}$-SPDA ($d=\lambda$). However, despite achieving the same diversity order as CAPA, both $\frac{\lambda}{4}$-SPDA and $\frac{\lambda}{2}$-SPDA result in higher OP values than CAPA. This performance gap originates from CAPA's larger array gain compared to SPDA. These findings are consistent with the theoretical analysis in Section \ref{Section_SPDA}.

{\figurename} {\ref{fig7a}} and {\figurename} {\ref{fig7b}} illustrate the diversity gain and array gain as functions of the target multiplexing gain within the DMT framework. {\figurename} {\ref{fig7a}} shows that for the same multiplexing gain, CAPA achieves a diversity gain equivalent to both $\frac{\lambda}{4}$-SPDA and $\frac{\lambda}{2}$-SPDA, while exceeding ${\lambda}$-SPDA. This is because SPDAs with antenna spacing at or below half-wavelength can fully capture angular-domain information and thus preserve the same DoFs as CAPA. In contrast, SPDAs with spacing greater than $\frac{\lambda}{2}$ fail to utilize the complete angular-domain information, leading to a reduction in diversity gain. 

Turning now to the array gain shown in {\figurename} {\ref{fig7b}}, we observe that when achieving the same multiplexing gain, CAPA always yields a greater array gain than SPDAs. Notably, $\frac{\lambda}{4}$-SPDA attains an array gain comparable to CAPA, which suggests that densely deployed SPDAs may serve as viable practical alternatives to CAPA implementations. We also observe that as the target multiplexing gain approaches its upper limit, i.e., $1$, the array gain tends to infinity. This can be explained as follows: when multiplexing gain approaches $1$, the corresponding diversity gain diminishes to zero, which requires an infinitely large array gain to sustain a low OP.
\subsubsection{Planar Arrays}
We now analyze the MISO/SIMO channel with a planar CAPA. {\figurename} {\ref{fig8a}} illustrates the ECC of different arrays as a function of transmit power. The analytical results for SPDAs follow a methodology similar to that used for their linear counterparts. The results indicate that CAPA achieves a superior ECC compared to SPDAs while maintaining the same high-SNR slope (multiplexing gain). {\figurename} {\ref{fig8b}} and {\figurename} {\ref{fig8c}} illustrate the diversity gain and array gain as functions of the target multiplexing gain within the DMT framework. As previously discussed, the spatial autocorrelation function of a planar CAPA contains approximately $\pi N_{{\mathsf{r}},x}N_{{\mathsf{r}},z}$ or $\pi N_{{\mathsf{t}},x}N_{{\mathsf{t}},z}$ significant eigenvalues, which define its DoFs. For reference, we include the theoretical upper bound for diversity gain, i.e., $4N_{{\mathsf{r}},x}N_{{\mathsf{r}},z}(1-r)$ or $4N_{{\mathsf{t}},x}N_{{\mathsf{t}},z}(1-r)$, where $r$ represents the multiplexing gain. {\figurename} {\ref{fig8b}} shows that CAPA achieves the same DMT as $\frac{\lambda}{4}$-SPDA and $\frac{\lambda}{2}$-SPDA, while outperforming ${\lambda}$-SPDA, with all cases bounded by the theoretical limit. {\figurename} {\ref{fig8c}} compares the array gain of different arrays and confirms that CAPA achieves the highest array gain among all considered case, which is consistent with the results observed in {\figurename} {\ref{fig7b}}.
\subsection{MIMO Channels}
We now shift to the MIMO channel, and for brevity, only the results for linear arrays are presented, where $L_{{\mathsf{t}},z}=\sqrt{A_{\mathsf{t}}}$ and $L_{{\mathsf{r}},z}=\sqrt{A_{\mathsf{r}}}$. {\figurename} {\ref{fig9a}} compares the ECC between the CAPA and SPDAs with different antenna spacings $(d_{\mathsf{r}},d_{\mathsf{t}})$. Here, $(d_{\mathsf{r}},d_{\mathsf{t}})$-SPDA represents an SPDA configuration with TX antenna spacing $d_{\mathsf{t}}$ and RX antenna spacing $d_{\mathsf{r}}$, with analytical results derived from \eqref{MIMO_Rate_SPDA_Basic}. Given our simulation setup, we have $L_{{\mathsf{r}},x}=L_{{\mathsf{t}},x}$, and thus the asymptotic ECC can be calculated using \eqref{Result1_Theorem_MIMO_CAPA_Multiplexing}. The findings show that CAPA achieves superior ECC performance compared to all SPDA configurations. Additionally, both $(\frac{\lambda}{4},\frac{\lambda}{4})$-SPDA and $(\frac{\lambda}{2},\frac{\lambda}{2})$-SPDA achieve the same high-SNR slope as CAPA while outperforming other SPDAs. This result aligns with the analysis in Section \ref{Section_SPDA}, as ensuring $\max\{d_{\mathsf{t}},d_{\mathsf{r}}\}\leq\frac{\lambda}{2}$ guarantees the capture of all angular-domain information in ${\mathbf{H}}_{\mathsf{mm}}$. 

{\figurename} {\ref{fig9b}} illustrates the DMT achieved by different arrays. As previously discussed, the number of spatial DoFs between the TX and RX is fully determined by its angular-domain counterpart. Since CAPA, $(\frac{\lambda}{4},\frac{\lambda}{4})$-SPDA, and $(\frac{\lambda}{2},\frac{\lambda}{2})$-SPDA captures all angular-domain information, they fully exploit the available DoFs and achieve the same DMT, outperforming other SPDAs. The results in {\figurename} {\ref{fig9b}} confirm this conclusion. We then move to the array gain illustrated in {\figurename} {\ref{fig9c}}. For $(\frac{\lambda}{4},\frac{\lambda}{4})$-SPDA, the upper and lower bounds are derived using \eqref{MIMO_Rate_SPDA_Basic} and Theorem \ref{Theorem_MIMO_CAPA_DMT}, while the remaining results follow from Corollary \ref{Corollary_MIMO_IID_Rayleigh_Array_Gain}. Notably, the array gain bounds for $(\frac{\lambda}{4},\frac{\lambda}{4})$-SPDA are nearly identical. Comparing the array gains achieved by CAPA, $(\frac{\lambda}{4},\frac{\lambda}{4})$-SPDA, and $(\frac{\lambda}{2},\frac{\lambda}{2})$-SPDA reveals that CAPA provides a higher array gain than SPDA when targeting the same multiplexing gain. This is because CAPA captures all spatial information with an AOR of $1$ and yields the highest possible array gain. For other SPDAs, the array gain is significantly larger than that of CAPA, especially as the target multiplexing gain approaches its upper limit. In this regime, the array gain tends to infinity and outperforms the finite array gain of CAPA. However, since these SPDAs cannot achieve the same multiplexing or diversity gain as CAPA, drawing a fair comparison between them remains challenging.

In summary, CAPA captures complete information from both the spatial and angular (or wavenumber) domains, thereby achieving the upper limits of array gain, diversity gain, and multiplexing gain. In contrast, SPDA with half-wavelength antenna spacing or less captures all angular-domain information but may fail to capture all spatial information. Therefore, it achieves the same DMT as CAPA but with a lower array gain. For SPDA with larger antenna spacing, neither angular-domain nor spatial-domain information is fully utilized, which yields reductions in array gain, diversity gain, and multiplexing gain.
\section{Conclusion}\label{Section: Conclusion} 
This paper has analyzed the performance of diversity and multiplexing in CAPA-based SIMO, MISO, and MIMO channels. We derived analytically tractable fading models between two CAPAs under a general non-parallel setup. Through high-SNR asymptotic analyses, we characterized the fundamental relationship between diversity gain, multiplexing gain, and array gain. Both theoretical analyses and numerical simulations demonstrated that CAPAs can achieve either a higher array gain or a better DMT than conventional SPDAs. These findings suggest that CAPAs outperform SPDAs in terms of SE, indicating its potential as a promising technology.
\begin{appendix}
\subsection{Proofs of Lemma \ref{Lemma_Autocorrelation_General} and Corollary \ref{Corollary_Autocorrelation_General_Linear}}\label{Proof_Lemma_Autocorrelation_General}
Using \eqref{Channel_Response_MISO_Random_Field}, we calculate the autocorrelation function of $h_{\mathsf{r}}({\mathbf{t}})$ as follows:
\begin{align}
&R_{h_{\mathsf{r}}}({\mathbf{t}},{\mathbf{t}}')
=\frac{1}{(2\pi)^4}\iiiint_{{\mathcal{D}}({\bm\kappa})\times{\mathcal{D}}({\bm\kappa}')}{\mathbbmss{E}}\{{\hat{\mathsf{H}}_a({\bm\kappa})}{\hat{\mathsf{H}}_a^{*}({\bm\kappa}')}\}\nonumber\\
&\times
{\rm{e}}^{{\rm{j}}(t_x\kappa_x+t_z\kappa_z-t_x'\kappa_x'-t_z'\kappa_z')}{\rm{d}}\kappa_x{\rm{d}}\kappa_z{\rm{d}}\kappa_x'{\rm{d}}\kappa_z',\label{Appendix_A_Result1}
\end{align}
where ${\bm\kappa}'=[\kappa_x',{\gamma}(\kappa_x',\kappa_z'),\kappa_z']^{\mathsf{T}}$. From \eqref{ZUCG_Gaussian_Random_Field_Origin}, $\hat{\mathsf{H}}_a({\bm\kappa})=\iint_{{\mathcal{D}}({\mathbf{k}})}{\mathsf{H}}_a({\mathbf{k}},{\bm\kappa}){\rm{d}}k_x{\rm{d}}k_z\overset{d}{=}{\hat{S}}^{\frac{1}{2}}({\bm\kappa}){\hat{W}}({\bm\kappa})$, where ${\hat{W}}({\bm\kappa})\sim{\mathcal{CN}}(0,1)$ is a ZUCG random field. It follows that
\begin{align}\label{Appendix_A_Result3}
{\hat{S}}({\bm\kappa})=\iint_{{\mathcal{D}}({\mathbf{k}})}S({\mathbf{k}},{\bm\kappa}){\rm{d}}k_x{\rm{d}}k_z
=\frac{2\pi k_0A_{s}^2(k_0)}{\gamma(\kappa_x,\kappa_z)},
\end{align}
with the last equality derived from \eqref{Angular_Domain_Power_Distribution_Isotropic_Scattering}. Therefore, we have
\begin{align}\label{Appendix_A_Result2}
{\mathbbmss{E}}\{{\hat{\mathsf{H}}_a({\bm\kappa})}{\hat{\mathsf{H}}_a^{*}({\bm\kappa}')}\}={\hat{S}}^{\frac{1}{2}}({\bm\kappa}){\hat{S}}^{\frac{1}{2}}({\bm\kappa}')\delta({\bm\kappa}-{\bm\kappa}').
\end{align}
By inserting \eqref{Appendix_A_Result2} into \eqref{Appendix_A_Result1} and applying the property of the Dirac delta function, the results in \eqref{Lemma_Autocorrelation_General_Result} follow immediately.

For the linear array, the autocorrelation function becomes
\begin{align}
R_{h_{{\mathsf{r}}_x}}(t_x,t_x')&=\frac{1}{(2\pi)^4}\int_{-k_0}^{k_0}\int_{-k_0}^{k_0}
{\mathbbmss{E}}\{\hat{\mathsf{H}}_{a_x}(\kappa_x)\hat{\mathsf{H}}_{a_x}^{*}(\kappa_x')\}\nonumber\\
&\times{\rm{e}}^{{\rm{j}}(t_x\kappa_x-t_x'\kappa_x')}{\rm{d}}\kappa_x{\rm{d}}\kappa_x'.\label{Appendix_A_Result4}
\end{align}
Similar to \eqref{Appendix_A_Result2}, we have
\begin{align}\label{Appendix_A_Result5}
{\mathbbmss{E}}\{\hat{\mathsf{H}}_{a_x}(\kappa_x)\hat{\mathsf{H}}_{a_x}^{*}(\kappa_x')\}
={\hat{S}}_x^{\frac{1}{2}}({\kappa}_x){\hat{S}}_x^{\frac{1}{2}}({\kappa}_x')\delta({\kappa}_x-{\kappa}_x'),
\end{align}
where
\begin{align}\label{Appendix_A_Result6}
{\hat{S}}_x({\kappa}_x)=\int_{-\sqrt{k_0^2-\kappa_x^2}}^{\sqrt{k_0^2-\kappa_x^2}}{\hat{S}}({\bm\kappa}){\rm{d}}\kappa_z
={2\pi k_0A_{s}^2(k_0)}\pi.
\end{align}
By substituting \eqref{Appendix_A_Result5} and \eqref{Appendix_A_Result6} into \eqref{Appendix_A_Result4}, the results in \eqref{Corollary_Autocorrelation_General_Linear_Result} can be immediately derived.
\subsection{Proof of Lemma \ref{Lemma_Linear_Random_Operator_Statistical_Equal}}\label{Proof_Lemma_Linear_Random_Operator_Statistical_Equal}
Since ${\overline{W}}(t_x')$ is a ZUCG random field over ${\mathcal{A}}_{{\mathsf{t}}_x}$, ${\overline{h}}_{{\mathsf{r}}_x}({{t}}_x)$ is a zero-mean Gaussian random field over ${\mathcal{A}}_{{\mathsf{t}}_x}$. Furthermore, 
\begin{equation}\nonumber
\begin{split}
&{\mathbbmss{E}}\{{\overline{h}}_{{\mathsf{r}}_x}({{t}}_x){\overline{h}}_{{\mathsf{r}}_x}^{*}({{t}}_x')\}
=\sum_{\ell_1,\ell_2=1}^{\infty}\iint_{{\mathcal{A}}_{{\mathsf{t}}_x}\times{\mathcal{A}}_{{\mathsf{t}}_x}}\sigma_{{\mathsf{r}}_x,\ell_1}^{\frac{1}{2}}\sigma_{{\mathsf{r}}_x,\ell_2}^{\frac{1}{2}}\phi_{{\mathsf{r}}_x,\ell_1}(t_x)\\
&\times\phi_{{\mathsf{r}}_x,\ell_2}^{*}({t}_x')\phi_{{\mathsf{r}}_x,\ell_1}^{*}(\hat{t}_x)
\phi_{{\mathsf{r}}_x,\ell_2}(\hat{t}_x'){\mathbbmss{E}}\{{\overline{W}}(\hat{t}_x){\overline{W}}^{*}(\hat{t}_x')\}{\rm{d}}\hat{t}_x
{\rm{d}}\hat{t}_x'.
\end{split}
\end{equation}
Using the fact that ${\mathbbmss{E}}\{{\overline{W}}(\hat{t}_x){\overline{W}}^{*}(\hat{t}_x')\}=\delta(\hat{t}_x-\hat{t}_x')$ and the property of the
Dirac delta function, we obtain
\begin{equation}
\begin{split}
&{\mathbbmss{E}}\{{\overline{h}}_{{\mathsf{r}}_x}({{t}}_x){\overline{h}}_{{\mathsf{r}}_x}^{*}({{t}}_x')\}
=\sum\nolimits_{\ell_1=1}^{\infty}\sum\nolimits_{\ell_2=1}^{\infty}\sigma_{{\mathsf{r}}_x,\ell_1}^{\frac{1}{2}}\sigma_{{\mathsf{r}}_x,\ell_2}^{\frac{1}{2}}\\
&\times\phi_{{\mathsf{r}}_x,\ell_1}(t_x)\phi_{{\mathsf{r}}_x,\ell_2}^{*}({t}_x')\int_{{\mathcal{A}}_{{\mathsf{t}}_x}}\phi_{{\mathsf{r}}_x,\ell_1}^{*}(\hat{t}_x)
\phi_{{\mathsf{r}}_x,\ell_2}(\hat{t}_x){\rm{d}}\hat{t}_x,
\end{split}
\end{equation}
which, together with \eqref{EVD_Linear_Random_Operator_Orthogonality}, yields
\begin{align}
{\mathbbmss{E}}\{{\overline{h}}_{{\mathsf{r}}_x}({{t}}_x){\overline{h}}_{{\mathsf{r}}_x}^{*}({{t}}_x')\}
&=\sum\nolimits_{\ell_1=1}^{\infty}\sigma_{{\mathsf{r}}_x,\ell_1}^{\frac{1}{2}}\sigma_{{\mathsf{r}}_x,\ell_1}^{\frac{1}{2}}\phi_{{\mathsf{r}}_x,\ell_1}(t_x)\phi_{{\mathsf{r}}_x,\ell_1}^{*}({t}_x')\nonumber\\
&=R_{h_{{\mathsf{r}}_x}}(t_x,t_x').
\end{align}
The above arguments imply that the zero-mean Gaussian random field ${\overline{h}}_{{\mathsf{r}}_x}({{t}}_x)$ has the same autocorrelation as the Gaussian random field $h_{{\mathsf{r}}_x}({{t}}_x)$. Therefore, ${\overline{h}}_{{\mathsf{r}}_x}({{t}}_x)\overset{d}{=}h_{{\mathsf{r}}_x}({{t}}_x)$, which completes the proof.
\subsection{Proof of Lemma \ref{Lemma_Linear_Random_Operator_Statistical_Equal_Cofficient}}\label{Proof_Lemma_Linear_Random_Operator_Statistical_Equal_Cofficient}
It follows from $\Phi_{{\mathsf{r}}_x,\ell}=\int_{{\mathcal{A}}_{{\mathsf{t}}_x}}\phi_{{\mathsf{r}}_x,\ell}^{*}(t_x'){\overline{W}}(t_x'){\rm{d}}t_x'$ that
\begin{equation}\label{Appendix_B_Result1}
\begin{split}
{\mathbbmss{E}}\{\Phi_{{\mathsf{r}}_x,\ell}\Phi_{{\mathsf{r}}_x,\ell'}^{*}\}&=
\iint_{{\mathcal{A}}_{{\mathsf{t}}_x}\times{\mathcal{A}}_{{\mathsf{t}}_x}}\phi_{{\mathsf{r}}_x,\ell}^{*}(t_x')\phi_{{\mathsf{r}}_x,\ell'}(\hat{t}_x')\\
&\times{\mathbbmss{E}}\{{\overline{W}}(t_x')
{\overline{W}}^{*}(\hat{t}_x')\}{\rm{d}}t_x'{\rm{d}}\hat{t}_x',
\end{split}
\end{equation}
which, together with the fact that ${\mathbbmss{E}}\{{\overline{W}}(\hat{t}_x){\overline{W}}^{*}(\hat{t}_x')\}=\delta(\hat{t}_x-\hat{t}_x')$ and \eqref{EVD_Linear_Random_Operator_Orthogonality}, yields ${\mathbbmss{E}}\{\Phi_{{\mathsf{r}}_x,\ell}\Phi_{{\mathsf{r}}_x,\ell'}^{*}\}=\delta_{\ell,\ell'}$. This result shows that $\Phi_{{\mathsf{r}}_x,\ell}\sim{\mathcal{CN}}(0,\delta_{l,l})\overset{d}{=}{\mathcal{CN}}(0,1)$. Moreover, since ${\mathbbmss{E}}\{\Phi_{{\mathsf{r}}_x,\ell}\Phi_{{\mathsf{r}}_x,\ell'}^{*}\}=\delta_{\ell,\ell'}=0$ for $\ell\ne \ell'$, it is found that $\Phi_{{\mathsf{r}}_x,\ell}$ and $\Phi_{{\mathsf{r}}_x,\ell'}$ are uncorrelated. Given that $\Phi_{{\mathsf{r}}_x,\ell}$ and $\Phi_{{\mathsf{r}}_x,\ell'}$ follow Gaussian distributions, their uncorrelation implies statistical independence. This completes the proof.
\subsection{MISO Theorem Proofs}\label{Proof_Theorem_MISO_OP_High_SNR_Asymptotic}
As $\overline{\gamma}\rightarrow\infty$, ${\hat{a}}_{\mathsf{r}}=\frac{2^R-1}{{\overline{\gamma}}}\rightarrow0$. Using the fact of $\lim_{x\rightarrow0}{\Upsilon(s,x)}\simeq\frac{x^s}{s}$ \cite[Eq. (8.354.1)]{gradshteyn2014table}, we derive
\begin{equation}\label{CDF_MISO_Asym_Simple}
\lim_{\overline{\gamma}\rightarrow\infty}F_{a_{\mathsf{r}}}({\hat{a}}_{\mathsf{r}})\simeq
\frac{\sigma_{{\mathsf{r}},\min}^{{\mathsf{DOF}}_{{\mathsf{r}}}}}{\prod_{\ell=1}^{{\mathsf{DOF}}_{{\mathsf{r}}}}\sigma_{{\mathsf{r}},\ell}}
\frac{\psi_0\left({{\hat{a}}_{\mathsf{r}}}/{\sigma_{{\mathsf{r}},\min}}\right)^{{\mathsf{DOF}}_{{\mathsf{r}}}}}{\Gamma({\mathsf{DOF}}_{{\mathsf{r}}}){\mathsf{DOF}}_{{\mathsf{r}}}}.
\end{equation}
Equation \eqref{MISO_OP_High_SNR_Asymptotic} follows directly from this result. We then proceed to analyze the ECC ${\mathcal{R}}_{\mathsf{r}}={\mathbbmss{E}}\{{\mathsf{C}}_{\mathsf{r}}\}$. The explicit expression in \eqref{MISO_ADR_Explicit} is obtained by substituting \eqref{PDF_Channel_Gain_MISO} into ${\mathcal{R}}_{\mathsf{r}}=\int_{0}^{\infty}\log_2(1+{\overline{\gamma}}x)f_{a_{\mathsf{r}}}(x){\rm{d}}x$ and calculating the resulting integral with the aid of \cite[Eq. (4.337.5)]{gradshteyn2014table}. Moreover, using the fact that $\lim_{x\rightarrow\infty}\frac{\log_2(1+ax)}{\log_2(ax)}=1$ ($a>0$), we get
\begin{align}
\lim_{\overline{\gamma}\rightarrow\infty}\frac{{\mathbbmss{E}}\{\log_2(1+\overline{\gamma}a_{\mathsf{r}})\}}
{{\mathbbmss{E}}\{\log_2(\overline{\gamma}a_{\mathsf{r}})}=1,
\end{align}
which implies that $\lim_{\overline{\gamma}\rightarrow\infty}{\mathcal{R}}_{\mathsf{r}}\simeq \log_2({\overline{\gamma}})+{\mathbbmss{E}}\{\log_2(a_{\mathsf{r}})\}$. By leveraging \cite[Eq. (4.352.1)]{gradshteyn2014table} to calculate the expectation ${\mathbbmss{E}}\{\log_2(a_{\mathsf{r}})\}$, we obtain \eqref{MISO_ADR_High_SNR_Asymptotic}. For the DMT analysis, we observe $\Pr({\mathsf{C}}_{\mathsf{r}}<r_{\mathsf{r}}\log_2(1+\overline{\gamma}))=F_{a_{\mathsf{r}}}\left(\frac{(1+\overline{\gamma})^{r_{\mathsf{r}}}-1}{\overline{\gamma}}\right)$. It follows that $\lim_{\overline{\gamma}\rightarrow\infty}\frac{(1+\overline{\gamma})^{r_{\mathsf{r}}}-1}{\overline{\gamma}}\simeq\frac{1}{\overline{\gamma}^{1-r_{\mathsf{r}}}}\rightarrow0$ for $r_{\mathsf{r}}\in(0,1)$. By using the asymptotic behaviors of $F_{\lVert{\mathbf{h}}_{\mathsf{t}}\rVert^2}(x)$ at $x\rightarrow0^{+}$, i.e., \eqref{CDF_MISO_Asym_Simple}, the final results follow immediately. 
\subsection{MIMO Theorem Proofs}\label{Proof_MIMO_Theorem}
When ${\mathsf{D}}_{\mathsf{r}}={\mathsf{D}}_{\mathsf{t}}$, we have
\begin{align}
\lim_{{\overline{\gamma}}\rightarrow\infty}{\mathbbmss{E}}\{{\mathsf{C}}_{\mathsf{mm}}\}&
\simeq{\mathbbmss{E}}\{\log_2\det({\overline{\gamma}}
{\mathbf{R}}^{\frac{1}{2}}{\tilde{\mathbf{H}}}
{\mathbf{T}}{\tilde{\mathbf{H}}}^{\mathsf{H}}{\mathbf{R}}^{\frac{1}{2}})\}.
\end{align}
For two square matrices $\mathbf{A}_1$ and $\mathbf{A}_2$, it holds that $\det(\mathbf{A}_1\mathbf{A}_2)=\det(\mathbf{A}_1)\det(\mathbf{A}_2)$. As a result, we have
\begin{align}
\det({\mathbf{R}}^{\frac{1}{2}}{\tilde{\mathbf{H}}}
{\mathbf{T}}{\tilde{\mathbf{H}}}^{\mathsf{H}}{\mathbf{R}}^{\frac{1}{2}})=\det({\mathbf{R}})\det({\mathbf{T}})\det({\tilde{\mathbf{H}}}^{\mathsf{H}}{\tilde{\mathbf{H}}}),
\end{align}
which, together with \cite[Eq. (2.12)]{tulino2004random}, yields \eqref{Result1_Theorem_MIMO_CAPA_Multiplexing}. We then consider the case of ${\mathsf{D}}_{\mathsf{t}}<{\mathsf{D}}_{\mathsf{r}}$, which yields
\begin{align}
\lim_{{\overline{\gamma}}\rightarrow\infty}{\mathbbmss{E}}\{{\mathsf{C}}_{\mathsf{mm}}\}&
\simeq{\mathbbmss{E}}\{\log_2\det({\overline{\gamma}}
{\mathbf{T}}^{\frac{1}{2}}{\tilde{\mathbf{H}}}^{\mathsf{H}}{\mathbf{R}}{\tilde{\mathbf{H}}}{\mathbf{T}}^{\frac{1}{2}})\}.
\end{align}
Note that $\det({\mathbf{T}}^{\frac{1}{2}}{\tilde{\mathbf{H}}}^{\mathsf{H}}{\mathbf{R}}{\tilde{\mathbf{H}}}{\mathbf{T}}^{\frac{1}{2}})=\det({\mathbf{T}})
\det({\tilde{\mathbf{H}}}^{\mathsf{H}}{\mathbf{R}}{\tilde{\mathbf{H}}})$ and
\begin{align}\label{Det_Equality}
{\tilde{\mathbf{H}}}^{\mathsf{H}}(\sigma_{\circ}^2{\mathbf{I}}_{{\mathsf{D}}_{\mathsf{r}}}){\tilde{\mathbf{H}}}\succeq{\tilde{\mathbf{H}}}^{\mathsf{H}}{\mathbf{R}}{\tilde{\mathbf{H}}}\succeq
{\tilde{\mathbf{H}}}^{\mathsf{H}}(\sigma_{\diamond}^2{\mathbf{I}}_{{\mathsf{D}}_{\mathsf{r}}}){\tilde{\mathbf{H}}}.
\end{align}
Since $\det(\cdot)$ is an increasing function over the cone of positive-definite Hermitian matrices, it follows that for $\mathbf{A}_1\succeq{\mathbf{0}}$ and $\mathbf{A}_2\succeq{\mathbf{0}}$, then ${\mathbf{A}}_1\succeq {\mathbf{A}}_2\Rightarrow\det({\mathbf{A}}_1)\geq\det({\mathbf{A}}_2)$. Using this property along with \cite[Eq. (2.12)]{tulino2004random}, we arrive at \eqref{Result2_Theorem_MIMO_CAPA_Multiplexing} for ${\mathsf{D}}_{\mathsf{t}}<{\mathsf{D}}_{\mathsf{r}}$, which also applies to the case where ${\mathsf{D}}_{\mathsf{t}}>{\mathsf{D}}_{\mathsf{r}}$. 

We now analyze the high-SNR OP and DMT. The high-SNR asymptotic OP of a Kronecker MIMO channel is characterized in \cite[Eq. (33)]{yang2020asymptotic}, from which \eqref{Result_Theorem_MIMO_CAPA_Diversity} is derived. For DMT, we first consider the case of ${\mathsf{D}}_{\mathsf{t}}\leq{\mathsf{D}}_{\mathsf{r}}$. It follows from \eqref{Det_Equality} that
\begin{subequations}
\begin{align}
{\mathsf{C}}_{\mathsf{mm}}&\leq \log_2\det({\mathbf{I}}_{{\mathsf{D}}_{\mathsf{t}}}+{\overline{\gamma}}\sigma_{\circ}^2
{\mathbf{T}}^{\frac{1}{2}}{\tilde{\mathbf{H}}}^{\mathsf{H}}{\tilde{\mathbf{H}}}{\mathbf{T}}^{\frac{1}{2}})\\
&=\log_2\det({\mathbf{I}}_{{\mathsf{D}}_{\mathsf{r}}}+{\overline{\gamma}}\sigma_{\circ}^2
{\tilde{\mathbf{H}}}{\mathbf{T}}{\tilde{\mathbf{H}}}^{\mathsf{H}})\\
&=\log_2\det({\mathbf{I}}_{{\mathsf{D}}_{\mathsf{r}}}+(\max\nolimits_{j}\varrho_{{\mathsf{t}},j}^2){\overline{\gamma}}\sigma_{\circ}^2
{\tilde{\mathbf{H}}}{\tilde{\mathbf{H}}}^{\mathsf{H}}).
\end{align}
\end{subequations}
Similarly, we can show that ${\mathsf{C}}_{\mathsf{mm}}\geq\log_2\det({\mathbf{I}}_{{\mathsf{D}}_{\mathsf{r}}}+(\min\nolimits_{j}\varrho_{{\mathsf{t}},j}^2){\overline{\gamma}}\sigma_{\diamond}^2
{\tilde{\mathbf{H}}}{\tilde{\mathbf{H}}}^{\mathsf{H}})$. Taken together, the high-SNR behavior of the considered system aligns with that of an i.i.d. Rayleigh MIMO channel, as described in \cite{zheng2003diversity} and \cite{ordonez2012array}. When $\varrho_{{\mathsf{r}},i}=\varrho_{{\mathsf{r}}}$ ($\forall i$) and $\varrho_{{\mathsf{t}},j}=\varrho_{{\mathsf{t}}}$ ($\forall j$), the channel reduces to an i.i.d. Rayleigh MIMO channel whose DMT and array gain are given in \cite{zheng2003diversity} and \cite{ordonez2012array}, respectively. These derivations extend directly to the case where ${\mathsf{D}}_{\mathsf{t}}>{\mathsf{D}}_{\mathsf{r}}$, leading to the final results.
\end{appendix}
\bibliographystyle{IEEEtran}
\bibliography{mybib}
\end{document}